\documentclass[aps, prd, twocolumn, amsmath, floats,floatfix, superscriptaddress, nofootinbib]{revtex4}

\usepackage{amssymb}
\usepackage{amsmath}
\usepackage{verbatim}
\usepackage{mathrsfs}
\usepackage{amsfonts}
\usepackage{latexsym}
\usepackage{epsfig}
\usepackage{color}
\usepackage{graphicx,subfigure}
\usepackage{units}
\usepackage{slashbox}
\usepackage{natbib}
\usepackage[normalem]{ulem}
\usepackage{amssymb}

\begin{document}


\def\CovDev{D}
\def\Res{{\mathcal R}}
\def\Gammaflat{\hat \Gamma}
\def\metricflat{\hat \gamma}
\def\Dflat{\hat {\mathcal D}}
\def\part_n{\partial_\perp}

\def\Lie{\mathcal{L}}
\def\A{\mathcal{X}}
\def\Aphi{\A_{\phi}}
\def\hAphi{\hat{\A}_{\phi}}
\def\E{\mathcal{E}}
\def\Ham{\mathcal{H}}
\def\M{\mathcal{M}}
\def\R{\mathcal{R}}
\def\p{\partial}

\def\hg{\hat{\gamma}}
\def\hA{\hat{A}}
\def\hD{\hat{D}}
\def\hE{\hat{E}}
\def\hR{\hat{R}}
\def\hcA{\hat{\mathcal{A}}}
\def\hDelt{\hat{\triangle}}

\def\na{\nabla}
\def\dif{{\rm{d}}}
\def\non{\nonumber}
\newcommand{\erf}{\textrm{erf}}

\renewcommand{\t}{\times}
\newcommand{\jc}[1]{{\textcolor{red}{[JCB: #1]}}}
\newcommand{\alert}[1]{{\textcolor{red}{#1}}}

\long\def\symbolfootnote[#1]#2{\begingroup%
\def\thefootnote{\fnsymbol{footnote}}\footnote[#1]{#2}\endgroup}


\title{Impact of the wave-like nature of Proca stars on their gravitational-wave emission
}

  \author{Nicolas Sanchis-Gual}
	\affiliation{Departamento de Astronom\'{i}a y Astrof\'{i}sica, Universitat de Val\`{e}ncia,
Dr. Moliner 50, 46100, Burjassot (Val\`{e}ncia), Spain}
	\affiliation{Departamento  de  Matem\'{a}tica  da  Universidade  de  Aveiro  and  Centre  for  Research  and  Development in  Mathematics  and  Applications  (CIDMA),  Campus  de  Santiago,  3810-183  Aveiro,  Portugal}
 \author{Juan Calder\'on~Bustillo}
	\affiliation{Instituto Galego de F\'{i}sica de Altas Enerx\'{i}as, Universidade de
Santiago de Compostela, 15782 Santiago de Compostela, Galicia, Spain}
	\affiliation{Department of Physics, The Chinese University of Hong Kong, Shatin, N.T., Hong Kong}
\author{Carlos Herdeiro} 
		\affiliation{Departamento  de  Matem\'{a}tica  da  Universidade  de  Aveiro  and  Centre  for  Research  and  Development in  Mathematics  and  Applications  (CIDMA),  Campus  de  Santiago,  3810-183  Aveiro,  Portugal}
\author{Eugen Radu}
	\affiliation{Departamento  de  Matem\'{a}tica  da  Universidade  de  Aveiro  and  Centre  for  Research  and  Development in  Mathematics  and  Applications  (CIDMA),  Campus  de  Santiago,  3810-183  Aveiro,  Portugal}
\author{Jos\'e A. Font}
	\affiliation{Departamento de Astronom\'{i}a y Astrof\'{i}sica, Universitat de Val\`{e}ncia,
Dr. Moliner 50, 46100, Burjassot (Val\`{e}ncia), Spain}
	\affiliation{Observatori Astron\`{o}mic, Universitat de Val\`{e}ncia,
C/ Catedr\'{a}tico Jos\'{e} Beltr\'{a}n 2, 46980, Paterna (Val\`{e}ncia), Spain}  

\author{Samson H. W. Leong}
\affiliation{Department of Physics, The Chinese University of Hong Kong, Shatin, N.T., Hong Kong}
\author{Alejandro Torres-Forn\'e}
\affiliation{Departamento de Astronom\'{i}a y Astrof\'{i}sica, Universitat de Val\`{e}ncia,
Dr. Moliner 50, 46100, Burjassot (Val\`{e}ncia), Spain}


\date{\today}


\begin{abstract} 
We present a systematic study of the dynamics and gravitational-wave emission of head-on collisions of spinning vector boson stars, known as Proca stars. To this aim we build a catalogue of about 800 numerical-relativity simulations of such systems. We find that the wave-like nature of bosonic stars has a large impact on the gravitational-wave emission. In particular, we show that the initial relative phase $\Delta \epsilon =\epsilon_1-\epsilon_2$ of the two complex fields forming the stars (or equivalently, the relative phase at merger) strongly impacts both the emitted gravitational-wave energy and the corresponding mode structure. This leads to a non-monotonic dependence of the emission on the frequency of the secondary star $\omega_2$, for fixed frequency  $\omega_1$ of the primary. This phenomenology, which has not been found for the case of black-hole mergers, reflects the distinct ability of the Proca field to interact with itself in both constructive and destructive manners. We postulate this may serve as a smoking gun to shed light on the possible existence of these objects.
\end{abstract}



\maketitle


\section{Introduction}
 
Gravitational waves (GWs) provide information about the strong-field regime of gravity and can potentially reveal the true nature and structure of astrophysical compact objects. Their analysis could help unveil the classical and quantum essence of black holes, as well as the interior of neutron stars through the dense-matter equation of state, a long-term open issue. Moreover, theoretical proposals for dark or ``exotic'' compact objects (ECOs)~\cite{Cardoso:2019} could be probed through the study of their GW signals as long as those could be distinguished from the signals produced by black holes and neutron stars. Such investigations require a deep understanding of the emitted GWs and, in particular, rely on theoretical waveform templates against which observational data can be compared. As an example, the detection of GWs from compact binary coalescences -- the sources so far observed by Advanced LIGO and Advanced Virgo~\cite{Abbott2016,TheLIGOScientific:2017qsa,Abbott:2017oio,LIGOScientific:2018mvr,Abbott:2020niy,abbott2020gw190521,abbott2021gwtc} -- and the source parameter inference thereof, rely on the matched filtering of the data to waveform templates (or approximants). This makes the production of waveform catalogues of {physically} motivated exotic compact objects an endeavour both well timed and worth-pushing. 

Amongst all proposed exotic objects that can reach a compactness comparable to that of black holes, bosonic stars stand out as one of the simplest and best-motivated models~\cite{Kaup:1968zz,Ruffini:1969qy}. Bosonic stars with masses in the astrophysical black hole range, from stellar-origin to supermassive objects, are made of ultralight fundamental bosonic fields that could account for (part of) dark matter. Triggered by this central open issue in theoretical physics -- the nature of dark matter -- the study of bosonic stars has earned quite some attention in recent years. From a particle physics perspective,  ultralight bosonic particles can emerge in the string axiverse~\cite{Arvanitaki:2009fg,Arvanitaki:2010sy} or in simple extensions of the Standard Model of particles~\cite{freitas2021ultralight}. Bosonic stars are asymptotically flat (although non-asymptotically flat generalizations exist), stationary and solitonic, $i.e.$ horizonless and everywhere regular equilibrium spacetime geometries, describing self-gravitating lumps of bosonic particles. In their simplest guise, they emerge by minimally coupling the complex, massive Klein-Gordon equation -- for scalar boson stars -- or the complex Proca equations  -- for vector boson stars, \textit{aka} Proca stars (PSs)~\cite{Brito:2015pxa} -- to Einstein's gravity.

Bosonic stars can be either static, in which case the simplest solutions are spherically symmetric (but see also~\cite{Herdeiro:2020kvf,sanchis2021multifield}) or spinning~\cite{Herdeiro:2019mbz} (thus stationary but non-static), in which case they have a non-spherical morphology which depends on the scalar or vector model. In all cases, the bosonic field oscillates periodically at a well-defined frequency $\omega$, which determines the mass, angular momentum (in spinning solutions) and compactness of the star.

The dynamical robustness of bosonic stars has been established for some models in well-identified regions of the parameter space (see~\cite{liebling2017dynamical}  for a review) making them viable dark-matter candidates. The case of non-spinning spherically symmetric bosonic stars is firmly established. The fundamental solutions (those with the minimum number of nodes of the bosonic field across the star) are perturbatively stable in a range of frequencies between the Newtonian limit (where they become non-compact) and the maximal-mass solution. Additionally, they exhibit a non-fine-tuned dynamical formation mechanism known as gravitational cooling~\cite{seidel1994formation,di2018dynamical}. On the contrary, the case of spinning bosonic stars has shown to be more subtle~\cite{sanchis2019nonlinear}. In particular, while the fundamental PS solutions have been found to be stable in the simplest model where the Proca field has only a mass term (no self-interactions), scalar boson stars are prone to non-axisymmetric perturbations that can trigger the development of instabilities akin to the bar-mode instability found in neutron stars~\cite{DiGiovanni:2020ror}, in the corresponding model without self-interactions~\cite{Siemonsen:2020hcg}.

The above findings support using the fundamental solutions of the simplest Proca model as a robust starting point to test the true nature of dark compact objects. In particular, this model appears as the most suitable choice to conduct dynamical studies aimed at gauging, through GW information, the potential astrophysical significance, if any, of an appealing ECO model. First, and promising, steps have recently been taken. Pursuing this route~\cite{,CalderonBustillo:2020srq}  found that waveforms from numerical-relativity simulations of head-on collisions of PSs can fit the signal GW190521 as good as those from quasi-circular binary-black-hole (BBH) mergers, even being slightly preferred from a Bayesian-statistics viewpoint. Moreover, the development of a larger numerical catalogue of PS mergers together with new data-analysis techniques \cite{psi4_tech}, have led to a more systematic study of several LIGO-Virgo-KAGRA (LVK) high-mass events in O3 under the PS collision scenario \cite{CalderonBustillo:2020srq} and to conduct the first population studies of these objects \cite{psi4_obs}.

The present paper complements those recent works. Here, we report on our catalogue of nearly 800 numerical-relativity simulations of head-on collisions of PSs  used to obtain the results presented in \cite{CalderonBustillo:2020srq,psi4_tech,psi4_obs}. Furthermore, we discuss additional numerical simulations we carried out to explore the impact of the wave-like nature of PSs in their GW emission. We find that the emission at merger dramatically depends on the relative phase of the complex field of each star. This has a major impact in both the net energy emission through GWs and the corresponding mode structure. Since this relative phase is an intrinsic parameter of PSs, absent in BBH mergers, the potential measurement of the GW modulation discussed in this work could serve as a smoking gun for the existence of PSs.

The remaining of this paper is  organized as follows. Section \ref{sec2} briefly describes the formalism needed to perform numerical simulations of PS mergers. The procedure we follow to obtain initial data for the simulations is outlined in Section \ref{sec3} as well as the specific numerical setups employed. We report and analyze our results in Section \ref{results}. Finally, our conclusions are presented in Section~\ref{Sect:conclusions} along with some remarks on possible pathways for future research. Henceforth, units with $G=c=1$ are used.

\section{Formalism}
\label{sec2}
We investigate the dynamics of a complex Proca field by solving numerically the Einstein-(complex, massive) Proca system, described by the action $\mathcal{S}=\int d^4x \sqrt{-g}\mathcal{L}$, where the Lagrangian density depends on the Proca potential $\mathcal{A}$ and field strength $\mathcal{F}=d\mathcal{A}$. It reads
\begin{equation}
\mathcal{L}=\frac{R}{16\pi}-\frac{1}{4}\mathcal{F}_{\alpha\beta}\bar{\mathcal 
{F}}^{\alpha\beta}-\frac{1 } {2}\mu^2\mathcal{A}_\alpha\bar{\mathcal{A}}^\alpha 
\ .
\label{model}
\end{equation}
Above, the bar denotes complex conjugation, $R$ is the Ricci scalar, and $\mu$ is the Proca-field mass. The stress-energy tensor of the Proca field is given by
\begin{eqnarray} \label{tmunu}
T_{\alpha\beta}&=& -\mathcal{F}_{\mu(\alpha}  \bar {\mathcal{F}}_{\beta)}^{\,\,\mu}-\frac{1}{4}g_{\alpha\beta}\mathcal{F}_{\mu\nu}\bar{\mathcal{F}}^{\mu\nu} 
\nonumber \\
&+& \mu^2 \left[
\mathcal{A}_{(\alpha}\bar{\mathcal{A}}_{\beta)}-\frac{1}{2}g_{\alpha\beta}\mathcal{A}_\mu\bar{\mathcal{A}}^{\mu}
\right]\, ,
\end{eqnarray}
where $g_{\alpha\beta}$ is the spacetime metric, with $g=\det g_{\alpha\beta}$, and the parenthesis denotes index symmetrization.
Using the standard 3+1 split (see $e.g.$~\cite{sanchis2017numerical}  for details) the Proca field is split into the following 3+1 quantities:
\begin{eqnarray}
\mathcal{A}_{\mu}&=&\mathcal{X}_{\mu}+n_{\mu}\mathcal{X}_{\phi},\\
\mathcal{X}_{i}&=&\gamma^{\mu}_{\,i}\mathcal{A}_{\mu},\\
\mathcal{X}_{\phi}&=&-n^{\mu}\mathcal{A}_{\mu},
\end{eqnarray}
where $n^{\mu}$ is the timelike unit vector, $\gamma^{\mu}_{\nu}=\delta^{\mu}_{\nu}+n^{\mu}n_{\nu}$ is the operator projecting spacetime quantities onto the spatial hypersurfaces, $\mathcal{X}_{i}$ is the vector potential, and $\mathcal{X}_{\phi}$ is the scalar potential.  
The fully non-linear Einstein-Proca system can be written as~\cite{sanchis2017numerical}:
\begin{eqnarray}
\label{eq:dtgamma}
\p_{t} \gamma_{ij} & = & - 2 \alpha K_{ij} + \Lie_{\beta} \gamma_{ij}
,\\
\label{eq:dtAi}
\p_{t} \A_{i}      & = & - \alpha \left( E_{i} + D_{i} \Aphi \right) - \Aphi D_{i}\alpha + \Lie_{\beta} \A_{i}
,\\
\label{eq:dtE}
\p_{t} E^{i}       & = &
        \alpha \left( K E^{i} + D^{i} Z + \mu^2\A^{i}
                + \epsilon^{ijk} D_{j} B_{k} \right)\nonumber\\ &&
        - \epsilon^{ijk} B_{j} D_{k}\alpha
        + \Lie_{\beta} E^{i},\\
\label{eq:dtKij}
\p_{t} K_{ij}      & = & - D_{i} D_{j} \alpha
        + \alpha \left( R_{ij} - 2 K_{ik} K^{k}{}_{j} + K K_{ij} \right)
\nonumber \\ & &
        + 2 \alpha \biggl( E_{i} E_{j} - \frac{1}{2} \gamma_{ij} E^{k} E_{k} 
        + B_{i} B_{j} \nonumber\\ &&
          - \frac{1}{2} \gamma_{ij} B^{k} B_{k} - \mu^{2} \A_{i} \A_{j}
          \biggl) + \Lie_{\beta} K_{ij} ,\\
\label{eq:dtAphi}
\p_{t} \Aphi  & = & - \A^{i} D_{i} \alpha
        + \alpha \left( K \Aphi - D_{i} \A^{i} - Z \right)\nonumber\\&&
        + \Lie_{\beta} \Aphi ,\\
\label{eq:dtZ}
\p_{t} Z          & = & \alpha \left( D_{i} E^{i} + \mu^{2} \Aphi - \kappa Z \right)
        + \Lie_{\beta} Z\,,
\end{eqnarray}
where $\alpha$ is the lapse function, $\beta$ is the shift vector, $\gamma_{ij}$ is the spatial metric, $K_{ij}$ is the extrinsic curvature (with $K=K^{i}_{\,\,i}$), $D_i$ is the covariant 3-derivative,  $\Lie_{\beta}$ is the Lie derivative (along the shift-vector direction), and $\kappa$ is a damping parameter that helps stabilize the numerical evolution. Moreover, the three-dimensional ``electric" $E^{i}$ and ``magnetic" $B^{i}$
fields are also introduced in the previous equations in analogy with Maxwell's theory:
\begin{equation}
E_{i}=\gamma^{\mu}_{i}\mathcal{F}_{\mu\nu}n^{\nu},\quad B_{i}=\gamma^{\mu}_{i}\star\mathcal{F}_{\mu\nu}n^{\nu}=\epsilon^{ijk}D_{i}\mathcal{X}_{k},
\end{equation}
with $E_{\mu}n^{\mu}=B_{\mu}n^{\mu}=0$ and $\epsilon^{ijk}$ the three-dimensional Levi-Civita tensor. The system of equations is closed by two
constraint equations, namely, the Hamiltonian constraint
and the momentum constraint, which  are given by:
\begin{eqnarray}
\label{eq:Hamiltonian}
\Ham  &=& R - K_{ij} K^{ij} + K^2 
        - 2 \bigl( E^{i} E_{i} + B^{i} B_{i}+ \nonumber\\
        && \mu^{2} \bigl( \Aphi^{2} 
        + \A^{i} \A_{i} \bigl)
            \bigl)= 0\,,\\
\label{eq:momentumConstraint}
\M_{i}  &=& D^{j} K_{ij} - D_{i} K 
        - 2 \bigl( \epsilon_{ijk} E^{j} B^{k} +\nonumber\\
         &&\mu^{2} \Aphi \A_{i}
            \bigl)
       = 0
\,.
\end{eqnarray}

\section{Initial Data and numerics}
\label{sec3}

\subsection{The stationary PS solutions}
Following the conventions in~\cite{Brito:2015pxa},
we consider an axially symmetric and stationary line element
\begin{eqnarray}
\label{ansatz}
ds^2=&-&e^{2F_0} dt^2+e^{2F_1}\left(dr^2+r^2 d\theta^2\right)\nonumber\\
&+&e^{2F_2}r^2 \sin^2\theta \left(d\varphi-\frac{W}{r} dt\right)^2\ , \qquad 
\end{eqnarray} 
where $F_0,F_1,F_2$, and $W $ are functions of $(r,\theta)$.
Here, $r,\theta,\varphi$
can be taken as spherical coordinates (in fact spheroidal), with the usual range,
while $t$ is the time coordinate. The spinning PS solutions of the Einstein--Proca system have been discussed in
\cite{Brito:2015pxa} with these conventions
and $e.g.$ in~\cite{Herdeiro:2016tmi}
for a slightly different version of (\ref{ansatz}) with $W/r\to W$.

\begin{figure}[t!]
\includegraphics[width=0.85\linewidth]{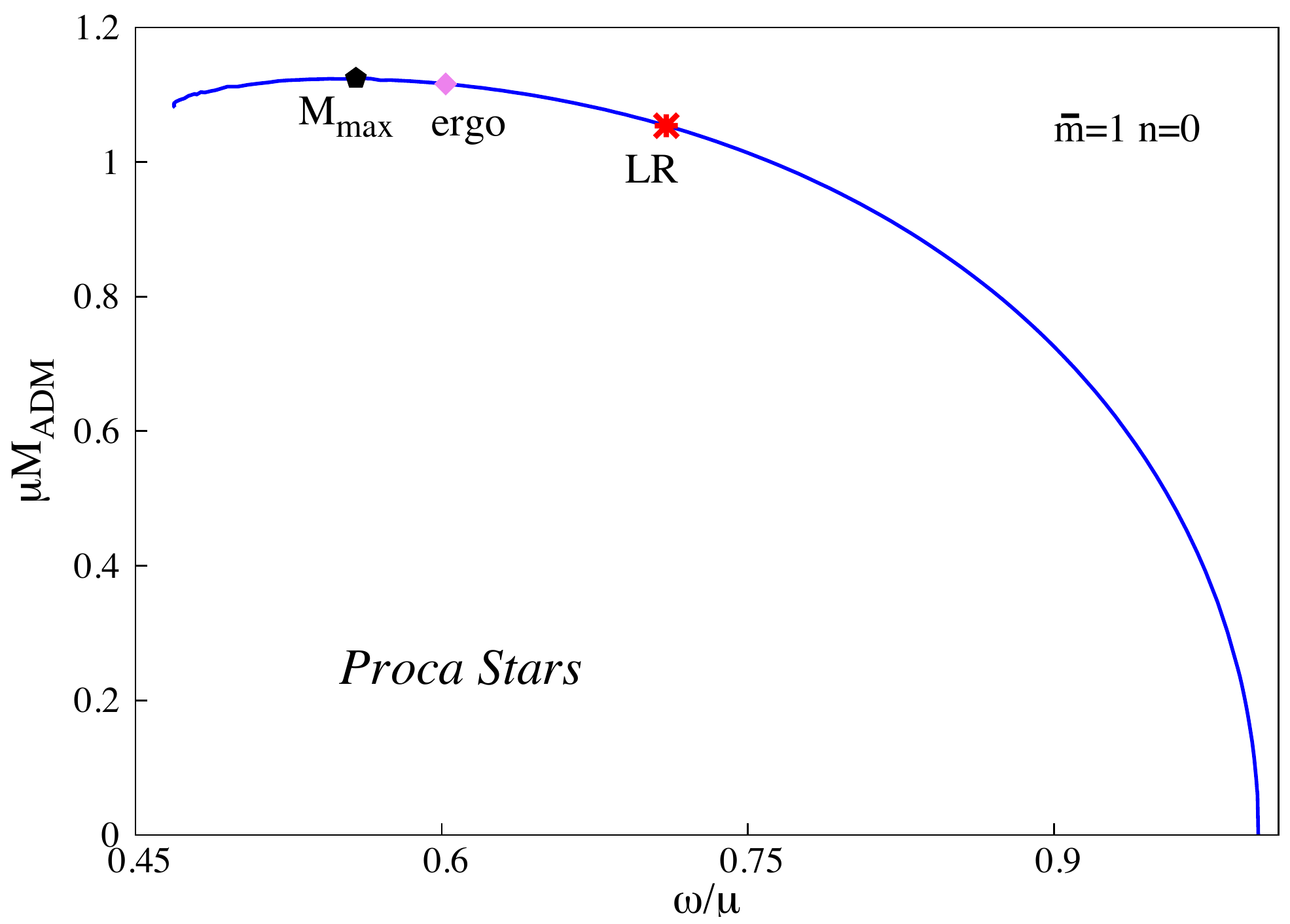}
\includegraphics[width=0.9\linewidth]{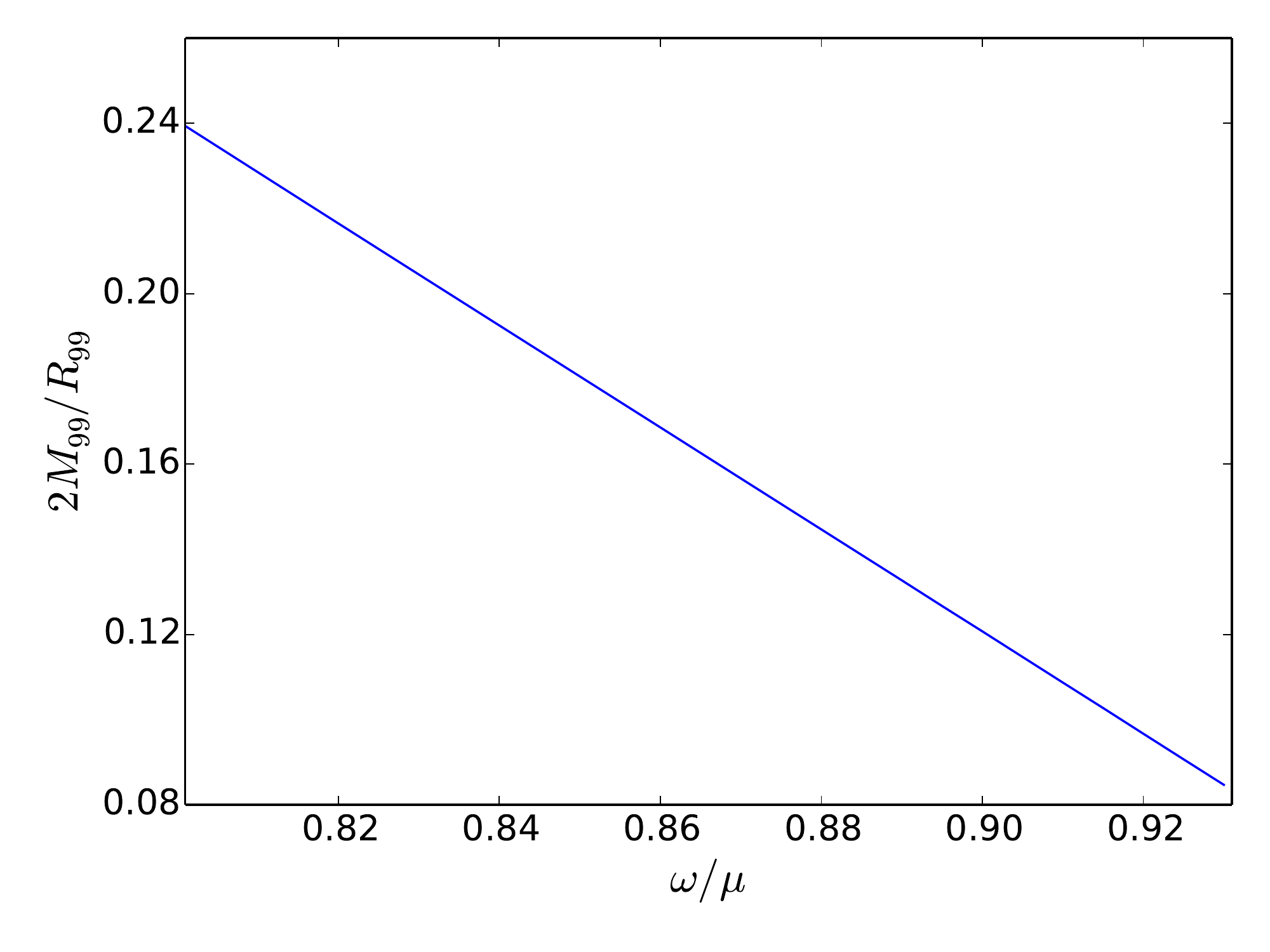}
\caption{Top panel: Sequence of equilibrium configurations of nodeless fundamental spinning $\bar{m}=1$ PSs. The PSs develop a pair of light rings for $\omega/\mu\lesssim 0.711$ and an ergo-region for $\omega/\mu\lesssim0.602$. The maximal mass is attained at $\omega/\mu\simeq 0.562$. Bottom panel: compactness of the Proca stars as function of the oscillation frequency $\omega/\mu$ in the range considered in this study.} 
\label{fig0}
\end{figure}

The ansatz for the Proca field is:
\begin{eqnarray}
\label{Paxial}
\mathcal{A}=\left(
\frac{H_1}{r}dr+H_2d\theta+i H_3 \sin \theta d\varphi + iVdt   
\right)
e^{i(\bar m\varphi-\omega t+\epsilon)},   \nonumber
\\
\end{eqnarray}
 with $\bar m\in \mathbb{Z}^+$ and $\epsilon$ is the initial phase of the star. The domain of existence and the compactness of the solutions of the Einstein-Proca equations describing the fundamental spinning PSs are shown in Fig.~\ref{fig0}. These solutions have $\bar{m}=1$ and are nodeless ($i.e.$ $\mathcal{A}_0$ has no nodes). The frequency range of the solutions of interest varies between $\omega/\mu=1$ (Newtonian limit) and $\omega/\mu\sim0.562$ (maximal-mass solution). As the latter is approached, the PS solutions become ultra-compact, $i.e.$ they develop a light ring \textit{pair}~\cite{cunha2017light} for $\omega/\mu\lesssim 0.711$. This creates a spacetime instability~\cite{Cunha:2022gde} which motivates us to avoid this region of the parameter space.  The compactness is defined as
 \begin{equation}
 \text{Compactness} = \frac{2M_{99}}{R_{99}}
 \end{equation}
 where $R_{99}$ is the perimetral radius that contains 99\% of the star’s mass, $M_{99}$. Bosonic stars do not have a surface with a discontinuity of the energy density occurs, { {\it i.e.} a surface outside which the energy density is zero (in contrast with a fluid star)}.
 We remark that for all PS solutions reported in the literature so far,
 the line element  (\ref{ansatz}) 
possesses a reflection symmetry, with respect to the $\theta=\pi/2$ plane.

The translation between the functions above, $F_0$, $F_1$, $F_2$, $W$, $V$, $H_1$, $H_2$, $H_3$, and 
the initial value for the metric and the 3+1 Proca field variables is given as follows:
\begin{eqnarray}
\alpha &=& e^{F_{0}}\ , \beta^{\varphi}=\frac{W}{r},\\
\gamma_{rr}&=&e^{2F_{1}},\gamma_{\theta\theta}=e^{2F_{1}}\,r^{2},\gamma_{\phi\phi}=e^{2F_{2}}\,r^{2}\sin^{2}\theta,\\
\mathcal{X}_{\phi}&=&-n^{\mu}\mathcal{A}_{\mu}\  , \label{propot}\\
\mathcal{X}_{i}&=&\gamma^{\mu}_{i}\mathcal{A}_{\mu}\ ,\\
E^{i}&=&-i\,\frac{\gamma^{ij}}{\alpha}\,\biggl(D_{j} (\alpha\mathcal{X}_{\phi})+\partial_{t}\mathcal{X}_{j}\biggl) \ .
\end{eqnarray}

\subsection{Binary head-on data}

\begin{figure}[t!]
\includegraphics[width=1\linewidth]{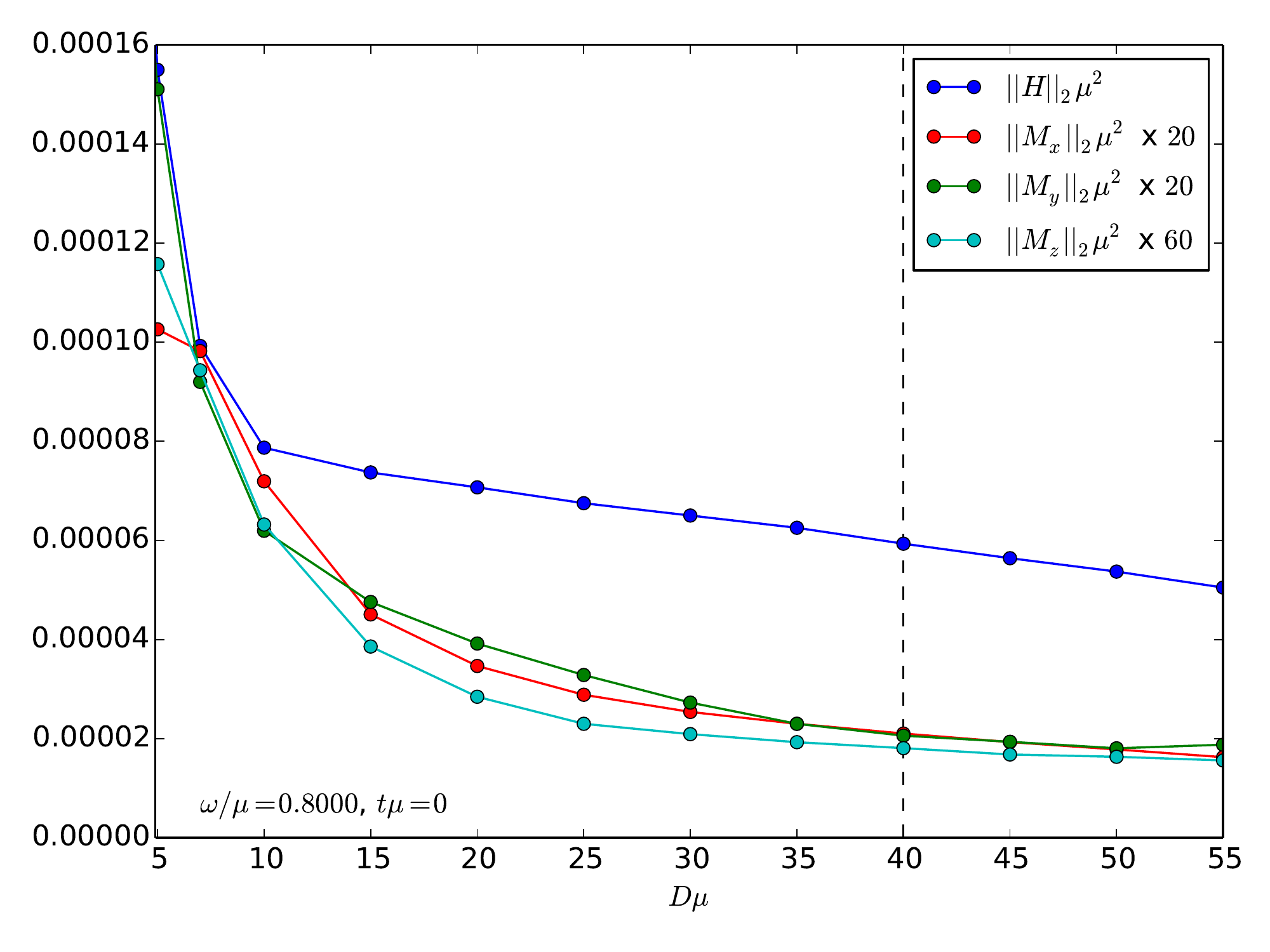}
\caption{Hamiltonian and momentum constraint violations of the initial data as function of the distance $D\mu$ for the equal-mass model with $\omega/\mu=0.8000$. The vertical black dashed line corresponds to our choice of the initial separation $D\mu=40$ for our set of simulations, in the roughly constant region of the $L_2$-norm.}
\label{figConstraints}
\end{figure}

As initial data for the head-on simulations we consider a superposition of two PSs with both stars described by the same Proca field following~\cite{palenzuela2007head, bezares2017final,sanchis2019head,CalderonBustillo:2020srq,jaramillo2022head,Bezares:2022obu,Sanchis-Gual:2022zsr} (see also~\cite{helfer2022malaise,croft2022gravitational}):
\begin{itemize}
\item $\mathcal{A}(x_{i}) = \mathcal{A}^{(1)}(x_{i}-x_{0}) + \mathcal{A}^{(2)}(x_{i}+x_{0})$,
\item $\gamma_{ij}(x_{i}) = \gamma_{ij}^{(1)}(x_{i}-x_{0}) + \gamma_{ij}^{(2)}(x_{i}+x_{0})-\gamma_{ij}^{\rm flat}(x_{i}),$
\item $\alpha(x_{i}) = \alpha^{(1)}(x_{i}-x_{0}) + \alpha^{(2)}(x_{i}+x_{0}) - 1$,
\end{itemize}
where superscripts $(1)$ and $(2)$ label the stars and $\pm x_{0}$ indicates their initial positions. The stars are initially separated by a coordinate distance $D\mu = \Delta x \mu =  40$ ($x_{0}\mu=\pm20$). We note that the solutions are not boosted and that these initial data introduce (small) constraint violations~\cite{palenzuela2007head}. Figure \ref{figConstraints} shows the dependence of the $L_{2}$-norm of the Hamiltonian and momentum constraints, Eqs.~(\ref{eq:Hamiltonian}) and (\ref{eq:momentumConstraint}), with $D$, at the initial time. The values of the $L_{2}$-norm are $\mathcal{O}(10^{-4})$ or better. The error decreases with separation, reaching a fairly constant value for $D\mu\gtrsim 20$ (particularly visible for the momentum constraint), see~\cite{palenzuela2008orbital}. 

Each star is defined by its oscillation frequency, $\omega_1/\mu$ and $\omega_2/\mu$. For the initial catalogue used in \cite{psi4_obs}, comprising $\sim 800$ initial models, we fix the phase difference $\Delta \epsilon$ between the stars to zero. Here, we also explore the impact of varying this relative phase on the gravitational-wave emission. Equal-mass cases correspond to $\omega_1=\omega_2=\omega$. Correspondingly, $\omega_1\neq\omega_2$ for unequal-mass binaries. Moreover, since we assume there is a single Proca field describing both stars, these also share a common value for the boson mass $\mu$. 

\subsection{Parameter space}

\begin{figure}[t!]
\includegraphics[width=0.99\linewidth]{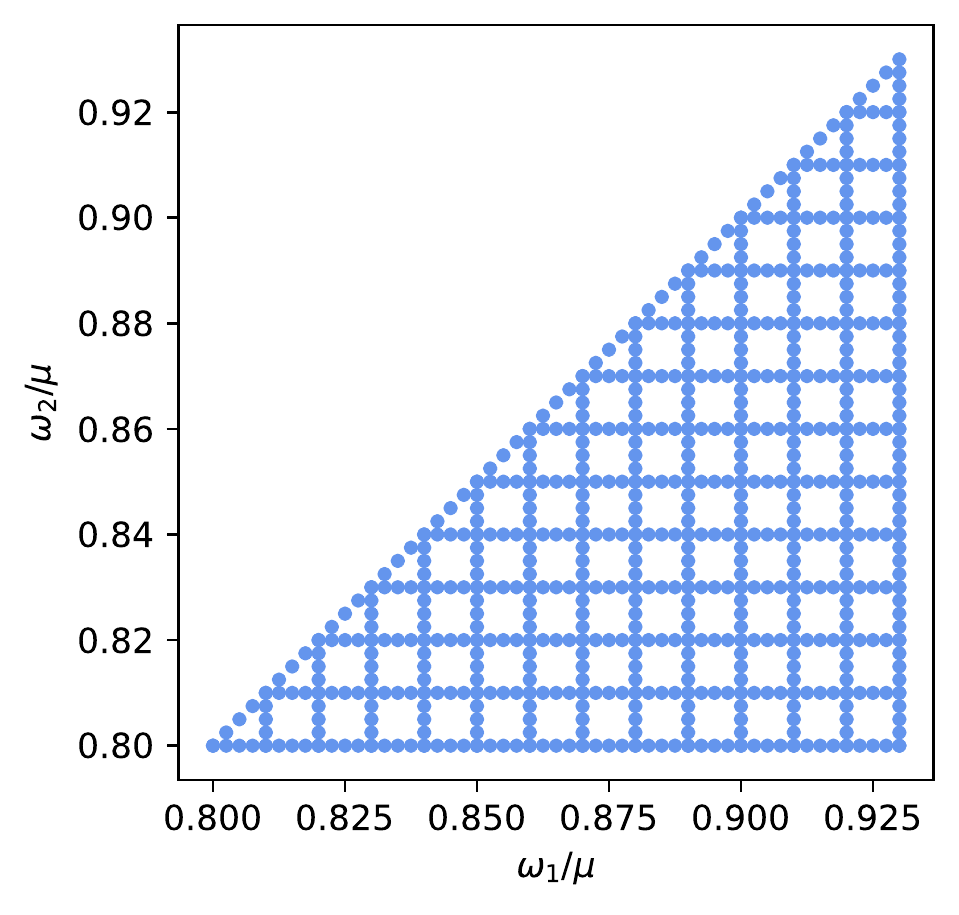}\\
\caption{Main dataset of PS binaries discussed in this work, labelled by the value of the frequencies $\omega_1/\mu$ and $\omega_2/\mu$ of each star. In all cases, the stars are released from rest at a distance of $D\mu=40$ and have the same initial phase.}
\label{catalog}
\end{figure}

Our main catalogue of 759 simulations is depicted in a compact way in Fig.~\ref{catalog}. Each axis in this plot labels the frequencies of the two stars, $\omega_1/\mu$ and $\omega_2/\mu$. For the equal-mass models, placed in the diagonal, we run simulations using a uniform grid in frequencies in the range $\omega_1/\mu = \omega_2/\mu = \omega/\mu \in [0.8000,0.9300]$ with $\Delta\omega = 0.0025$. For the unequal-mass cases, we fix the oscillation frequency of the primary star, $\omega_1/\mu$, and then vary the frequency of the secondary star, $\omega_2/\mu$. Both frequencies range from 0.8000 to 0.9300 with a resolution of $\Delta\omega_1/\mu=0.01$ for $\omega_1/\mu$ and of $\Delta\omega_2/\mu=0.0025$ for $\omega_2/\mu$. As mentioned before, in all of these cases the two stars have null relative phase $\Delta \epsilon=0$ at the start of the simulation. 

As we show below, the initial set of simulations revealed unexpected non-trivial interactions between the stars described by the same Proca field due to their wave-like nature. As a result, we also build an additional set of models to study the impact of the relative phases of the star both in the dynamics and on the GW emission. The effect of this parameter is studied in two (implicit and explicit) ways. First, for some selected cases we vary the initial star-separation at which the simulation is started keeping $\Delta \epsilon = 0$ which, for the cases with $\omega_1 \neq \omega_2$ translates into a varying relative phase at merger. This, however, also causes a variation in the velocity of the two stars at merger whose effect mixes with that of the varying phase. Therefore, in order to explicitly isolate the impact of the relative phase change, for a few selected cases of Fig.~\ref{catalog} we explicitly vary $\Delta \epsilon$ in a uniform grid $\Delta \epsilon \in [0,2\pi]$ with step $\delta\Delta\epsilon=\pi/6$. 

\subsection{Numerics} 
\label{sec:numerics}
To carry out the numerical evolutions we use the publicly available \texttt{Einstein Toolkit}~\cite{toolkit2012open,loffler2012f}, which uses the \texttt{Cactus} framework and mesh refinement. The method-of-lines is employed to integrate the time-dependent differential equations. In particular, we use a fourth-order Runge-Kutta scheme for this task. The left-hand-side of the Einstein equations is solved using the \texttt{MacLachlan} code~\cite{brown2009turduckening,reisswig2011gravitational}, which is based on the 3+1 Baumgarte-Shapiro-Shibata-Nakamura (BSSN) formulation.
On the other hand, the Proca evolution equations, Eqs.~(\ref{eq:dtgamma})-(\ref{eq:dtZ}), are solved using the code described and available in~\cite{Zilhao:2015tya, ZilhaoWitekCanudaRepository, Canuda_zenodo_3565474}. We extended the code to take into account a complex field~\cite{sanchis2019head,sanchis2019nonlinear}. Technical details, assessment of the code, and convergence tests can be found in~\cite{Zilhao:2015tya,sanchis2019head,sanchis2019nonlinear}.
We use a fixed numerical grid with 7 refinement levels, with the following structure $\lbrace(320, 48, 48, 24, 24, 6, 2)/\mu$, $(4, 2, 1, 0.5, 0.25, 0.125, 0.0625)/\mu\rbrace$, where the first set of numbers indicates the spatial domain of each level and the second set indicates the resolution. 
The simulations are performed using equatorial-plane symmetry.

To extract gravitational radiation we employ the Newman-Penrose (NP) formalism~\cite{Newman:1961qr} as described in~\cite{Zilhao:2015tya}.
We compute the NP scalar $\Psi_{4}$ expanded into spin-weighted spherical
harmonics of spin weight $s = -2$. 

 \begin{figure}[t!]
\includegraphics[width=1\linewidth]{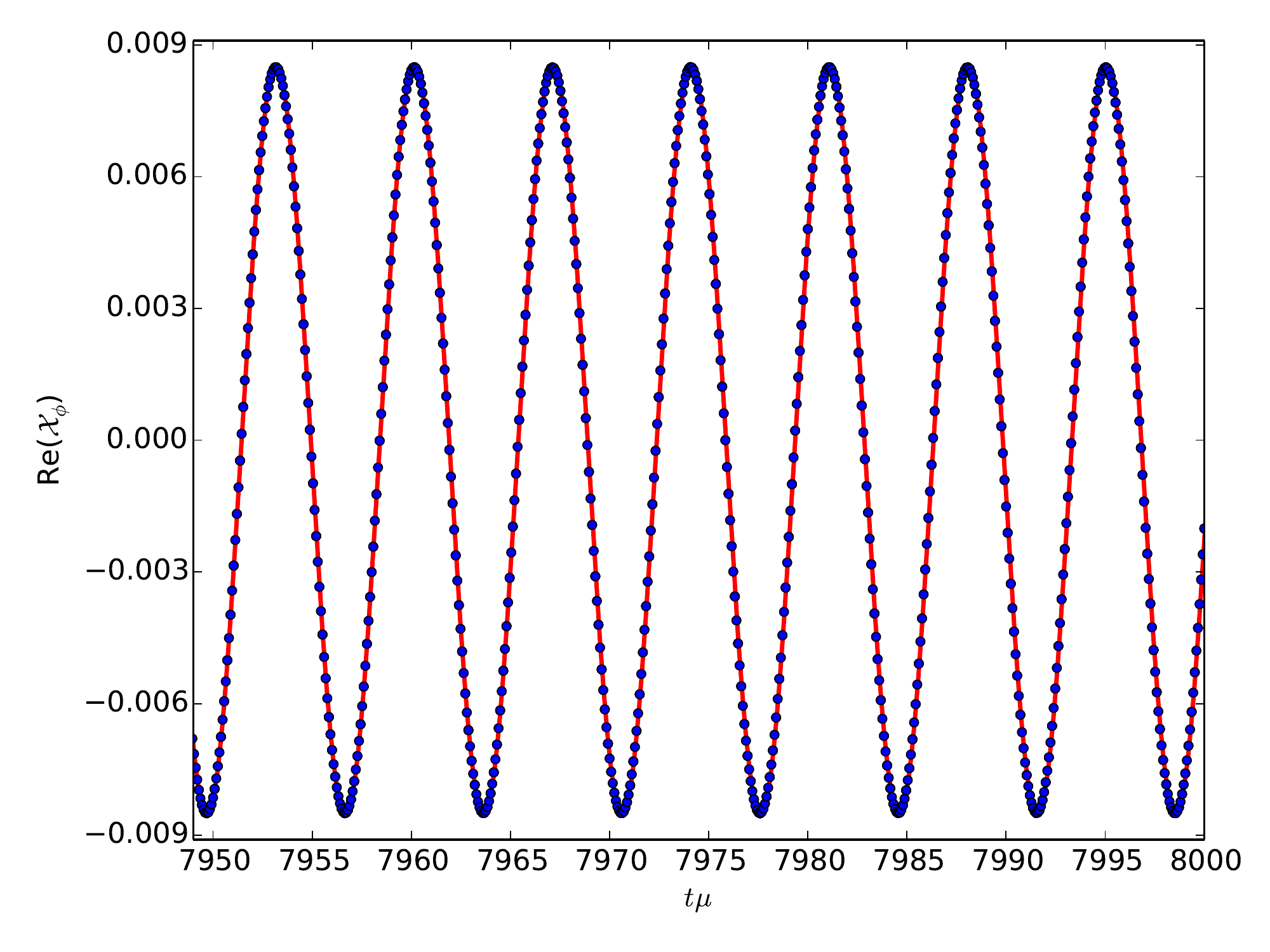}
\includegraphics[width=1\linewidth]{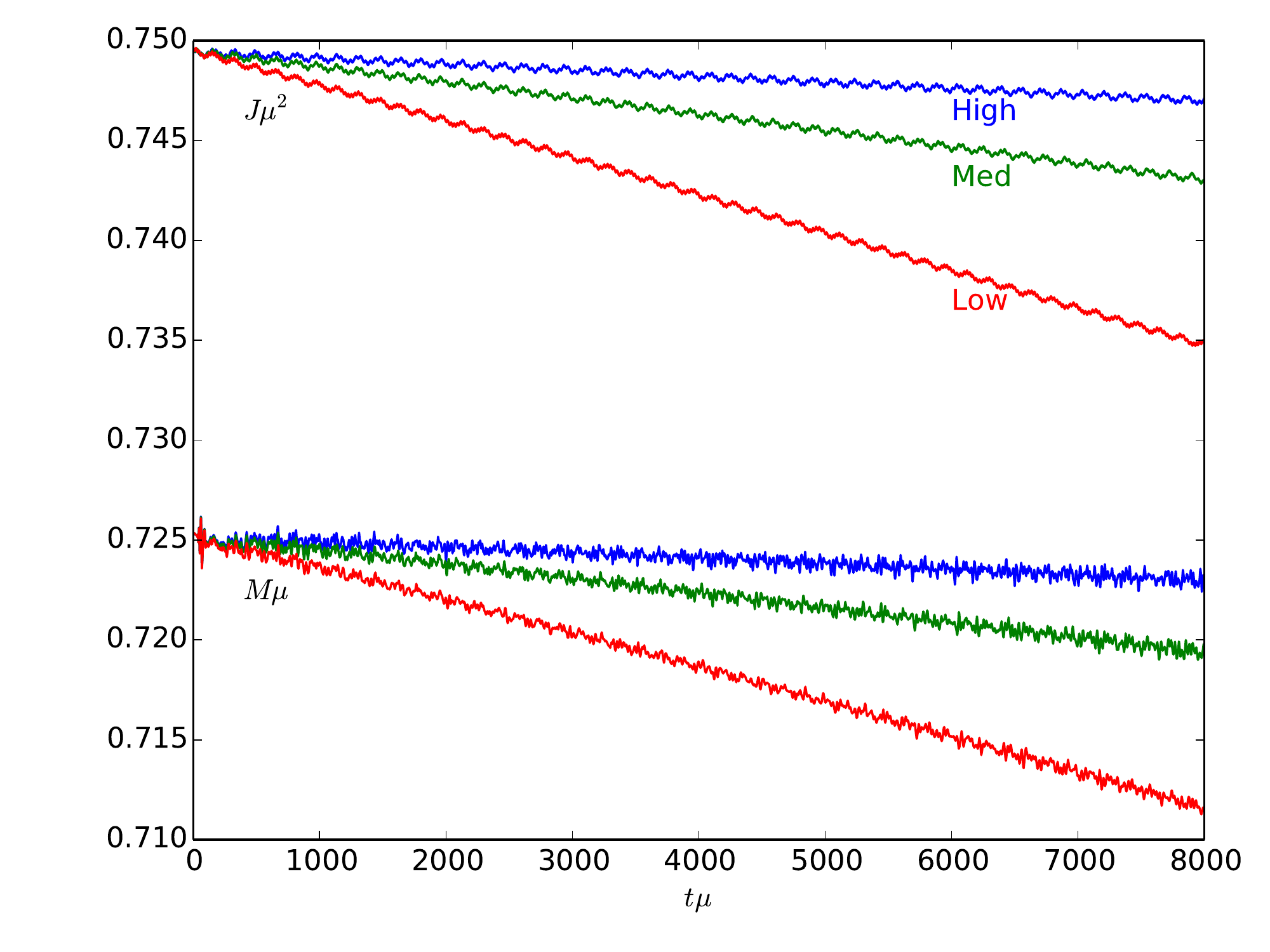}
\includegraphics[width=1\linewidth]{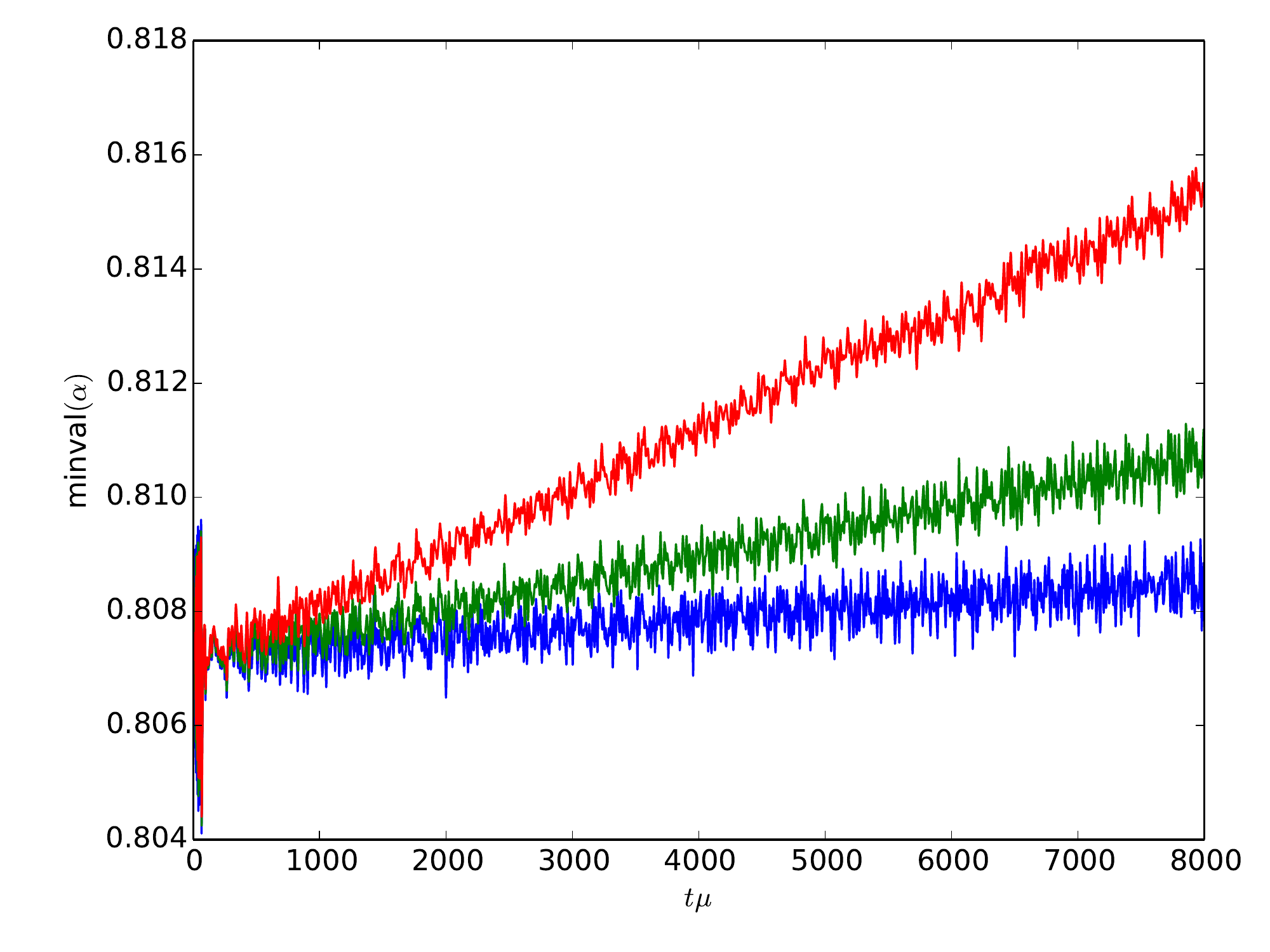}
\caption{Top panel: Evolution of the amplitude of the real part of $\mathcal{X}_{\phi}$. The solid red line corresponds to the analytical value $\cos{\omega t}$ with $\omega=0.90$ and the blue circles to the numerical solution. Middle panel: Evolution of the total Proca energy and angular momentum for the model with $\omega/\mu=0.90$ for three different resolutions. Bottom panel: same as the middle panel but for the minimum value of the lapse function $\alpha$. }
\label{figSingleStar}
\end{figure}

\section{Results}
\label{results}

We have performed 759 simulations of head-on collisions of spinning PSs starting at rest at fixed initial distance, $D\mu=40$. We explore both equal-mass and unequal-mass cases to produce a first systematic study of the GW signals emitted in collisions of these objects. Stationary fundamental bosonic stars are described by the oscillation frequency $\omega/\mu$ of the field, which determines the dimensionless mass $M\mu$ and angular momentum of the star $J\mu^2$, besides its compactness. Further specifying the  boson particle mass $\mu$ determines the corresponding physical quantities $M,J$ (see below). Thus,  $\mu$ can be set as a fundamental scale of the system and all quantities can be simply rescaled. Alternatively, we can trivially rescale the simulations to any fixed total mass, which in turn determines the mass of the boson. We also remark that in contrast with black holes, the angular momentum of PSs is quantized by the relation $J=\bar{m}Q$, where $Q$ is the Noether charge of the star, which counts the number of bosonic particles. This means that an infinitesimal loss/gain in angular momentum must be accompanied by a corresponding loss/gain of particles.

We restrict to the case of mergers of dynamically stable $\bar m=1$ spinning PSs.  For our range of frequencies the PS models 
 have masses and angular momentum that vary from $(\omega/\mu, M\mu, J\mu^2)$=$(0.9300,0.622,0.637)$ to $(0.8000,0.946,1.008)$. All of these mergers lead to a post-merger remnant that is compact enough to collapse into a Kerr black hole. Therefore, our waveform catalogue is well suited for the analysis of LVK GW events under the PS merger scenario.

\begin{figure}[t!]
$\omega_1/\mu=0.8300$
\begin{tabular}{ p{0.24\linewidth}  p{0.24\linewidth} p{0.24\linewidth} p{0.24\linewidth} }
 $\omega_2/\mu = 0.8300$ &  $\omega_2/\mu = 0.8600$ &  $\omega_2/\mu = 0.8750$ &  $\omega_2/\mu = 0.9100$
\end{tabular}
\includegraphics[width=0.24\linewidth]{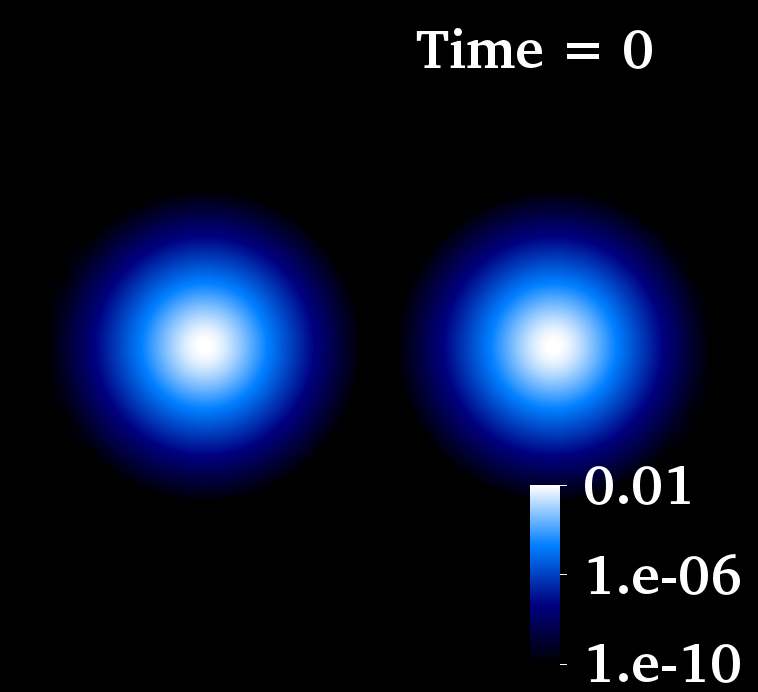}
\includegraphics[width=0.24\linewidth]{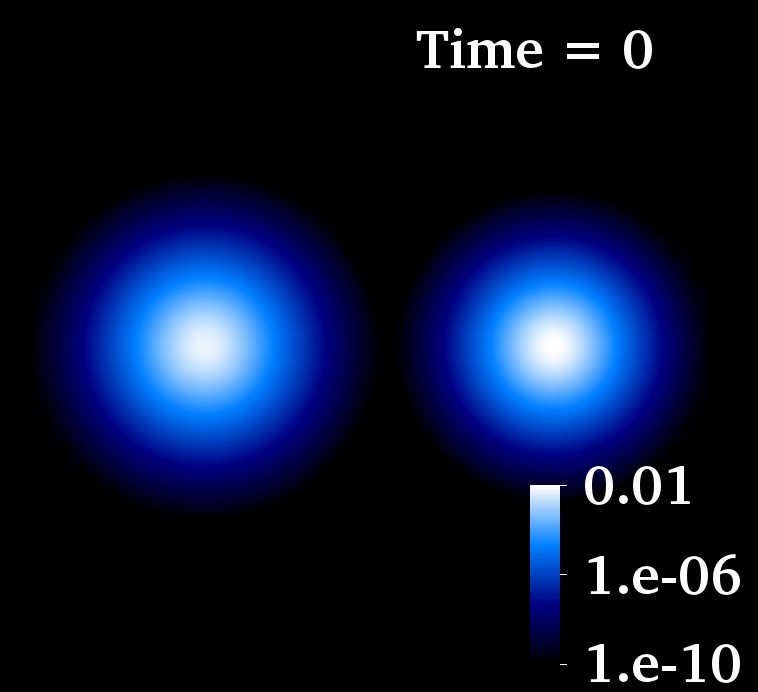}
\includegraphics[width=0.24\linewidth]{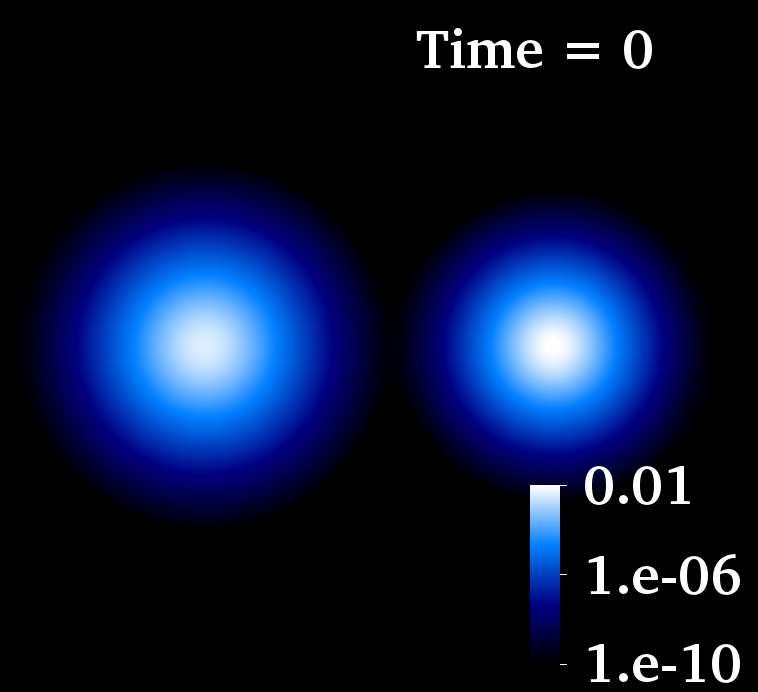}
\includegraphics[width=0.24\linewidth]{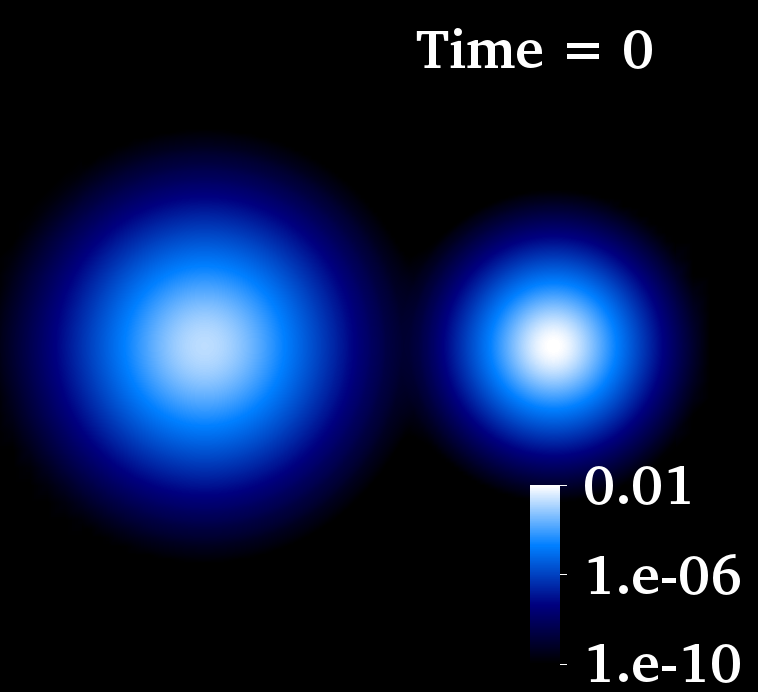}\\%
\includegraphics[width=0.24\linewidth]{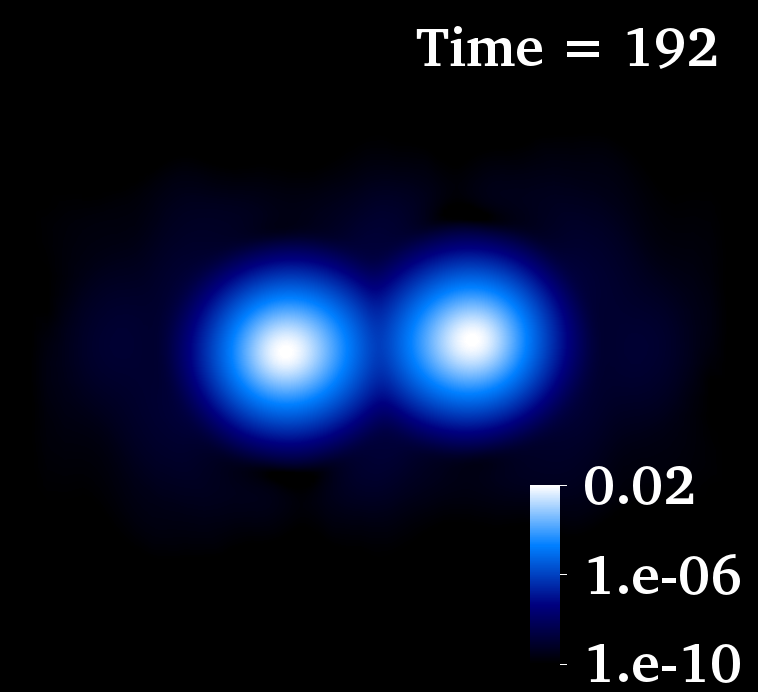}
\includegraphics[width=0.24\linewidth]{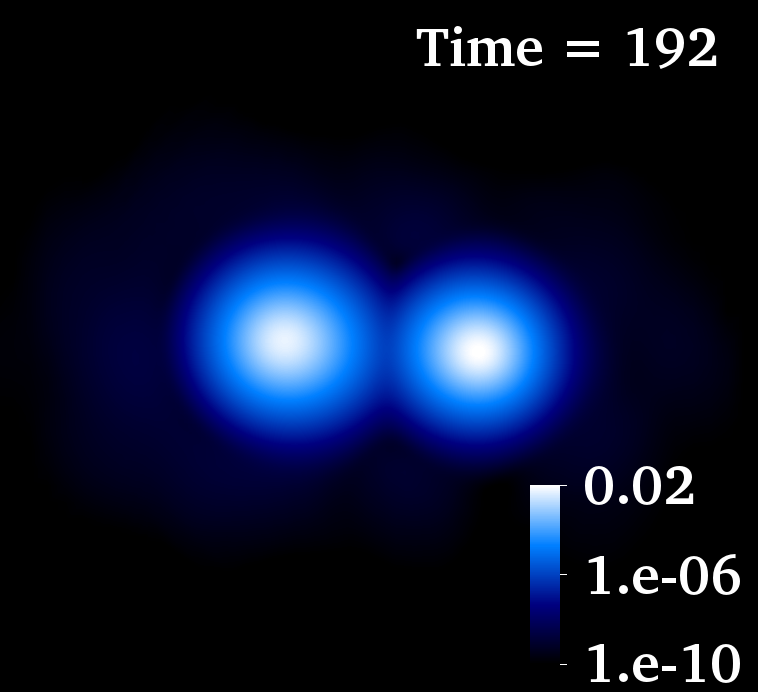}
\includegraphics[width=0.24\linewidth]{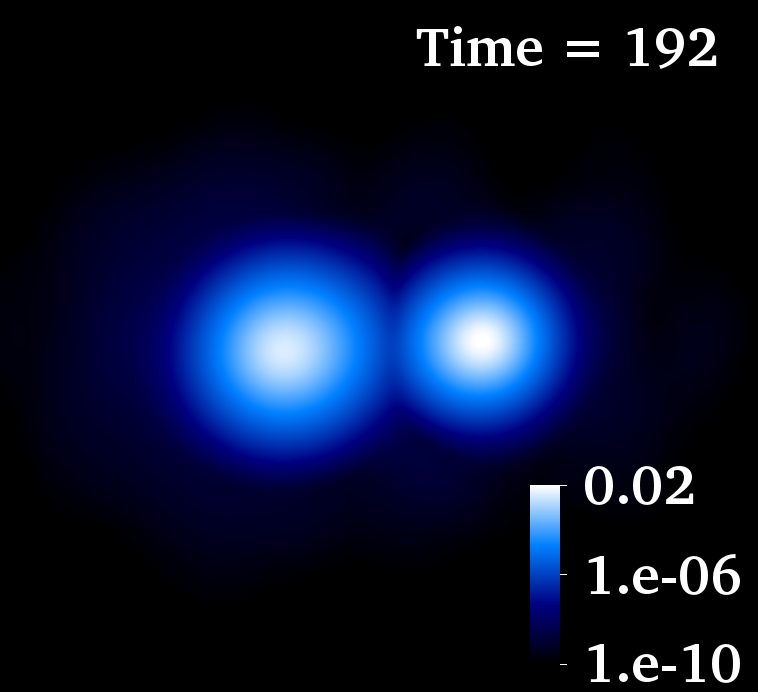}
\includegraphics[width=0.24\linewidth]{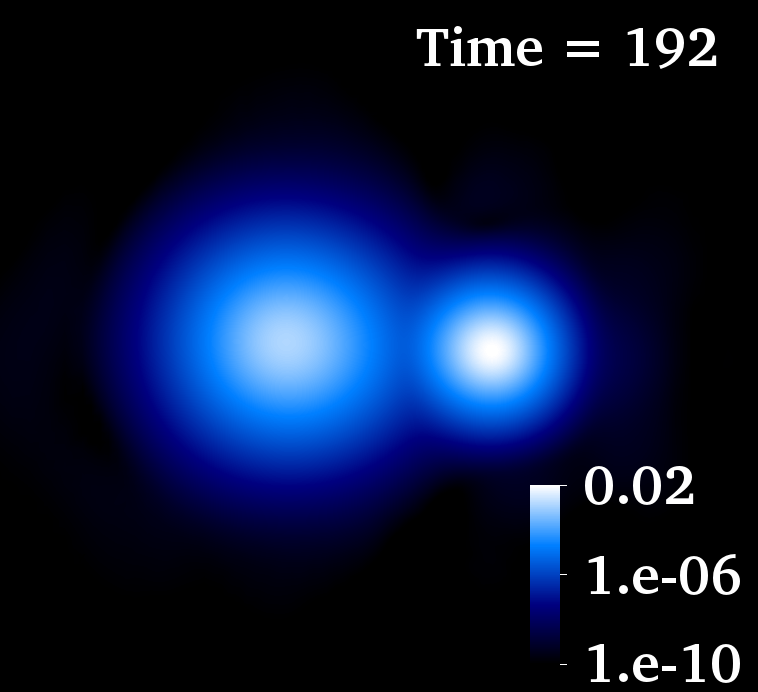}\\
\includegraphics[width=0.24\linewidth]{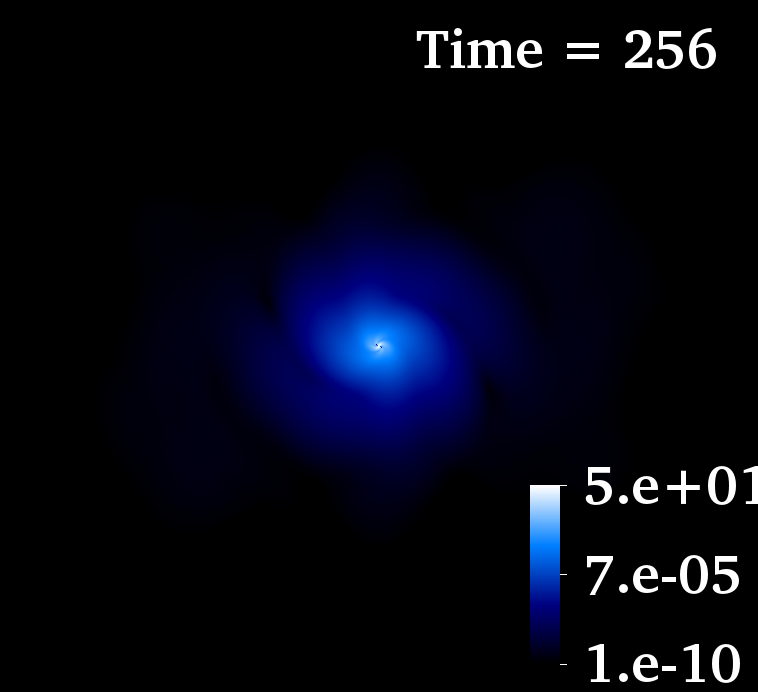}
\includegraphics[width=0.24\linewidth]{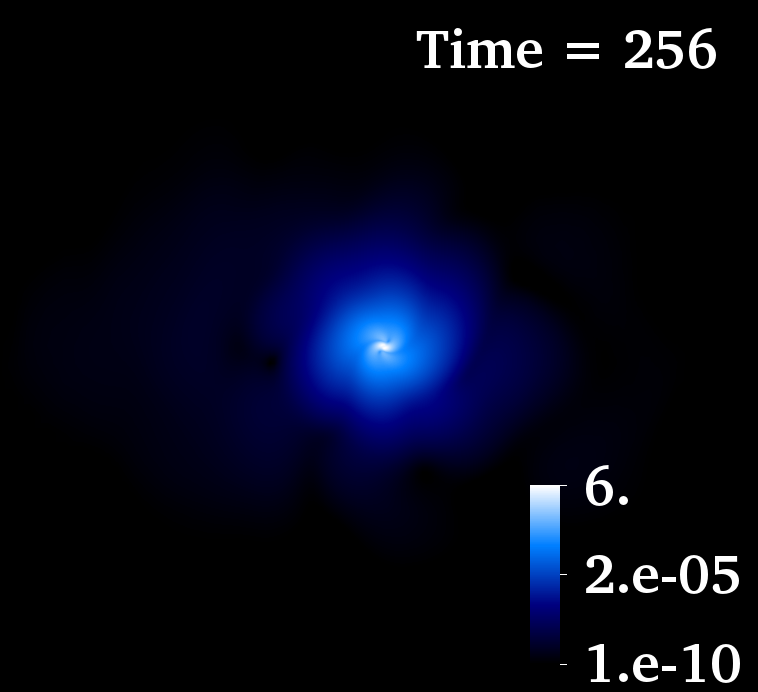}
\includegraphics[width=0.24\linewidth]{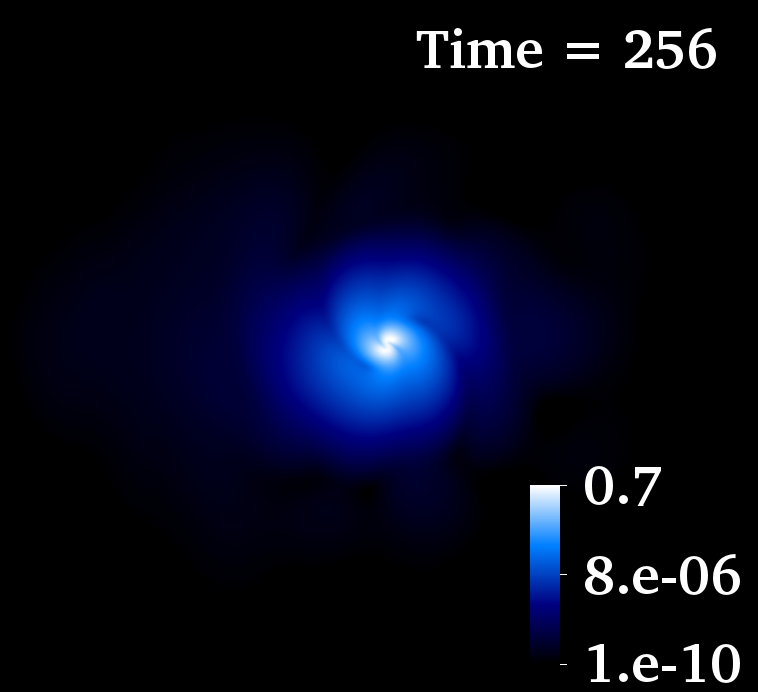}
\includegraphics[width=0.24\linewidth]{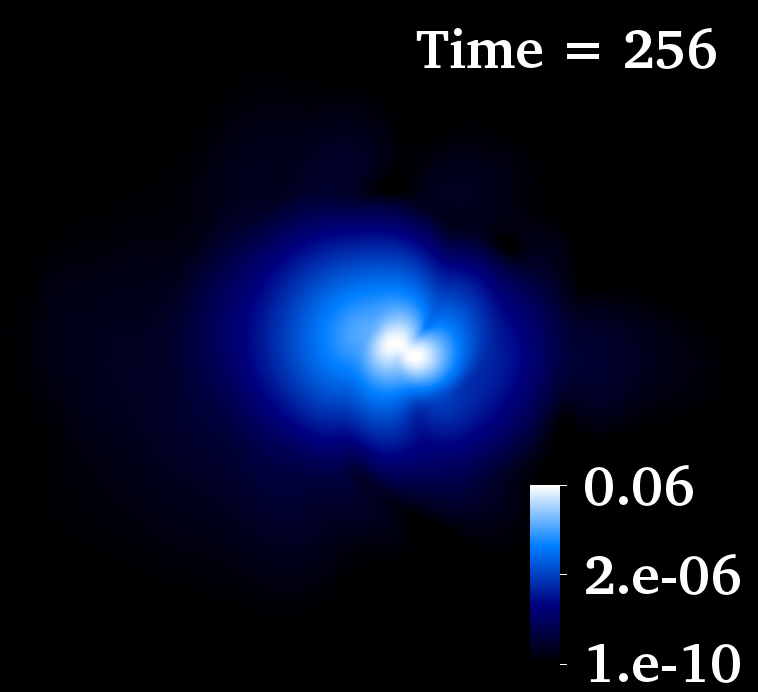}\\
\includegraphics[width=0.24\linewidth]{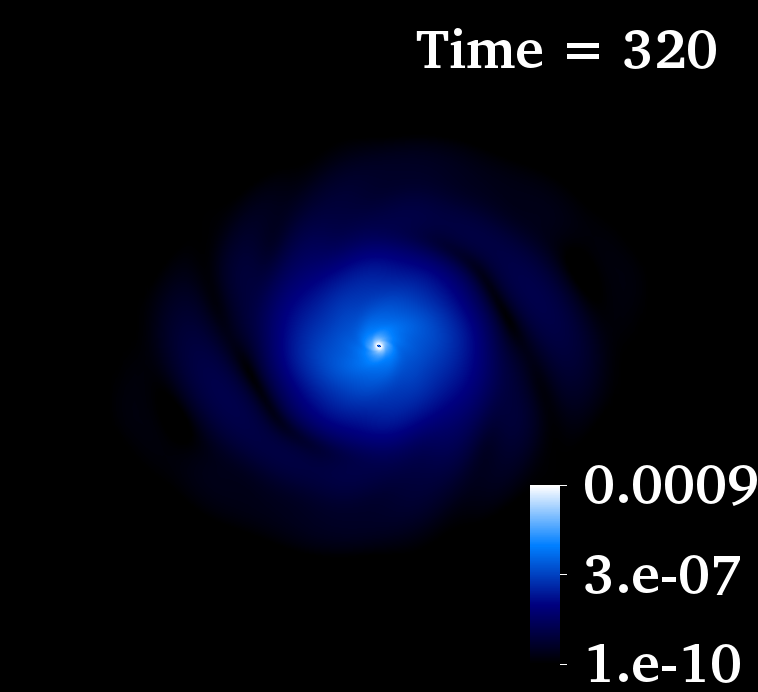}
\includegraphics[width=0.24\linewidth]{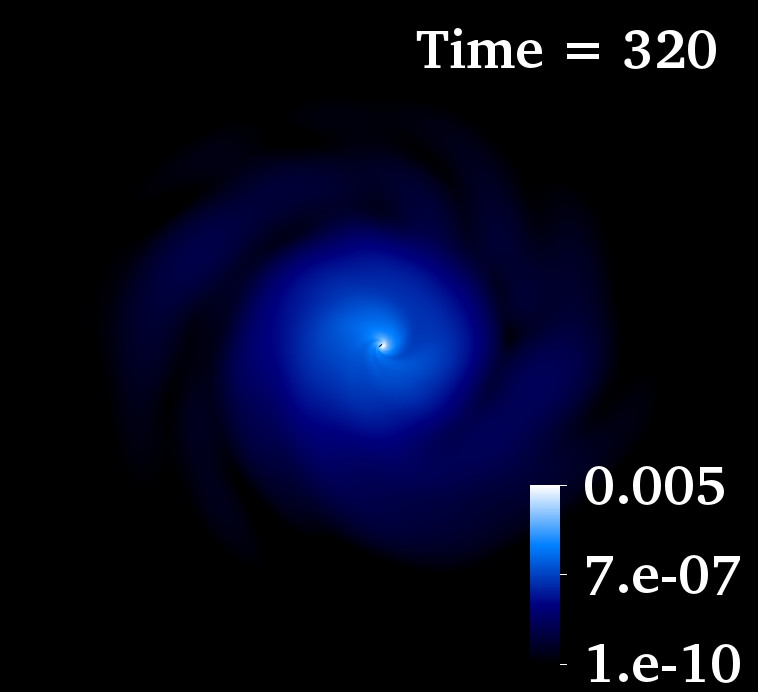}
\includegraphics[width=0.24\linewidth]{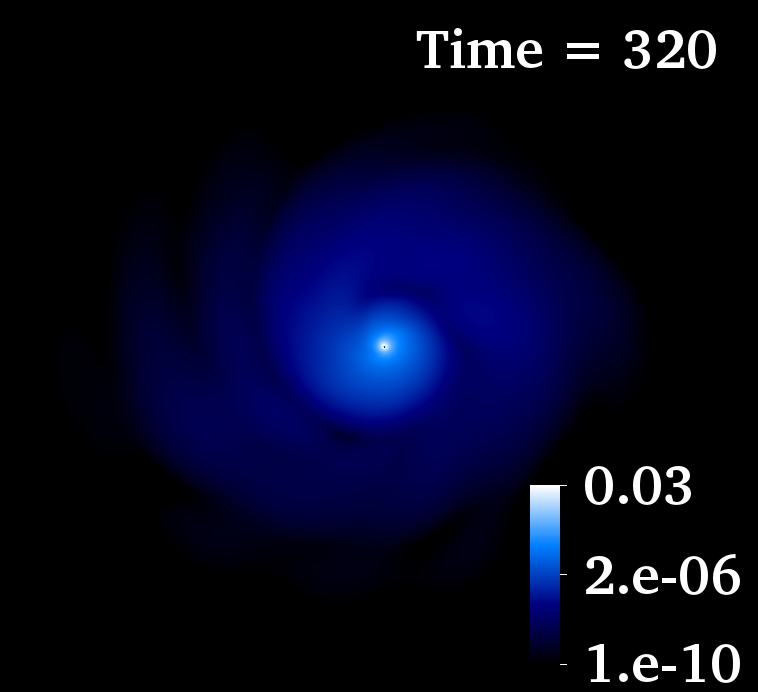}
\includegraphics[width=0.24\linewidth]{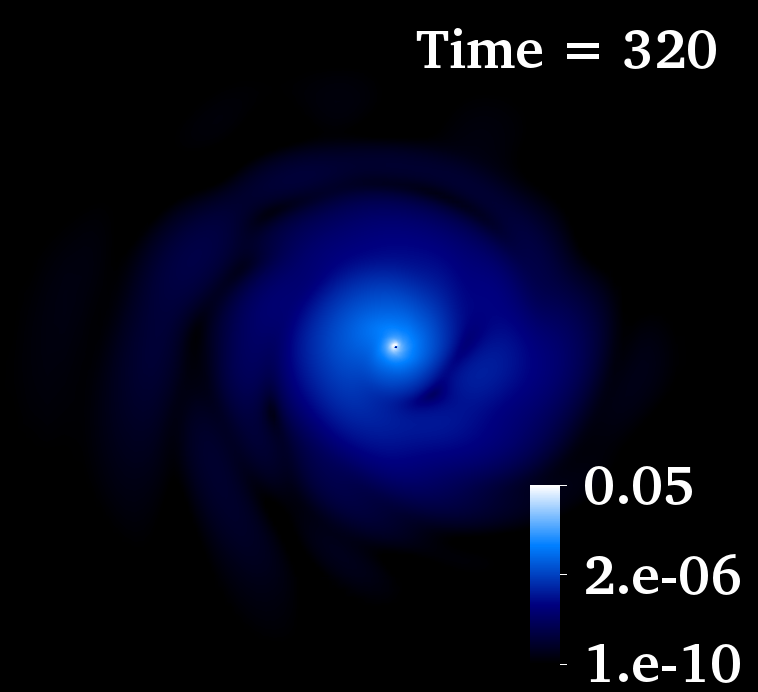}\\
\includegraphics[width=0.24\linewidth]{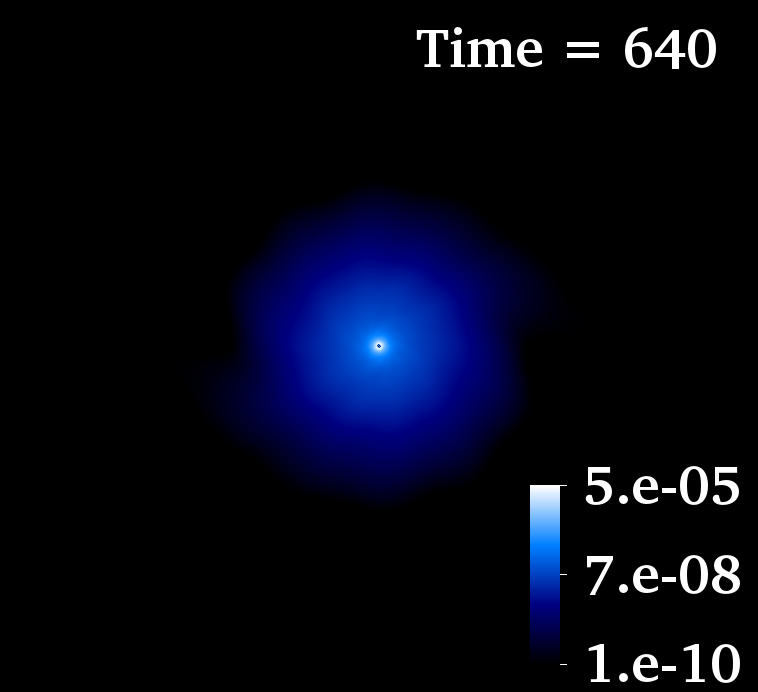}
\includegraphics[width=0.24\linewidth]{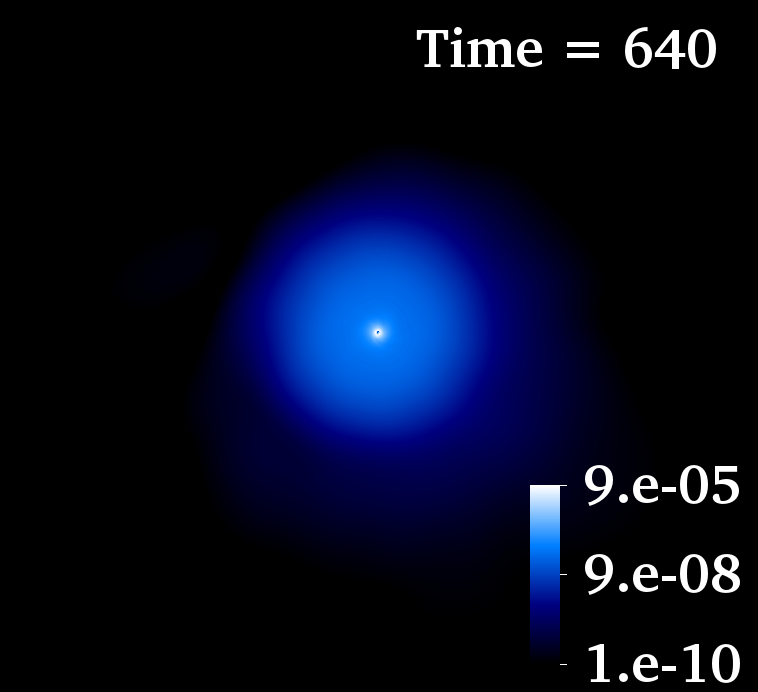}
\includegraphics[width=0.24\linewidth]{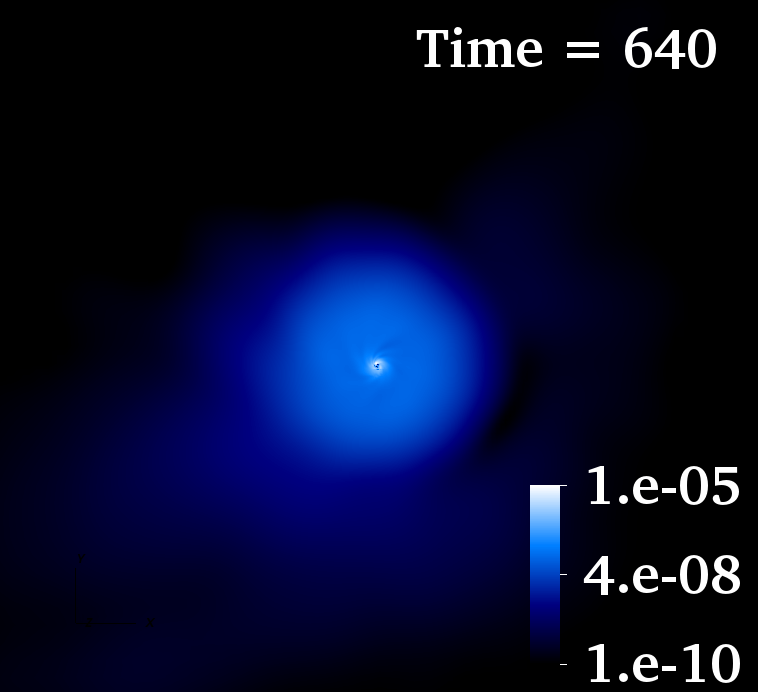}
\includegraphics[width=0.24\linewidth]{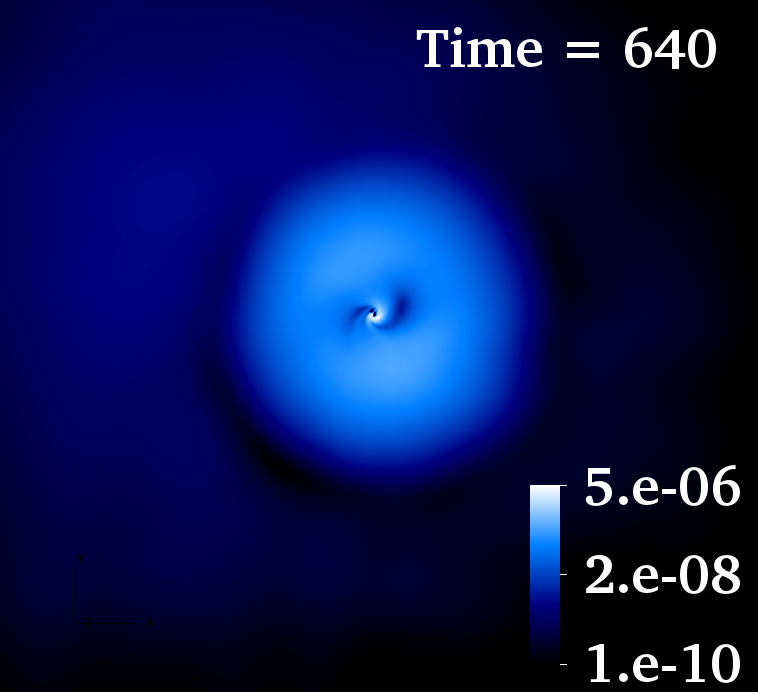}\\
\caption{Equatorial ($xy$) plane snapshots of the energy density in logscale during the evolution of the collisions of spinning PSs for different models with fixed $\omega_1/\mu=0.8300$ and varying the frequency $\omega_2/\mu$ of the secondary star. Time runs from top to bottom and is given in code units with $G=c=\mu=1$.}
\label{fig2D083}
\end{figure}
\begin{figure}[t!]
$\omega_1/\mu=0.9100$
\begin{tabular}{ p{0.24\linewidth}  p{0.24\linewidth} p{0.24\linewidth} p{0.24\linewidth} }
 $\omega_2/\mu = 0.8500$ &  $\omega_2/\mu = 0.8800$ &  $\omega_2/\mu = 0.8950$ &  $\omega_2/\mu = 0.9100$
\end{tabular}
\includegraphics[width=0.24\linewidth]{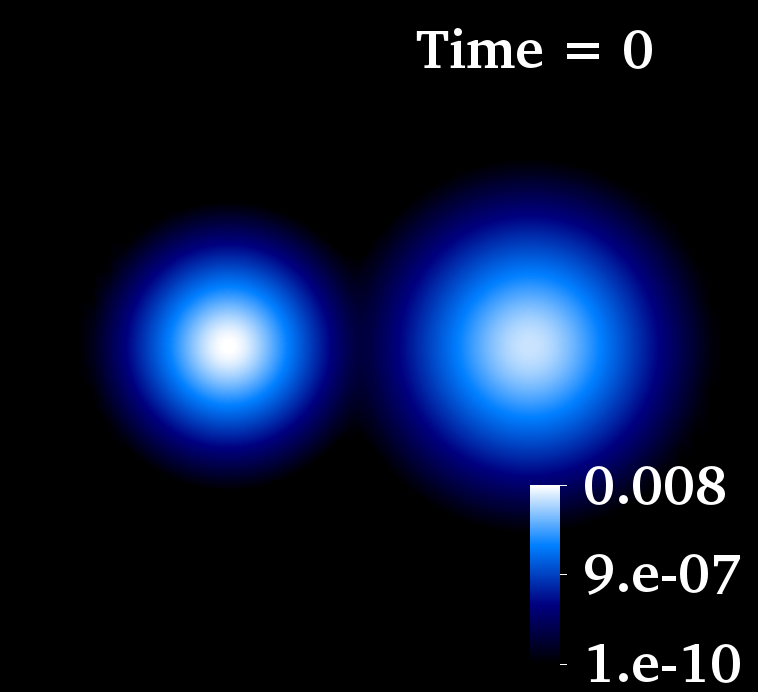}
\includegraphics[width=0.24\linewidth]{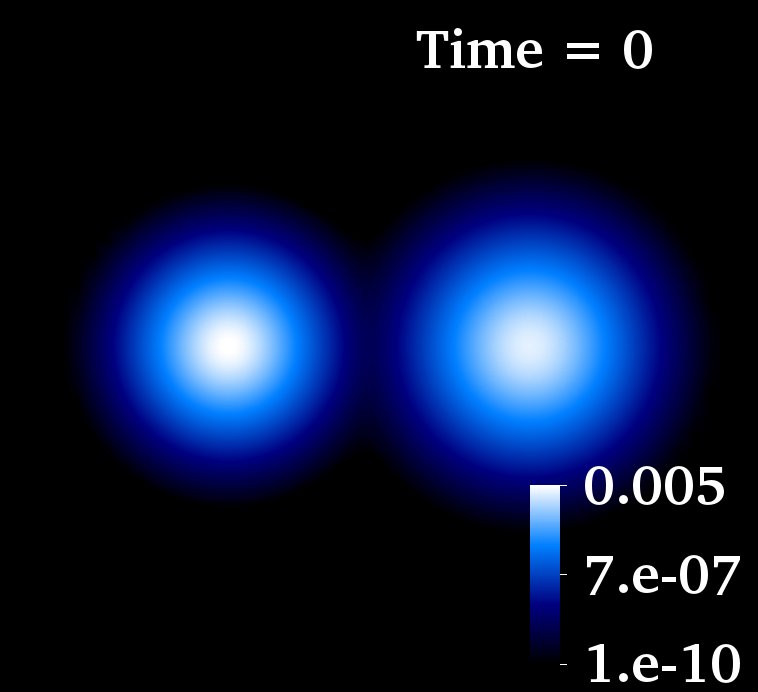}
\includegraphics[width=0.24\linewidth]{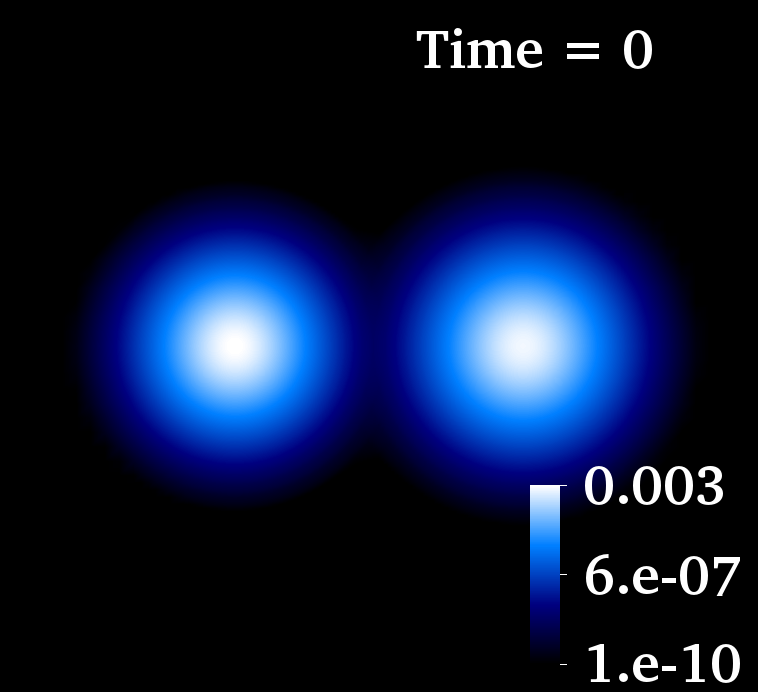}
\includegraphics[width=0.24\linewidth]{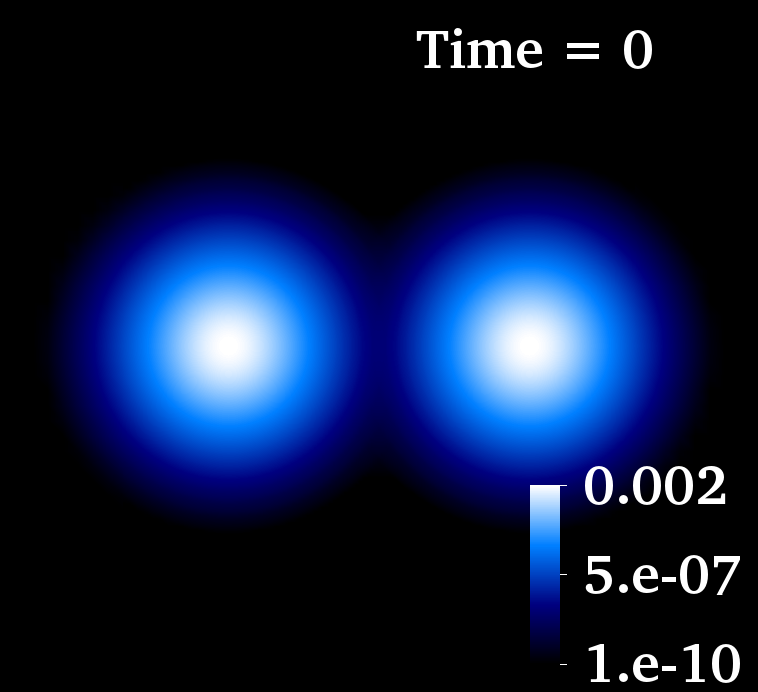}\\%
\includegraphics[width=0.24\linewidth]{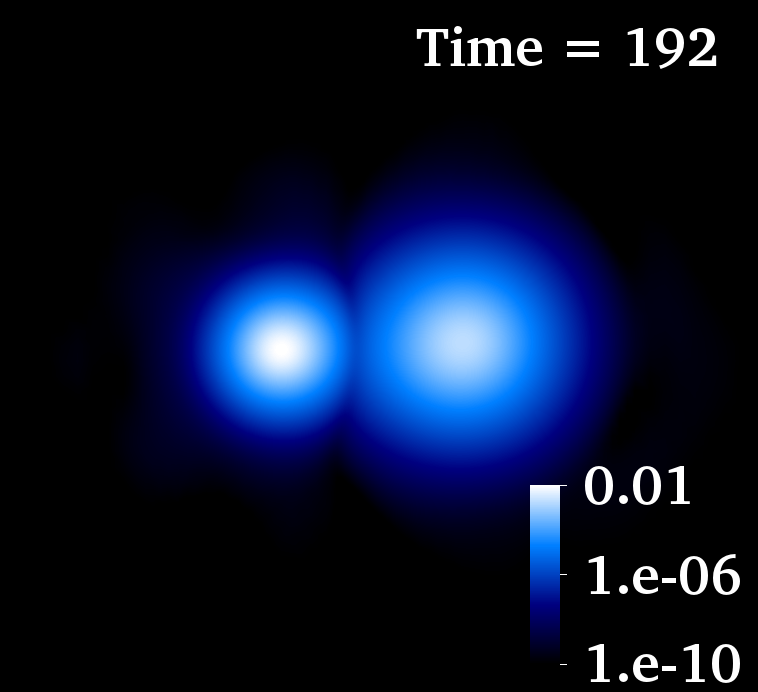}
\includegraphics[width=0.24\linewidth]{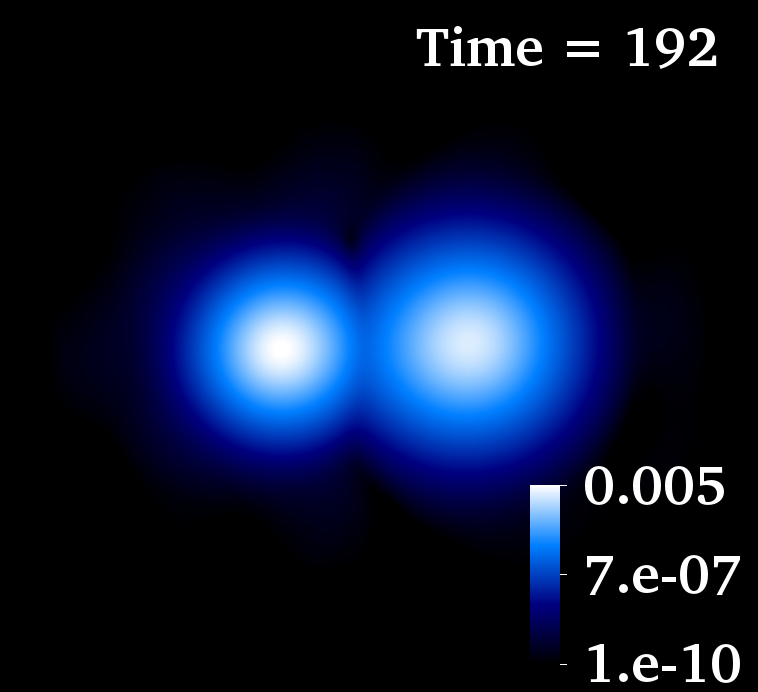}
\includegraphics[width=0.24\linewidth]{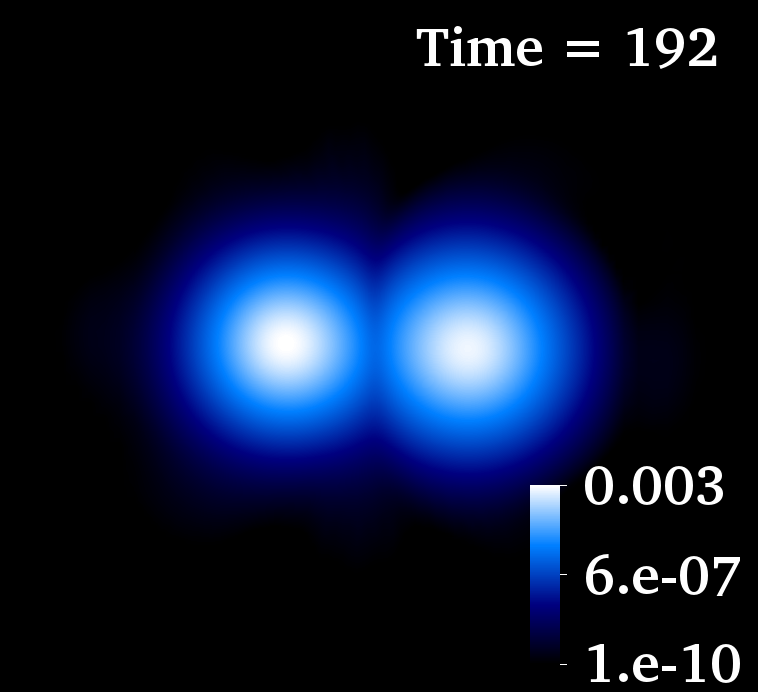}
\includegraphics[width=0.24\linewidth]{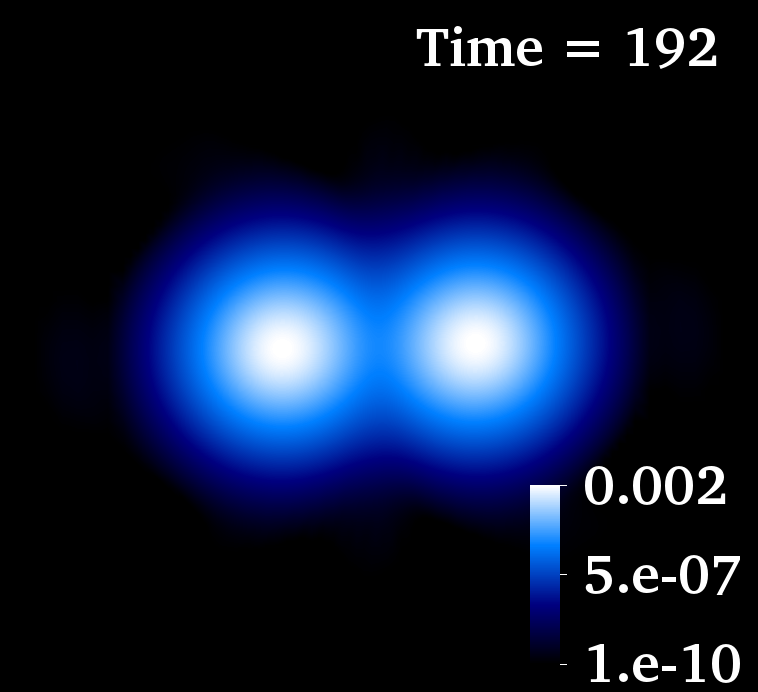}\\
\includegraphics[width=0.24\linewidth]{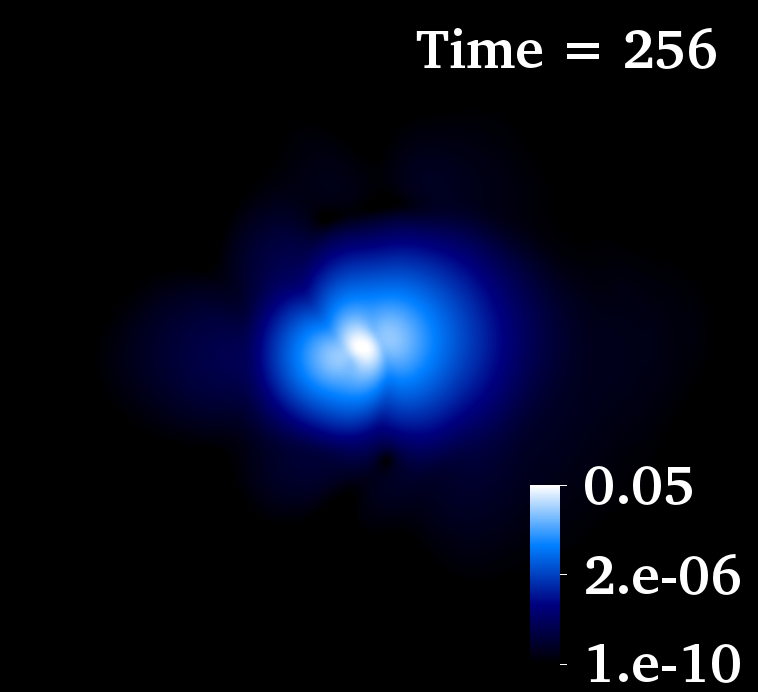}
\includegraphics[width=0.24\linewidth]{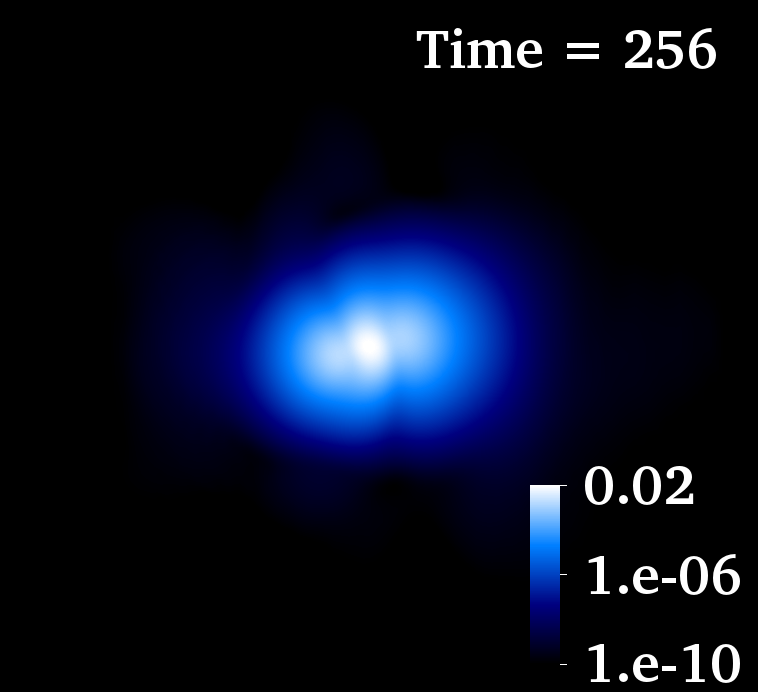}
\includegraphics[width=0.24\linewidth]{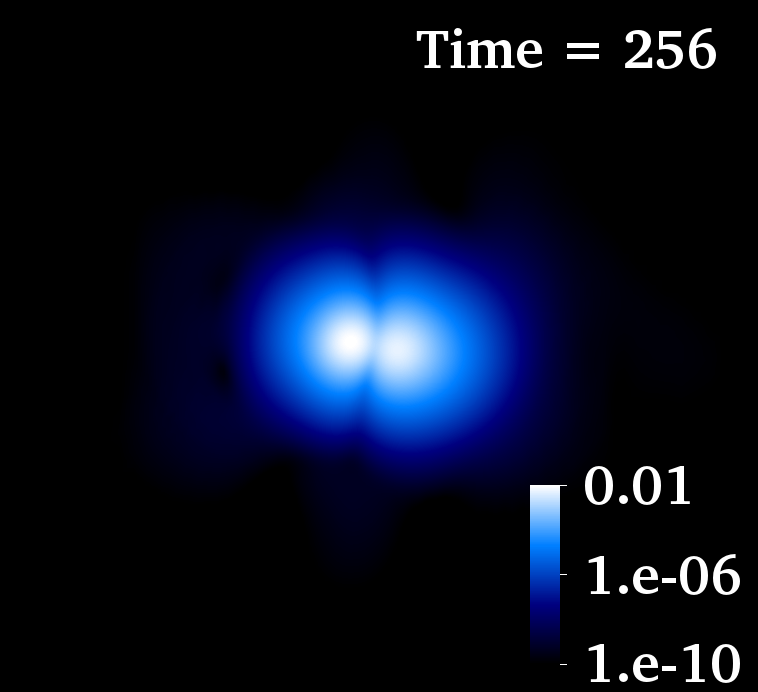}
\includegraphics[width=0.24\linewidth]{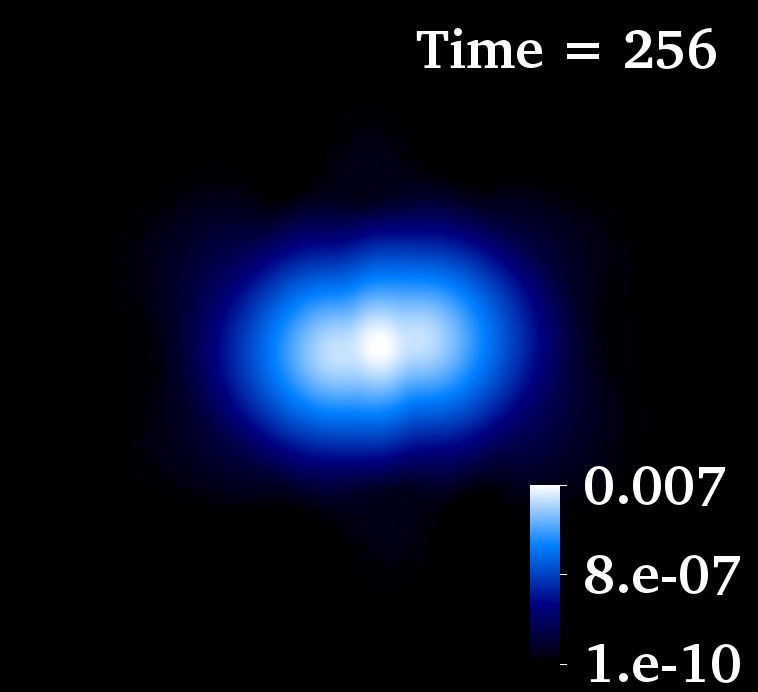}\\
\includegraphics[width=0.24\linewidth]{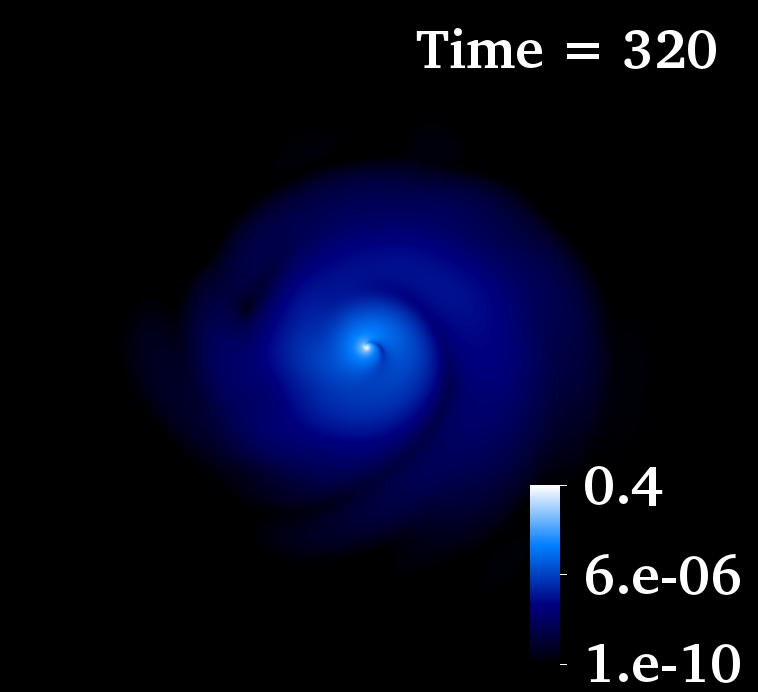}
\includegraphics[width=0.24\linewidth]{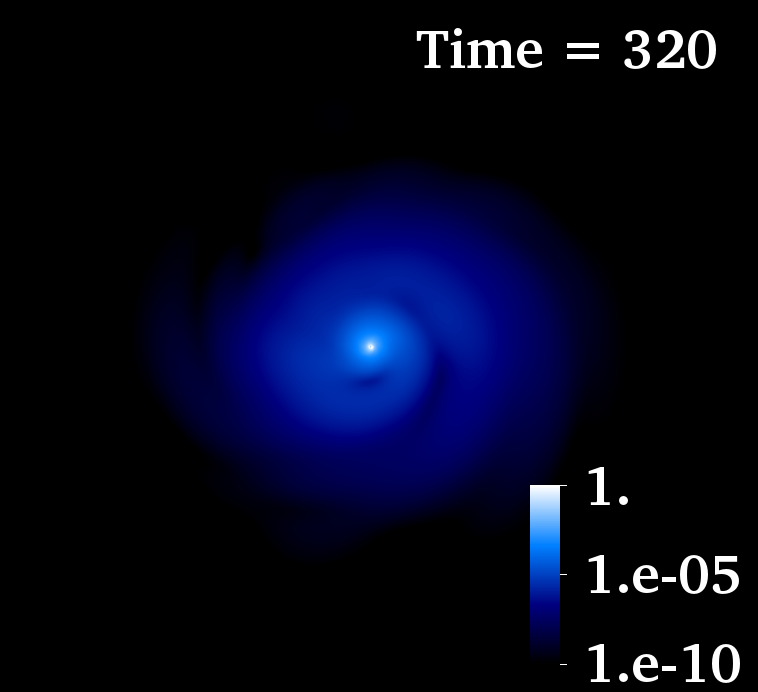}
\includegraphics[width=0.24\linewidth]{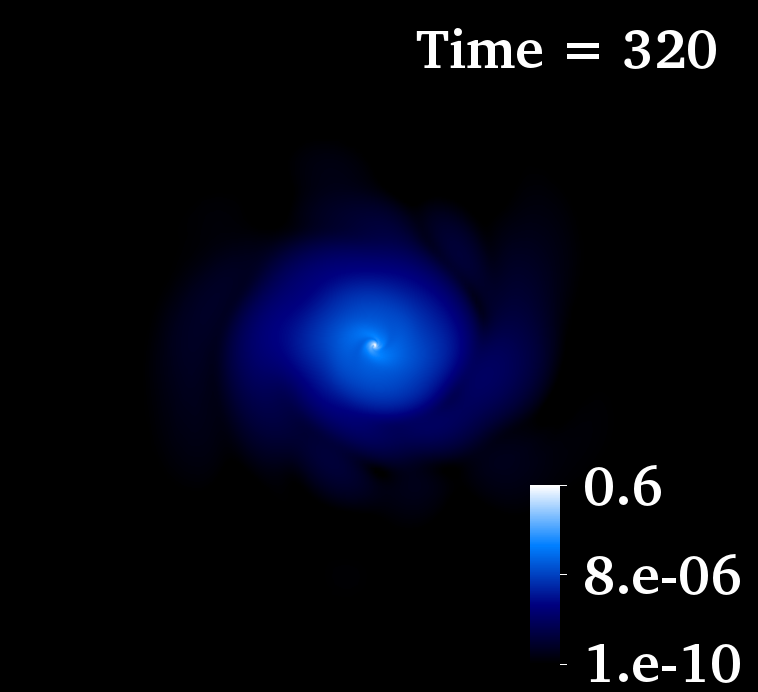}
\includegraphics[width=0.24\linewidth]{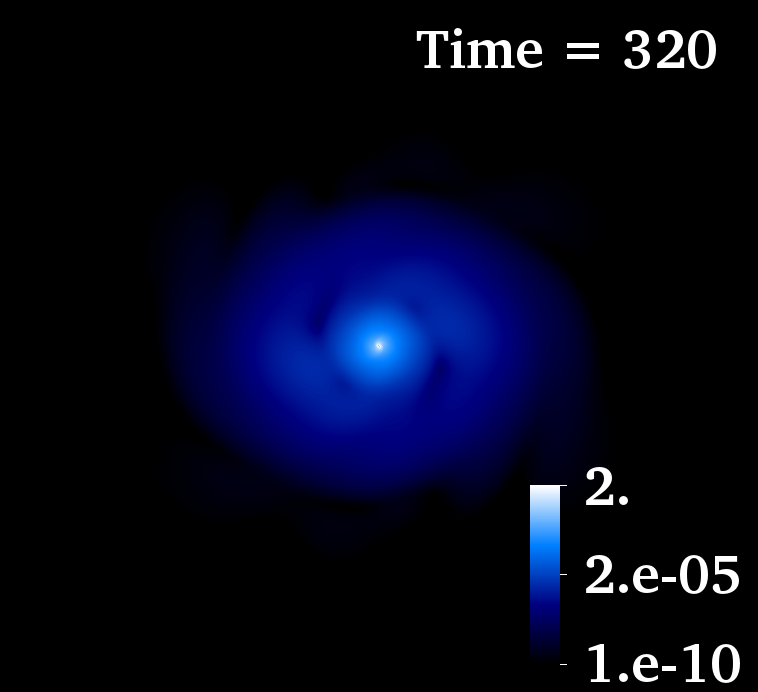}\\
\includegraphics[width=0.24\linewidth]{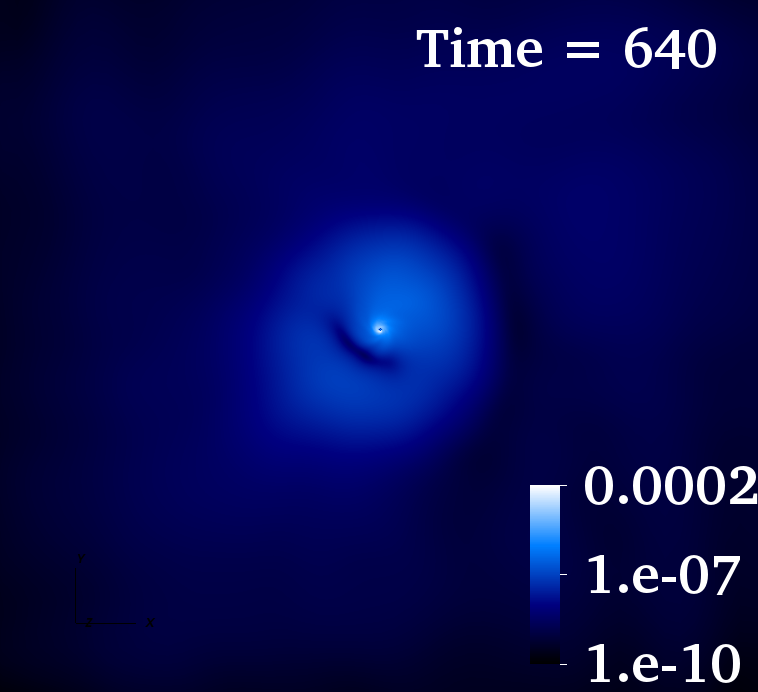}
\includegraphics[width=0.24\linewidth]{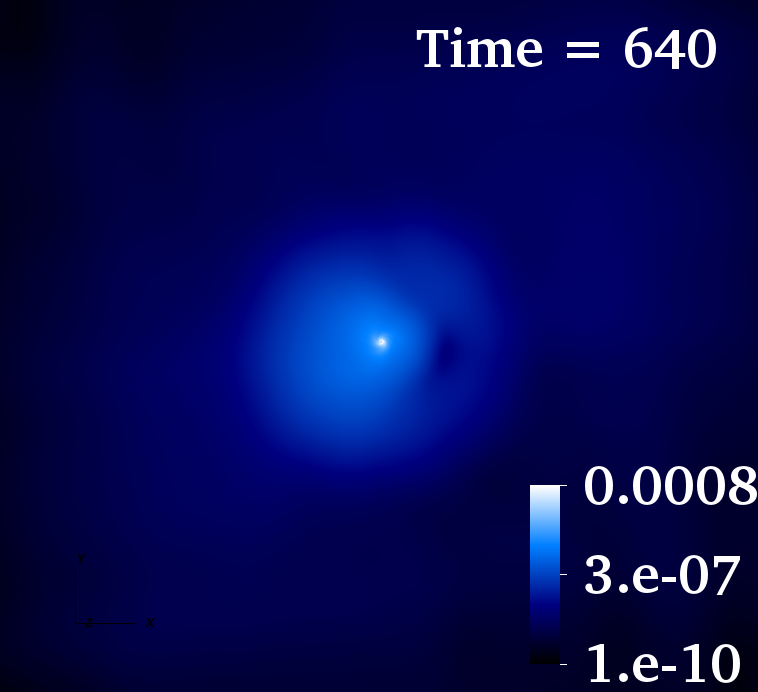}
\includegraphics[width=0.24\linewidth]{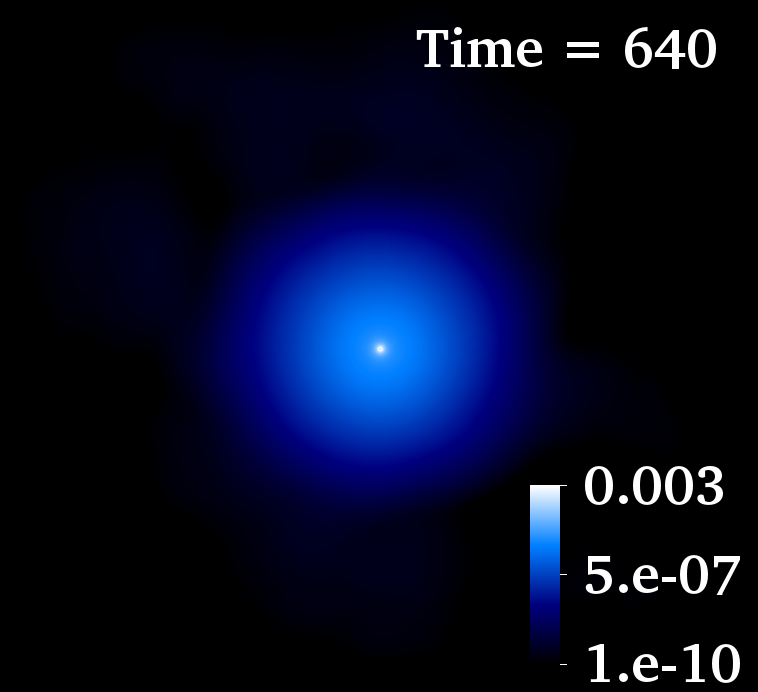}
\includegraphics[width=0.24\linewidth]{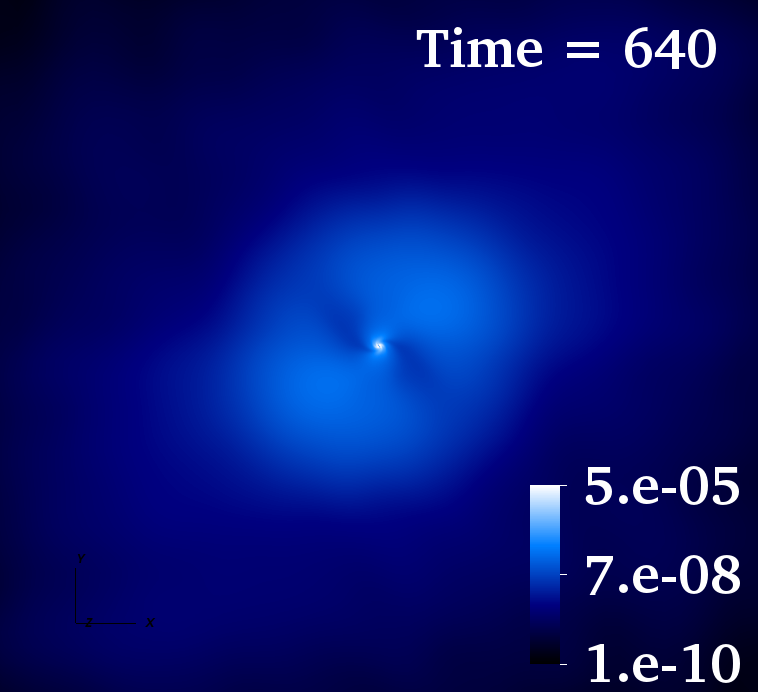}\\
\caption{Same as Fig.~\ref{fig2D083} but for PS models with $\omega_1/\mu=0.9100$ and different values of $\omega_2/\mu$.}
\label{fig2D091}
\end{figure}

\subsection{Single star}
 The dynamical robustness and formation of spinning PSs were addressed in~\cite{sanchis2019nonlinear,DiGiovanni:2020ror}. Here we illustrate the stability properties of these objects. We consider the case of a single isolated spinning PS. We fix the oscillation frequency to $\omega/\mu=0.90$, the mass to $M\mu=0.726$ and the angular momentum $J\mu^2=0.750$. We evolve the star up to a time $t\mu=8000$.  The top panel of Fig.~\ref{figSingleStar} shows the evolution of the amplitude of the real part of $\mathcal{X}_{\phi}$ at the end of the simulation together with the analytical value. The numerical result is in excellent agreement with the analytical estimate. In the middle panel we show, for three different resolutions, namely a fixed grid with four refinement levels and $dx=\lbrace0.8, 0.4, 0.2\rbrace/\mu$ in their finest level, the time evolution of the Proca energy and angular momentum given by the Komar integrals:
\begin{align}
M&=-\int_{\Sigma}drd\theta 
d\varphi\left(2T^t_t-T_\alpha^\alpha\right)\alpha\sqrt{\gamma} \ ,
\label{energy}\\
J&=\int_{\Sigma}drd\theta 
d\varphi\,T^{t}_{\varphi}\alpha\sqrt{\gamma} \ .
\label{angmom}
\end{align}   
In addition, we show in the bottom panel the minimum value of the lapse function $\alpha$. At the highest resolution, the deviations of the final mass, angular momentum, and lapse function with respect to the initial values are less than 0.4\% at $t\mu=8000$. The resolution is comparable to the merger case, for which we added more refinement levels at the centre of the grid to take into account black hole formation. The initial deviations come from interpolation errors from the initial data computed in a compactified grid to the Cartesian grid used for the numerical simulations. The convergence order of our code under grid resolution is found to be around 2.5. 

\subsection{Head-on mergers of Proca stars}

We now move to study head-on collisions of PSs and the corresponding GW emission. Figs.~\ref{fig2D083} and~\ref{fig2D091} show the energy density of the Proca field at the equatorial plane ($z$=0) for two families of collisions respectively characterised by primary-star frequencies, namely $\omega_1/\mu=0.8300$ and $\omega_1/\mu=0.9100$, and four illustrative secondary-star frequencies $\omega_2/\mu$. These figures exemplify the dynamics of all PS binaries in our dataset. In particular, we note  that the collisions are not strictly head-on since the objects do not follow a straight line. Instead, the trajectories of both stars are curved due to the frame-dragging induced by the stars spins. All mergers lead to the formation of a Kerr black hole with a faint Proca field remnant around the horizon, therefore storing a small fraction of the initial Proca mass and angular momentum~\cite{sanchis2020synchronized,Bezares:2022obu}. The final black holes not always form promptly as for some values of the PS parameters the collisions exhibit the formation of a transient hypermassive PS.

\begin{figure*}[t!]
\includegraphics[width=0.335\linewidth]{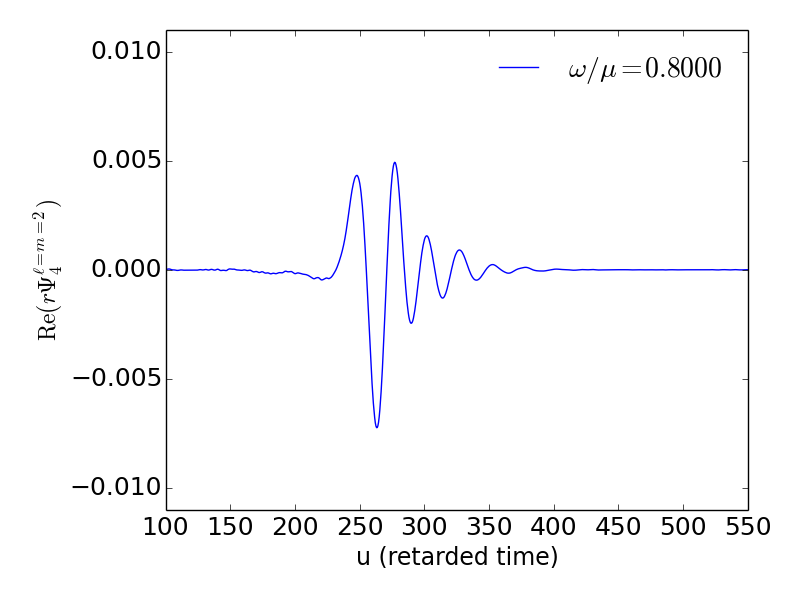}\includegraphics[width=0.335\linewidth]{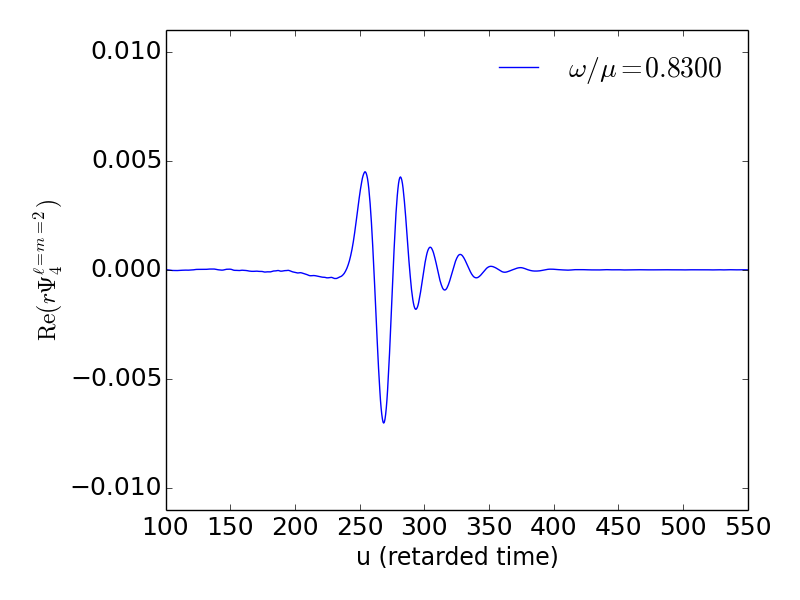}\includegraphics[width=0.335\linewidth]{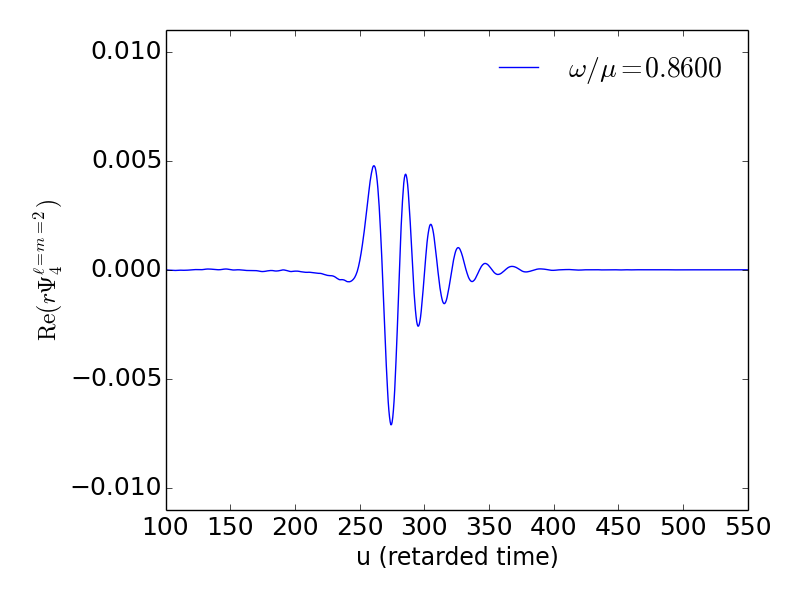}\\
\includegraphics[width=0.335\linewidth]{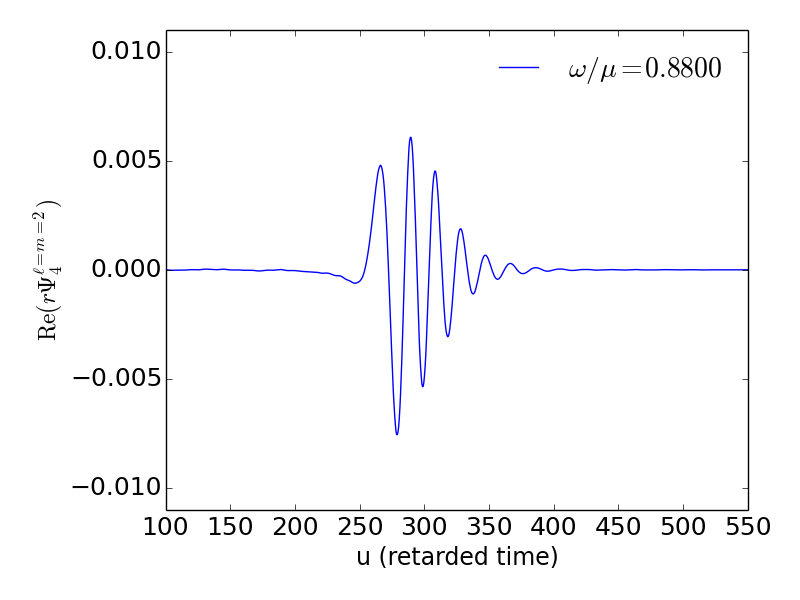}\includegraphics[width=0.335\linewidth]{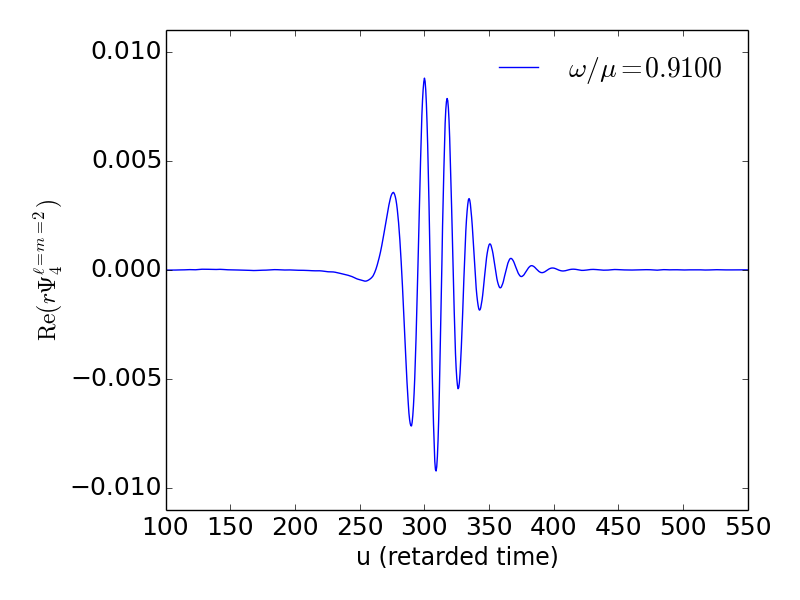}\includegraphics[width=0.335\linewidth]{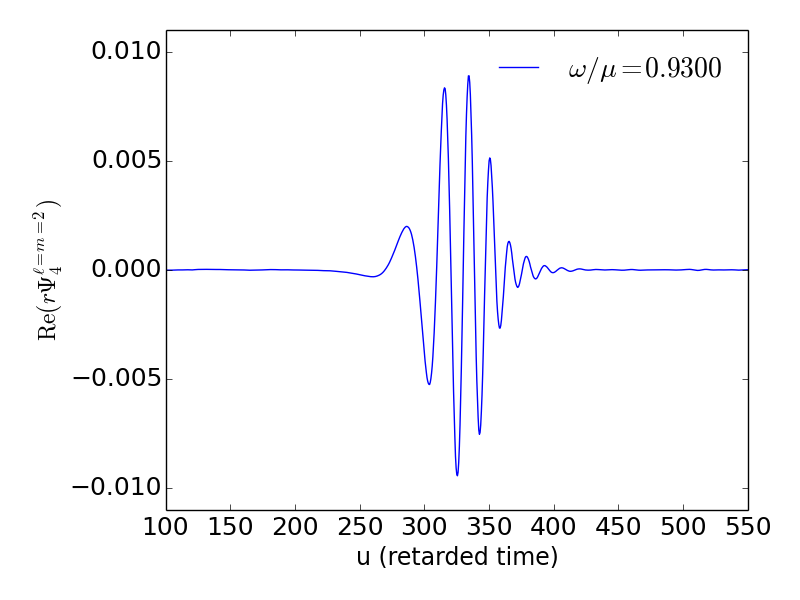}\\
\caption{$\ell= m=2$ mode of the $r\Psi_{4}^{\ell m}$ for six equal-mass PS collisions of increasing frequency. For an animation of the full set of GW signals from equal-mass collisions see~\cite{GWmovies}.}
\label{figGWequal}
\end{figure*}

The collisions produce a burst of GWs, similar to the signals from head-on collisions of black holes~\cite{bustillo2021confusing,CalderonBustillo:2020srq}. We note that the gravitational waveform sourced by head-on collisions is fundamentally different from that produced in orbital binary mergers. First, it is obviously much shorter as there is no inspiral phase preceding the merger. Second, the radiated energy is significantly lower (only around a 0.2\% of the initial energy of the system, when in orbital mergers it reaches a few percent) due to the slow velocities of the two objects at merger, caused by the fact that we release the stars from rest at very short distances. Third, while the GW emission from orbital mergers is vastly dominated by the quadrupole $\ell=2$,  $m=\pm 2$ modes, that from head-on mergers exhibits an $(\ell,m)=(2,0)$ mode, equally dominating~\cite{palenzuela2007head,sanchis2019head,CalderonBustillo:2020srq}. Fig.~\ref{figGWequal} shows the dominant $\ell= m=2$ mode of the  Newman-Penrose scalar $\Psi_4$ in the equal-mass case, for six different PS models. The frequency of the GWs increases with increasing $\omega/\mu$, {\it i.e.}, with decreasing mass and compactness of the PSs. The morphology of the waveforms changes as well: the less compact the stars, the longer the pre-collapse signal before black-hole formation, which corresponds to the peak emission and it is followed by the ringdown phase. For high $\omega/\mu$ collisions, the transient hypermassive PS that results from the merger has a total mass that is closer (as $\omega/\mu$ grows) to the maximum mass that defines the linear stability limit of such objects, therefore surviving for a longer time as it emits GWs before collapsing to a black hole.

\begin{figure*}[t!]
\includegraphics[width=0.33\linewidth]{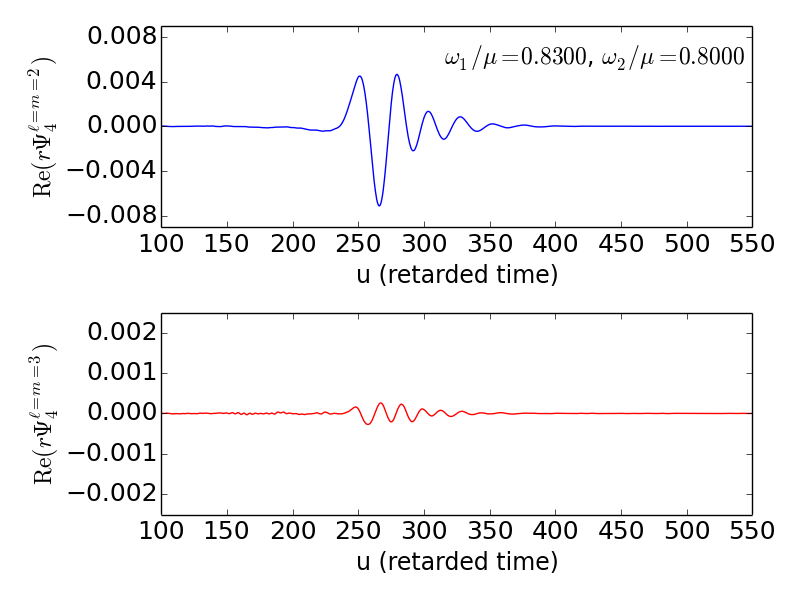}
\includegraphics[width=0.33\linewidth]{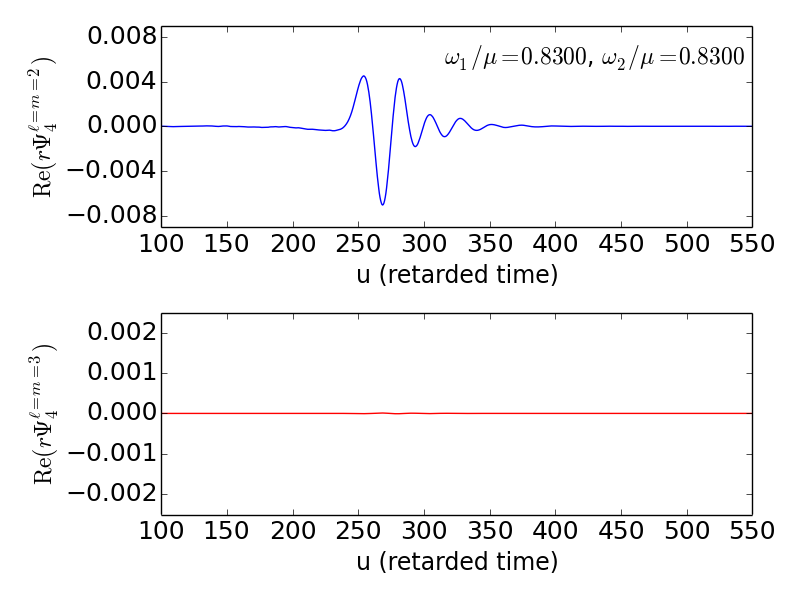}\includegraphics[width=0.328\linewidth]{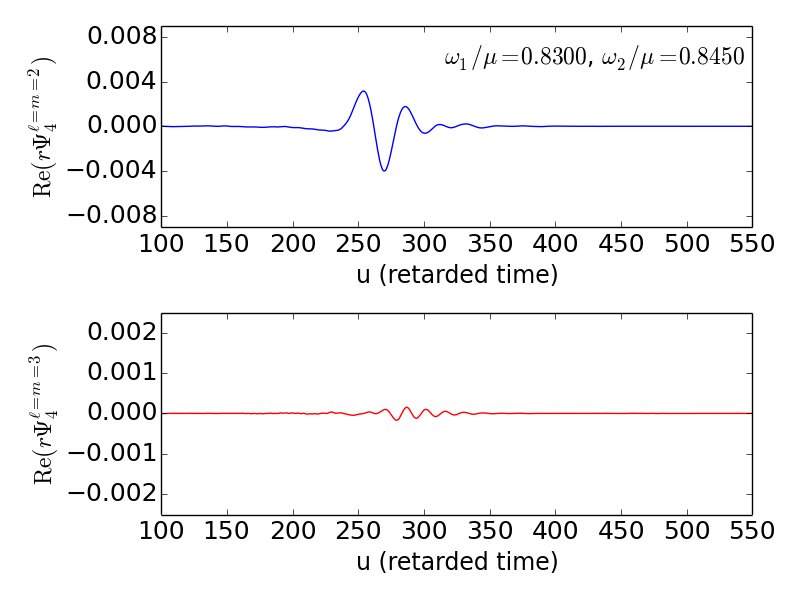}\\\includegraphics[width=0.33\linewidth]{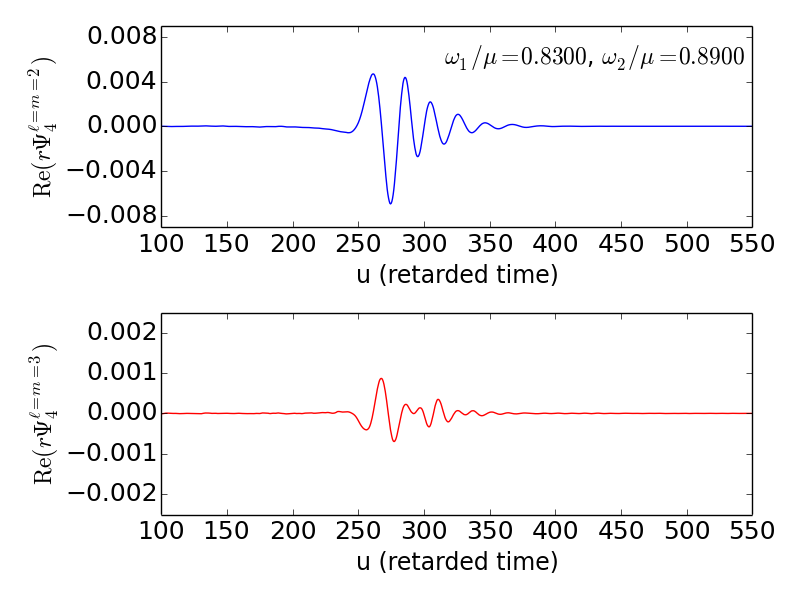}\includegraphics[width=0.328\linewidth]{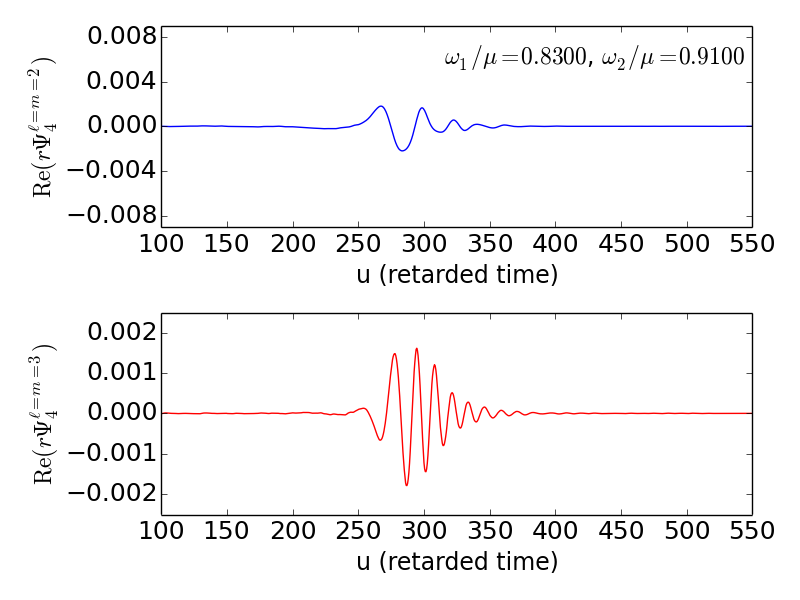}\includegraphics[width=0.33\linewidth]{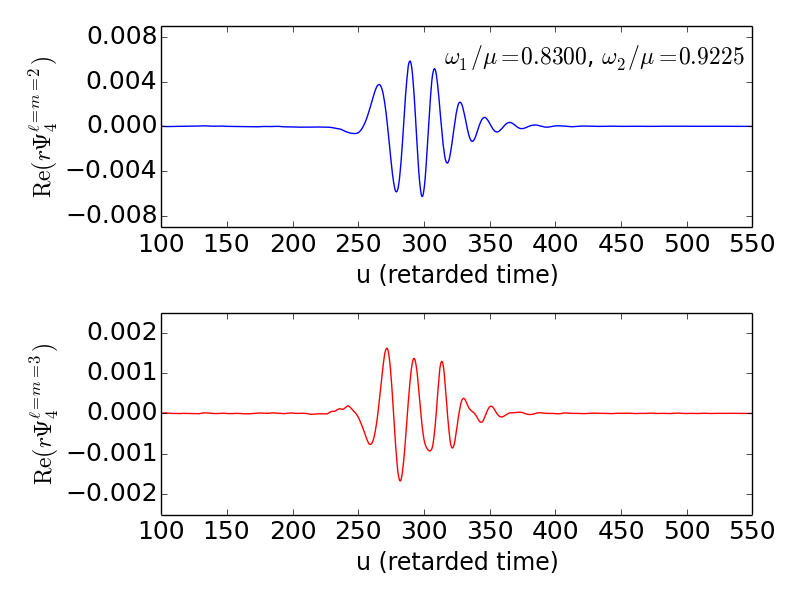}
\\
\caption{$\ell= m=2$ (blue lines) and $\ell= m=3$ (red lines) modes of $r\Psi_{4}^{\ell m}$ for six unequal-mass PS collisions with fixed $\omega_1/\mu=0.8300$. For animations of the full set of GW signals from unequal-mass collisions see~\cite{GWmovies}.}
\label{figGWunequal}
\end{figure*}

Fig.~\ref{figGWunequal} shows the $\ell= m=2$ and $\ell= m=3$ modes of $\Psi_4$ for one equal-mass and five unequal-mass PS binary mergers, with fixed $\omega_1/\mu=0.8300$ and varying $\omega_2/\mu$. The waveforms look similar to those for the equal-mass cases in terms of shape, duration, and frequency. However, they also exhibit important differences. First, while in equal-mass collisions odd-$m$ modes (e.g.~the $\ell= m=3$ mode) are almost completely suppressed (modulo numerical noise) due to the symmetries of the problem compared to the dominant $\ell=m=2$ (see top middle panel of Fig.~\ref{figGWunequal}), these are triggered for unequal-mass systems and can have a significant contribution (see also~\cite{Bezares:2022obu}). In addition, and most importantly, the morphology of the $\ell=m=2$ mode manifests a clear non-monotonic dependence on the frequency of the secondary star $\omega_2/\mu$ for fixed $\omega_1/\mu$. In particular, the waveform amplitude varies periodically as we increase $\omega_2/\mu$ from 0.8000 to 0.9300. For example, for a value of $\omega_1/\mu=0.8300$, we find that the amplitude maxima correspond to $\omega_2/\mu$ equal to 0.8000, 0.8300, 0.8600, 0.8900, and 0.9225, while the minima are found when $\omega_2/\mu$ is equal to 0.8150, 0.8450, 0.8750, and 0.9100. This effect is not present in mergers of other types of compact objects as binary black holes or binary neutron stars.

\begin{figure*}[t!]
\includegraphics[width=0.48\linewidth]{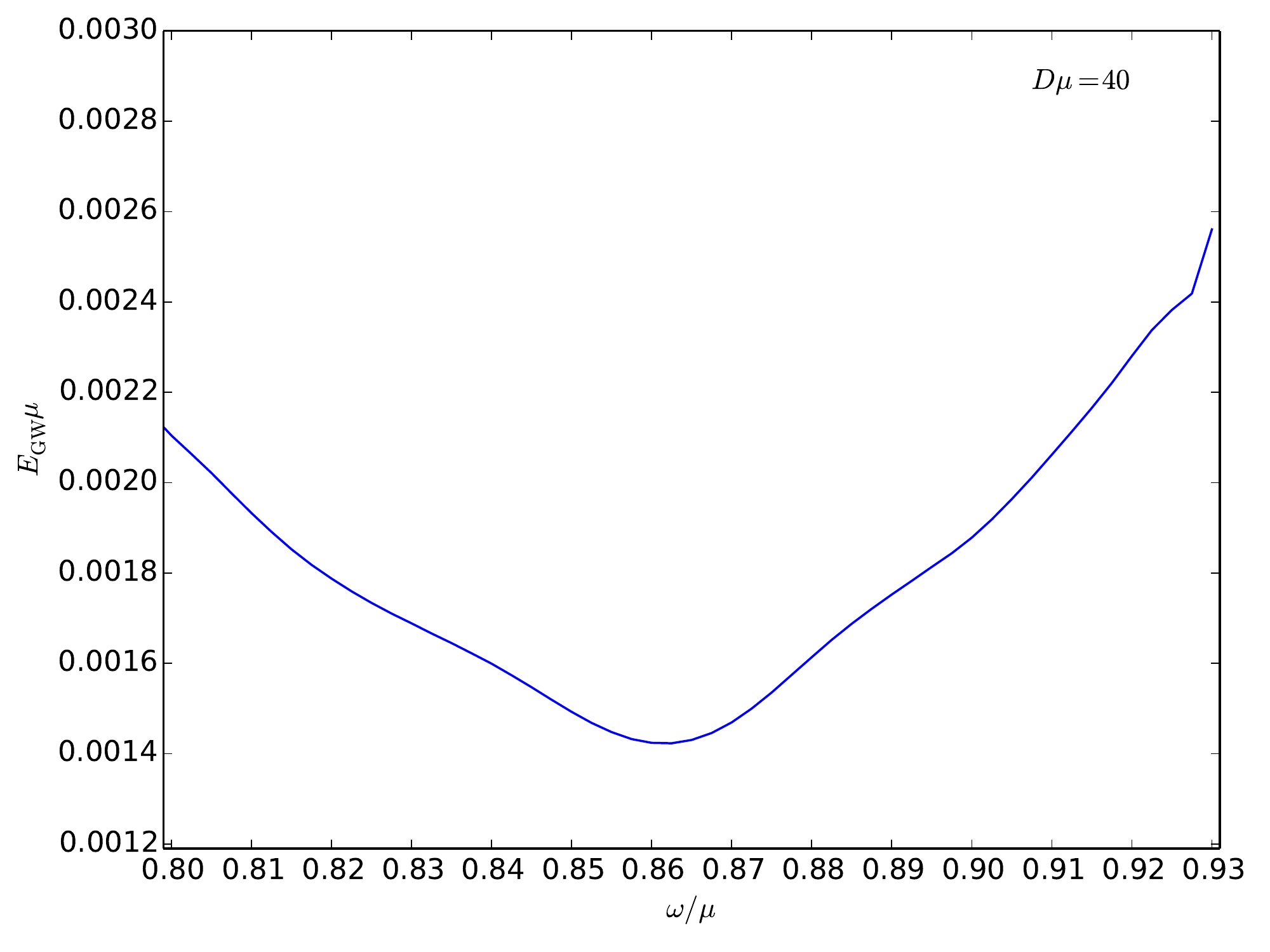}
\includegraphics[width=0.48\linewidth]{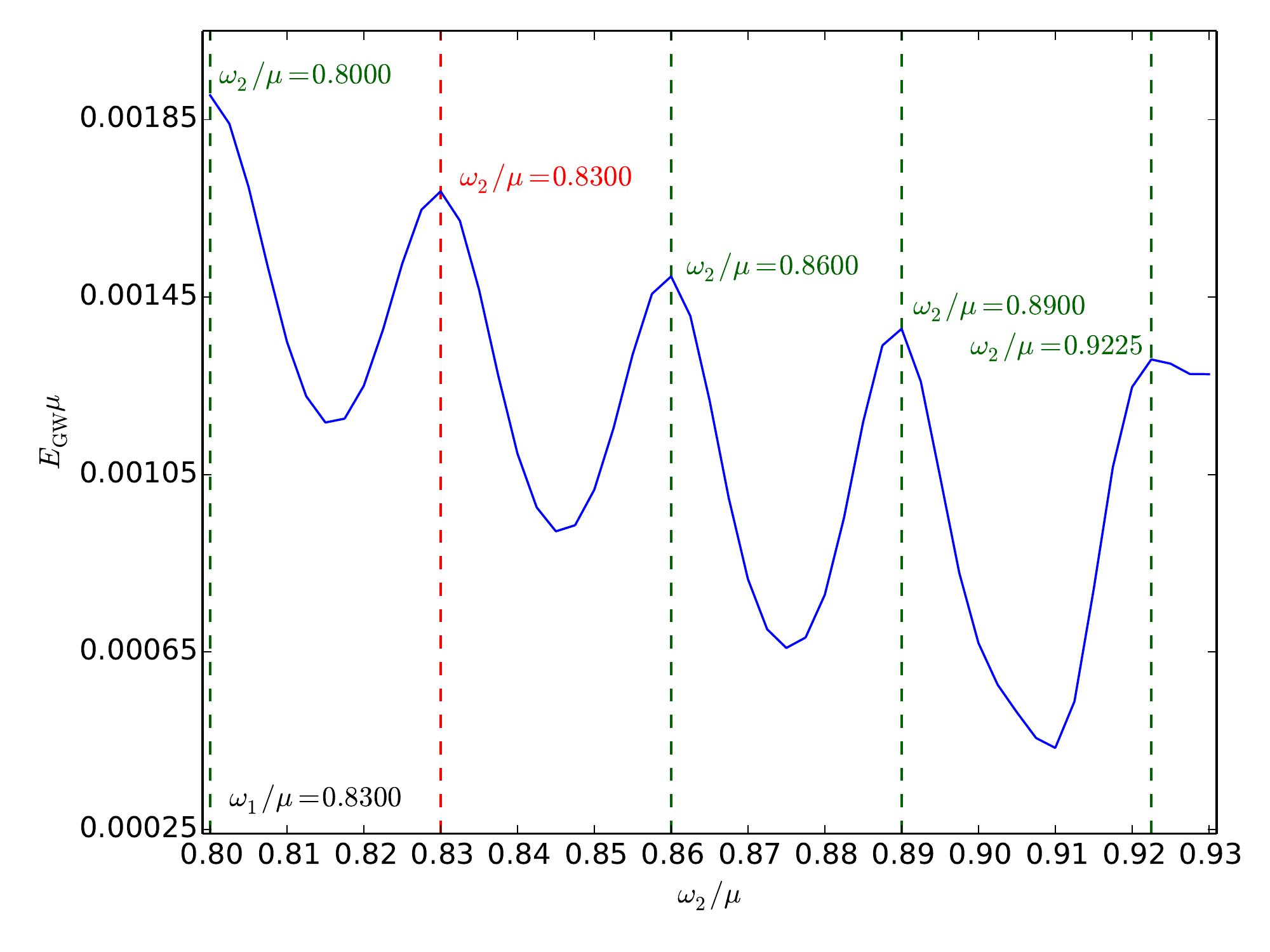}\\
\includegraphics[width=0.48\linewidth]{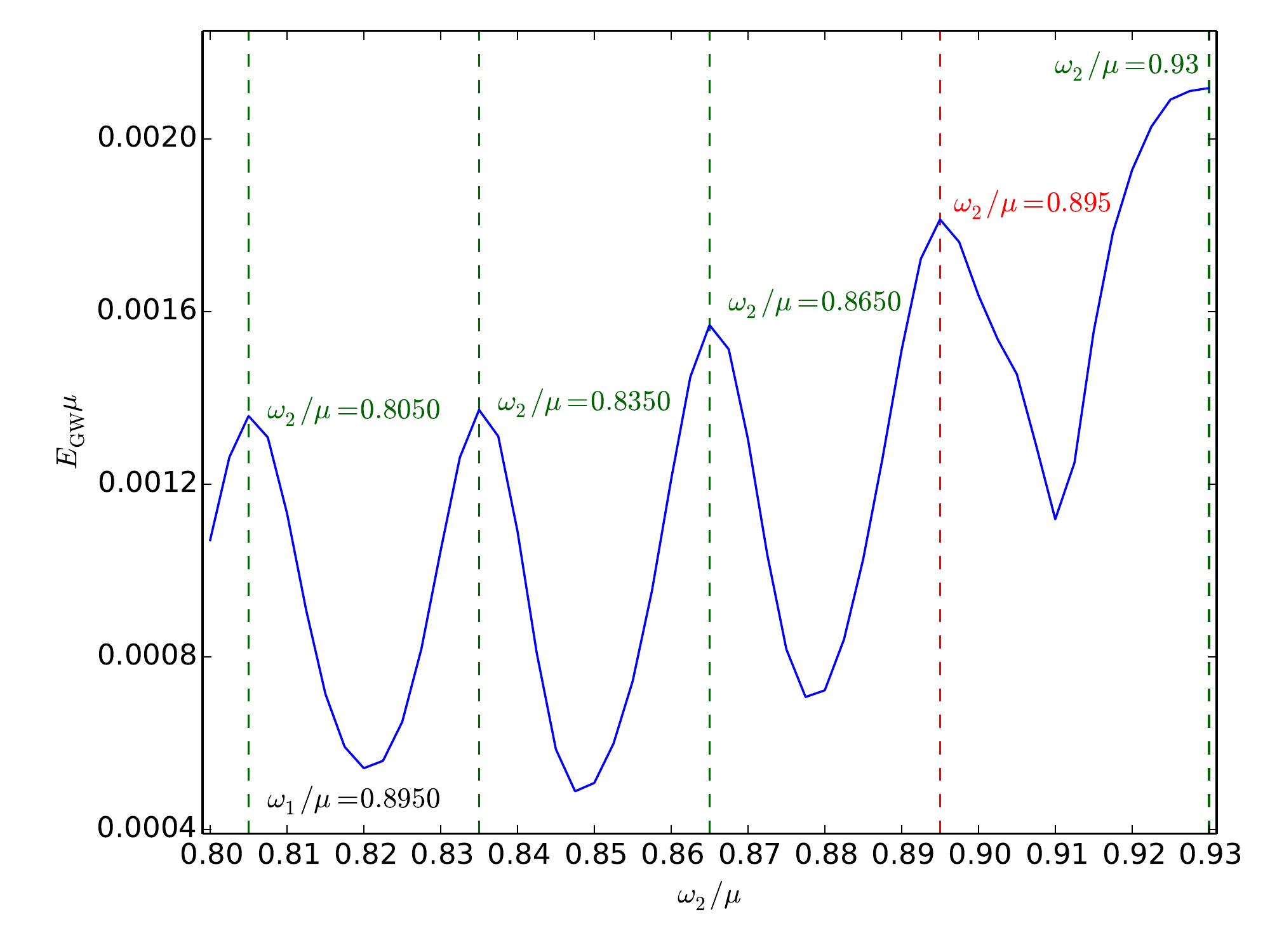}
\includegraphics[width=0.48\linewidth]{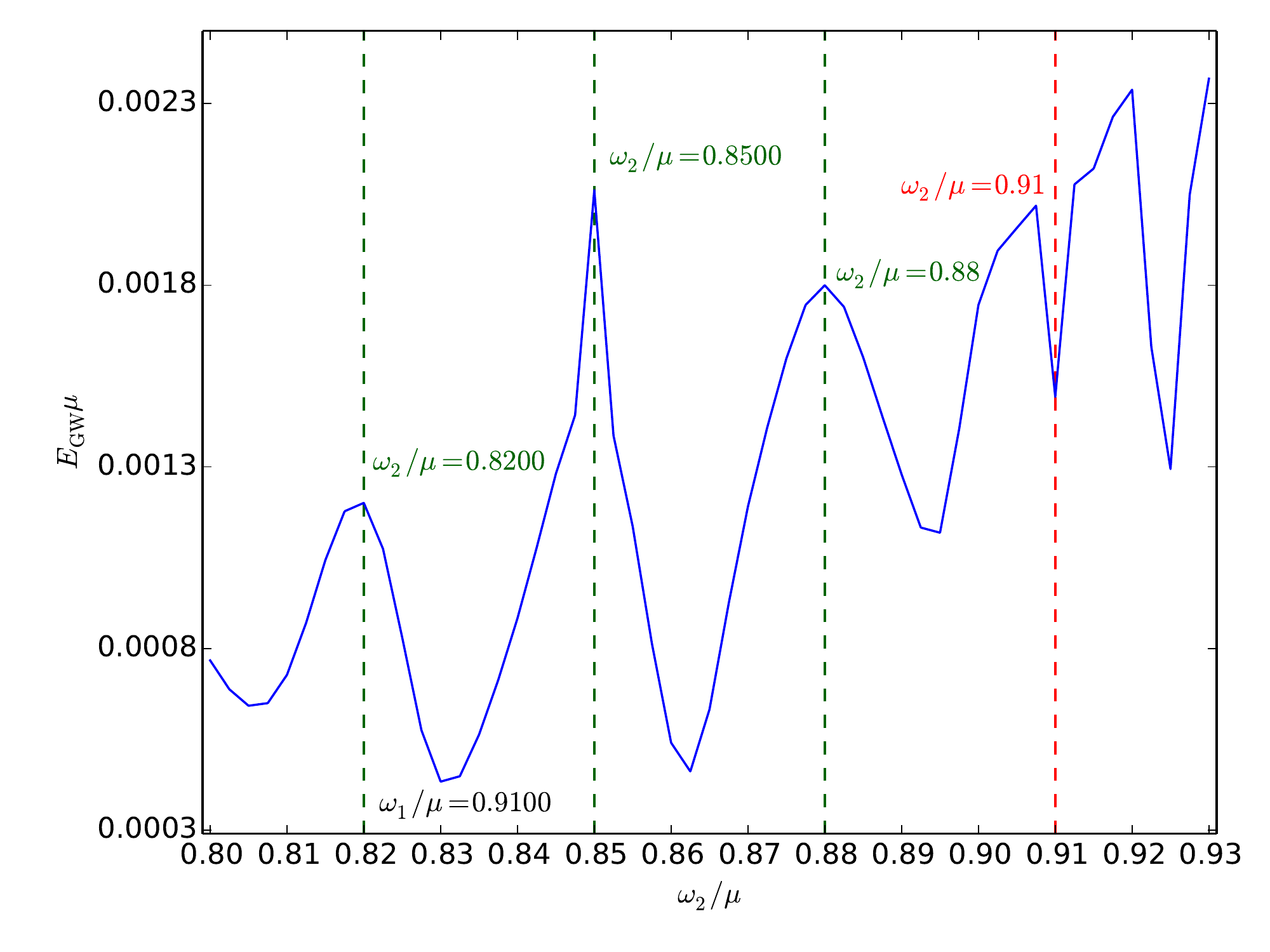}

\caption{Total GW energy as a function of $\omega/\mu$ in the equal-mass case (top left panel), and three unequal-mass cases: $\omega_1/\mu=0.8300$ (top right panel), $\omega_1/\mu=0.895$ (bottom left panel) and $\omega_1/\mu=0.91$ (bottom right panel). }
\label{figGWenergy}
\end{figure*}

The non-trivial dependence of the gravitational radiation with $\omega_2/\mu$ for fixed $\omega_1/\mu$ becomes more evident when studying the total emitted energy from the GW luminosity, given by
\begin{equation}
L_{\rm GW}=\frac{dE}{dt}=\lim_{r\rightarrow\infty}\frac{r^{2}}{16\pi}\sum^{\infty}_{l=2}\sum^{l}_{m=-l}\biggl|\int^{t}_{-\infty}dt'\Psi_{4}^{lm}\biggl|^{2}\label{eq:LGW}.
\end{equation}
Fig.~\ref{figGWenergy} shows the total GW energy as a function of $\omega/\mu$ or $\omega_2/\mu$ for the equal-mass (top left panel) and three illustrative unequal-mass cases, corresponding to fixed values of $\omega_1/\mu=\lbrace0.8300,0.8950,0.9100\rbrace$ (top right, bottom left, and bottom right panels of Fig.~\ref{figGWenergy}, respectively). In the equal-mass case the emitted energy decreases for decreasing $\omega/\mu$ reaching a minimum at $\omega/\mu\sim0.8625$ and increasing onwards. While naively one would expect that the emitted energy would primarily depend on the total mass and compactness of the stars, the described trend depends in a non-trivial way on the dynamics of the binary system, the trajectories followed by the stars due to frame-dragging, and the masses and angular momentum of the PSs. 

On the other hand, the unequal-mass cases yield interesting results already hinted above. We find that the GW energy displays a distinctive oscillatory pattern as a function of $\omega_2/\mu$ for fixed $\omega_1/\mu$. As Fig.~\ref{figGWenergy} shows, the energy maxima are located at intervals of $\Delta\omega_{\rm{max}} /\mu=(\omega_1-\omega_2)/\mu\sim  \,k\,0.03$ and the minima are located at intervals of $\Delta\omega_{\rm{min}} /\mu\sim\,(2k+1)\,0.015$ with $k\in\mathbb{Z}$. The value of these two intervals between maxima or minima, $\Delta\omega_{\rm{min}} /\mu$ and $\Delta\omega_{\rm{max}} /\mu$, are completely independent of $\omega_1/\mu$. This result can be explained by the wave-like nature of PSs and their fundamental oscillation frequency, which leads to an interference between the different frequencies in the unequal-mass case. The interference behaviour was already found in equal-mass head-on collisions of scalar boson stars with a non-zero initial phase difference~\cite{lai2004numerical,palenzuela2007head,choi2010dynamics,bezares2017final}, but its impact on the GW emission was not systematically explored. 

\subsection{The role of the relative phase at merger}

To explain the GW emission pattern, we assume that at the time of the collision we have a linear superposition of both stars (same Proca field) oscillating at different frequencies. Then, removing the $\bar m\varphi$-dependence which will not affect the interference and the initial phase $\epsilon$, it can be shown that 
\begin{equation}\label{interference}
\begin{split}
\rm{Re}(\mathcal{A})&\sim\cos(\omega_1\, t) + \cos(\omega_2\, t)
\\&=2\cos\biggl(\frac{(\omega_1+\omega_2)}{2}t\biggl)\cos\biggl(\frac{(\omega_1-\omega_2)}{2}t\biggl)\,,\\
\rm{Im}(\mathcal{A})&\sim\sin(\omega_1\, t) + \sin(\omega_2\, t)
\\&=2\sin\biggl(\frac{(\omega_1+\omega_2)}{2}t\biggl)\cos\biggl(\frac{(\omega_1-\omega_2)}{2}t\biggl).
\end{split}
\end{equation}
Therefore, the complex amplitude of the Proca field will be given by 
\begin{equation}\label{amplitude}
\begin{split}
|\mathcal{A}|^2&={\rm Re}(\mathcal{A})^2+{\rm Im}(\mathcal{A})^2\sim4\cos^2\biggl(\frac{(\omega_1-\omega_2)}{2}t\biggl)
\\&=2\bigl[1+\cos\bigl((\omega_1-\omega_2)t\bigl)\bigl].
\end{split}
\end{equation}
\begin{figure}[t!]
\includegraphics[width=1\linewidth]{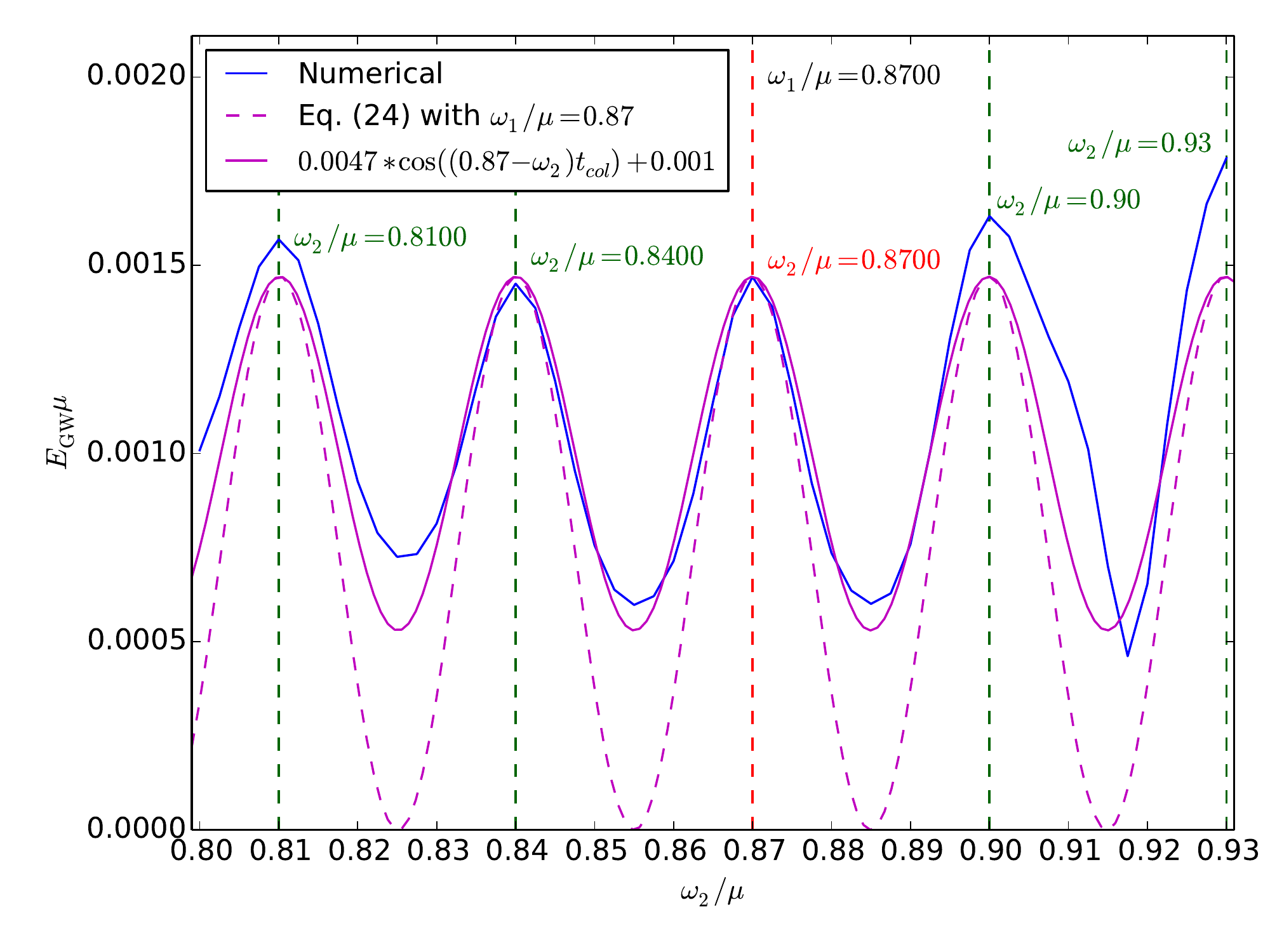}
\caption{Total GW energy for PS head-on collisions with fixed $\omega_1/\mu=0.8700$ and varying $\omega_2/\mu$. The magenta lines correspond to the behaviour of the estimated square of the Proca field amplitude computed from Eq.~(\ref{amplitude}) (dashed line) and from the formula $0.0047\cos((0.87-\omega_2)t_{\rm{col}})+0.001$ (solid line)  as a function of $\omega_2/\mu$ at $t_{\rm{col}}=210$. We fit the analytic expressions to the peak of the model with $\omega_1/\mu=\omega_2/\mu=0.8700$. }
\label{figGWenergy2}
\end{figure}
\begin{figure*}[t!]
\includegraphics[width=0.45\linewidth]{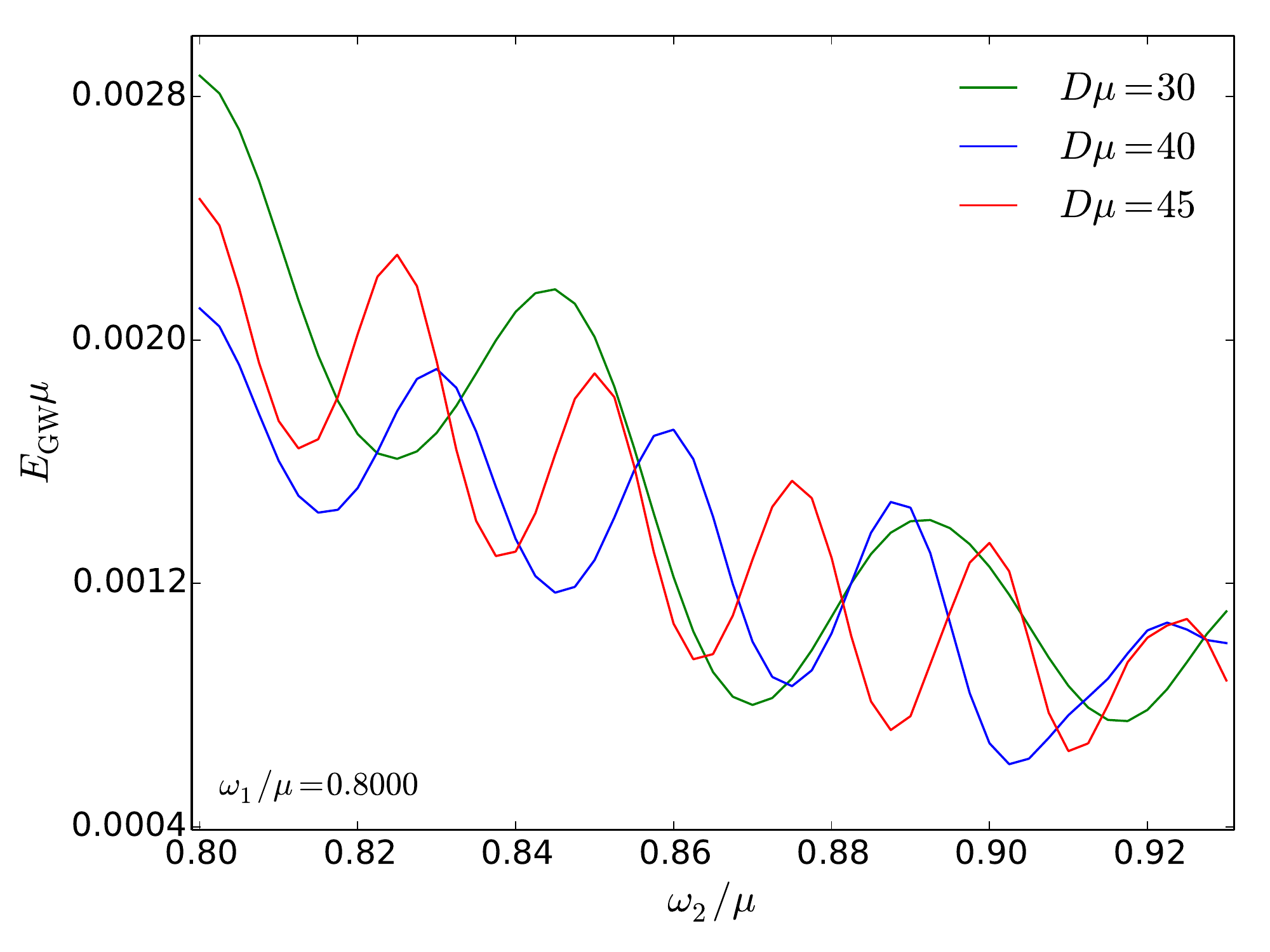}
\includegraphics[width=0.45\linewidth]{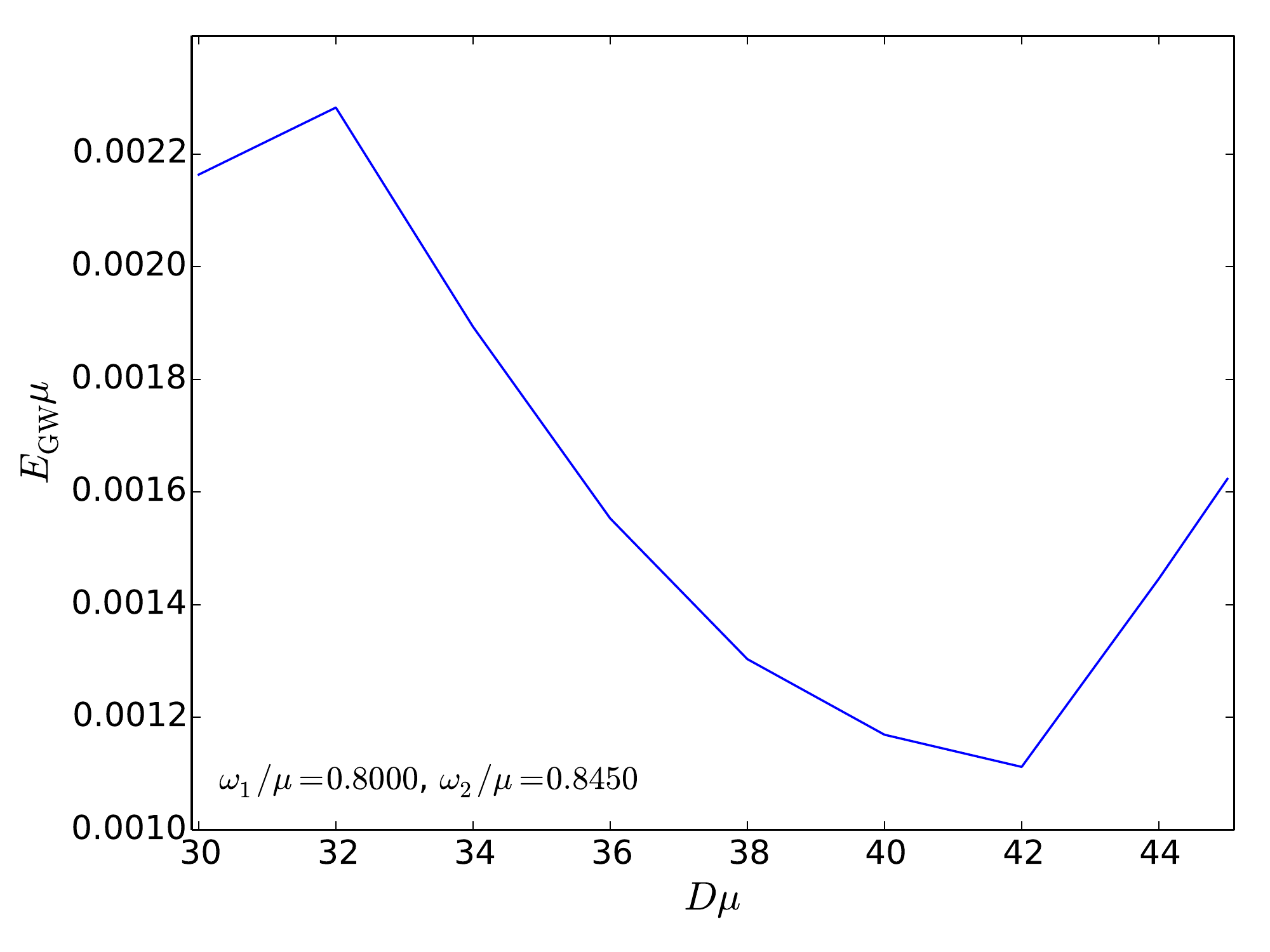}\\
\includegraphics[width=0.473\linewidth]{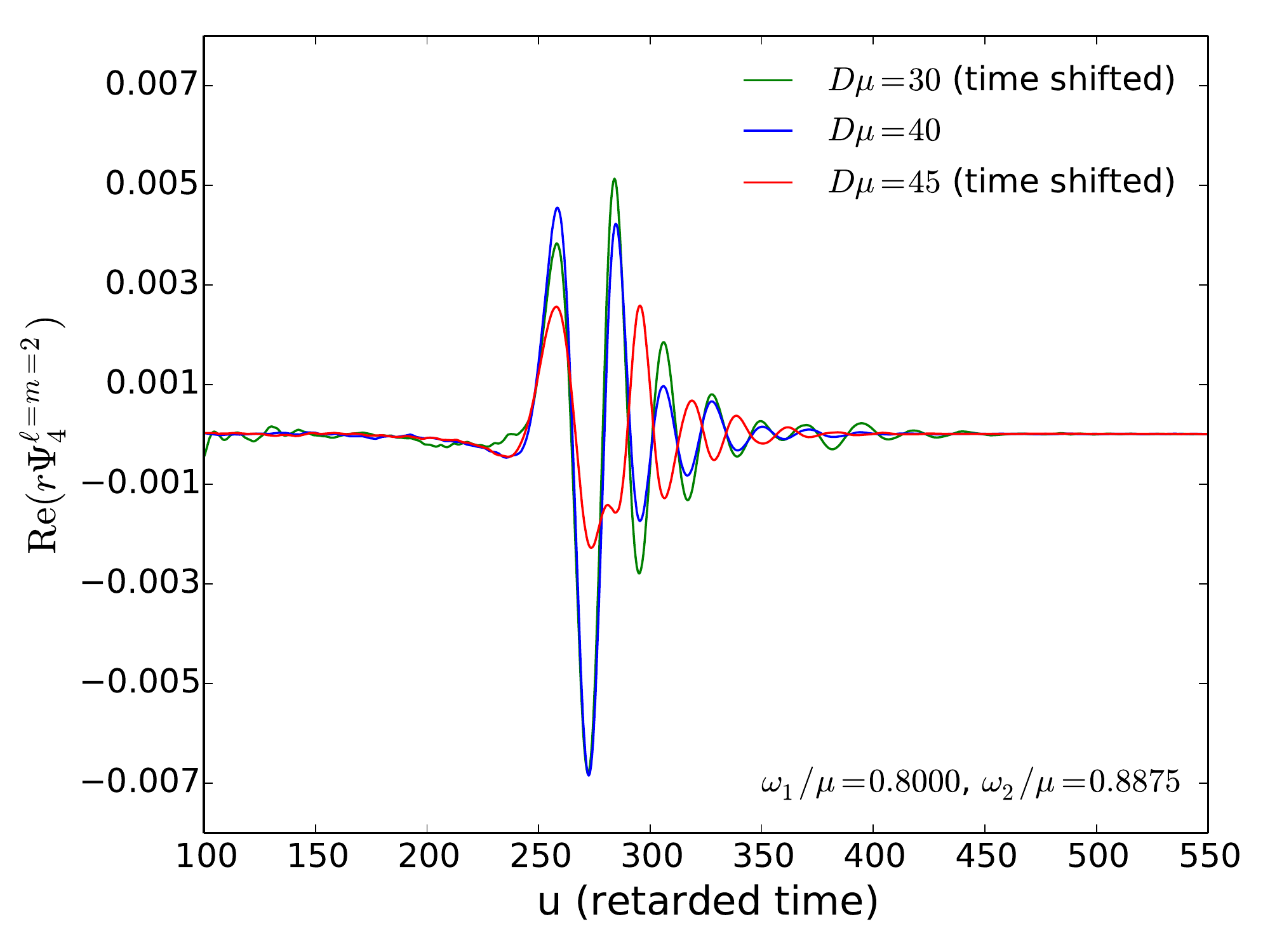}\hspace{-0.4cm}
\includegraphics[width=0.473\linewidth]{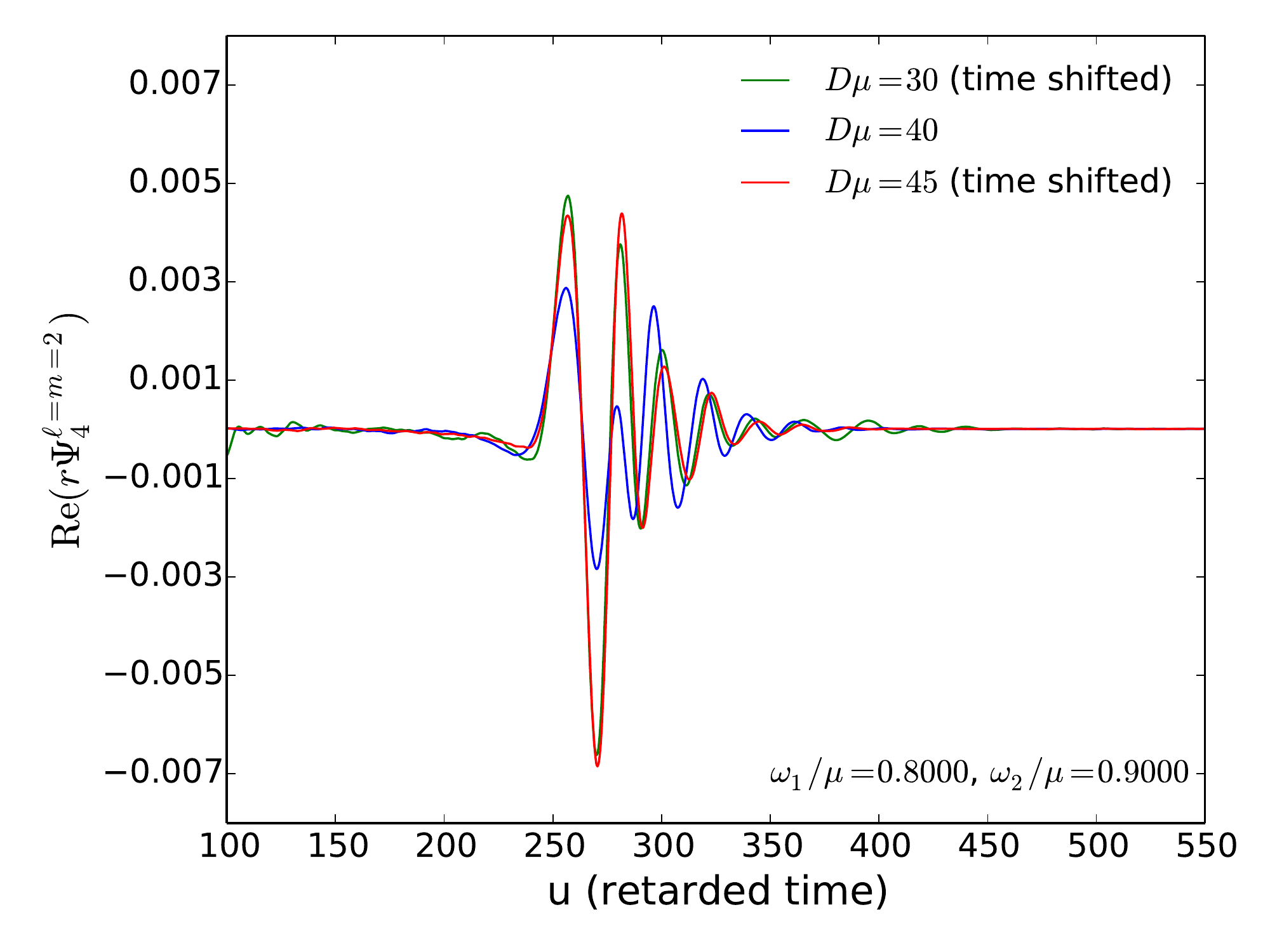}
\caption{Top left panel: Total GW energy for PS head-on collisions with fixed $\omega_1/\mu=0.8000$ and varying $\omega_2/\mu$ for three different initial distances $D\mu=[30,40,45]$. Top right panel: Total GW energy for the unequal-mass case with $\omega_1/\mu=0.8000$ and $\omega_2/\mu=0.8450$ as a function of the distance $D\mu$. Bottom left panel: $\ell=m=2$ waveforms for $\omega_1/\mu=0.8950$. Bottom right panel:  $\ell=m=2$ waveforms for $\omega_1/\mu=0.9100$.} 
\label{figDistances1}
\end{figure*}

Since the initial separation between the stars is the same for all cases, $D\mu=40$, the time of the collision is also approximately the same, $t_{\rm{col}}\mu\sim210$. This is precisely the time at which the maximum (constructive interference) for the envelope in Eqs.~(\ref{interference}) and~(\ref{amplitude}) is reached
\begin{equation}\label{cosinus}
\begin{split}
1+\cos\bigl((\omega_1-\omega_2)t_{\rm{col}}\bigl)_{\rm{max}}&=2\\
\Rightarrow (\omega_1-\omega_2)_{\rm{max}}t_{\rm{col}}&=2k\pi,
\end{split}
\end{equation}
 if $\Delta\omega_{\rm{max}}/\mu=(\omega_1-\omega_2)/\mu\sim k\,0.03$. On the other hand, the minimum (destructive interference) for the same time $t_{\rm{col}}$ is found for
\begin{equation}\label{minimum}
\begin{split}
1+\cos\bigl((\omega_1-\omega_2)t_{\rm{col}}\bigl)_{\rm{min}}&=0\\
\Rightarrow (\omega_1-\omega_2)_{\rm{min}}t_{\rm{col}}&=(2k+1)\pi,
\end{split}
\end{equation}
which gives $\Delta\omega_{\rm{min}}/\mu=(\omega_1-\omega_2)/\mu\sim (2k+1)\,0.015$. This simple linear analysis explains the periodicity between maxima and minima observed in Fig.~\ref{figGWenergy}, which therefore depends on the initial distance between the stars. This analysis, however, must be regarded as an approximation since the emission also depends on other factors such as the dynamics of the collision, the radius of the stars and the time of merger, which could give rise to some additional features in the GW energy, as hinted by the bottom right panel of Fig.~\ref{figGWenergy}. Thus, we anticipate that an increase in $D\mu$ will increase $t_{\rm{col}}$ and will decrease both  $\Delta\omega_{\rm{max}}$ and $\Delta\omega_{\rm{min}}$. Accordingly, if the whole merger takes more time to reach the collapse, the factor $\Delta\omega/\mu$ will be low enough so that its period will be longer than the life of the transient hypermassive PS.  Depending on the amplitude of the envelope, the GW emission could be critically affected. 

\begin{figure}[t!]
\includegraphics[width=0.24\linewidth]{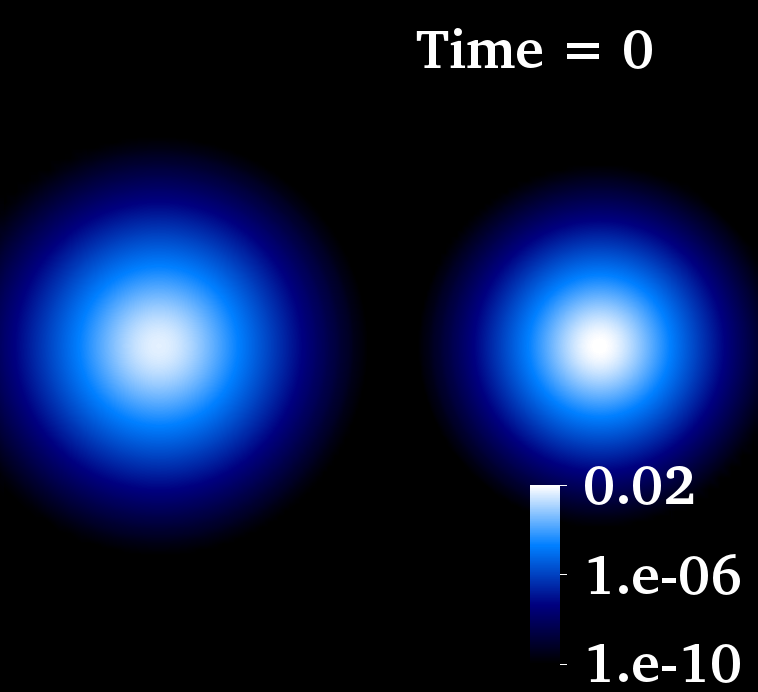}
\includegraphics[width=0.24\linewidth]{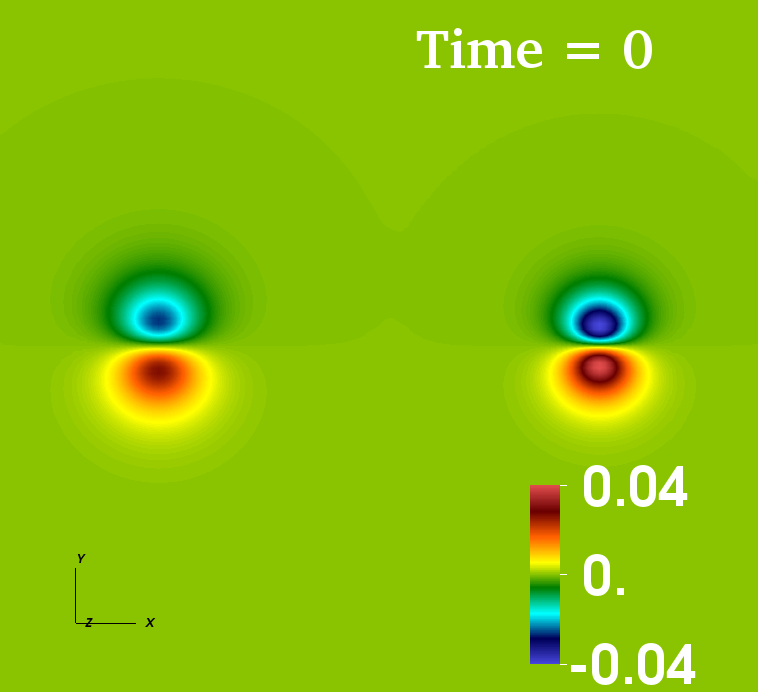}
\includegraphics[width=0.24\linewidth]{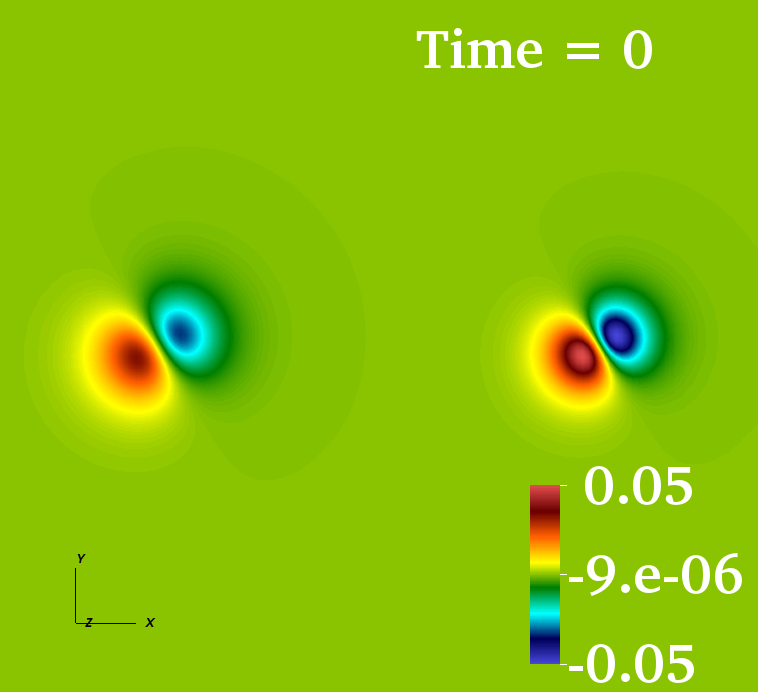}
\includegraphics[width=0.24\linewidth]{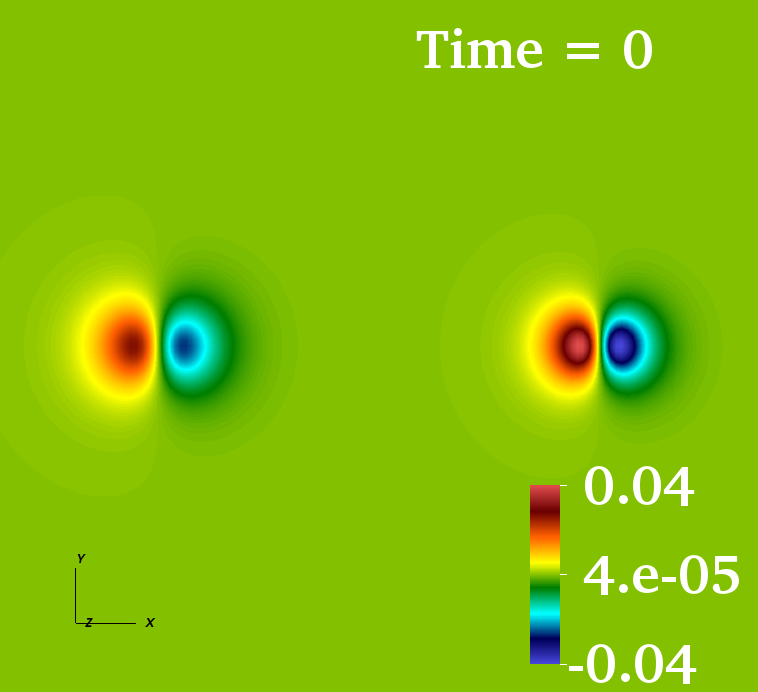}\\%
\includegraphics[width=0.24\linewidth]{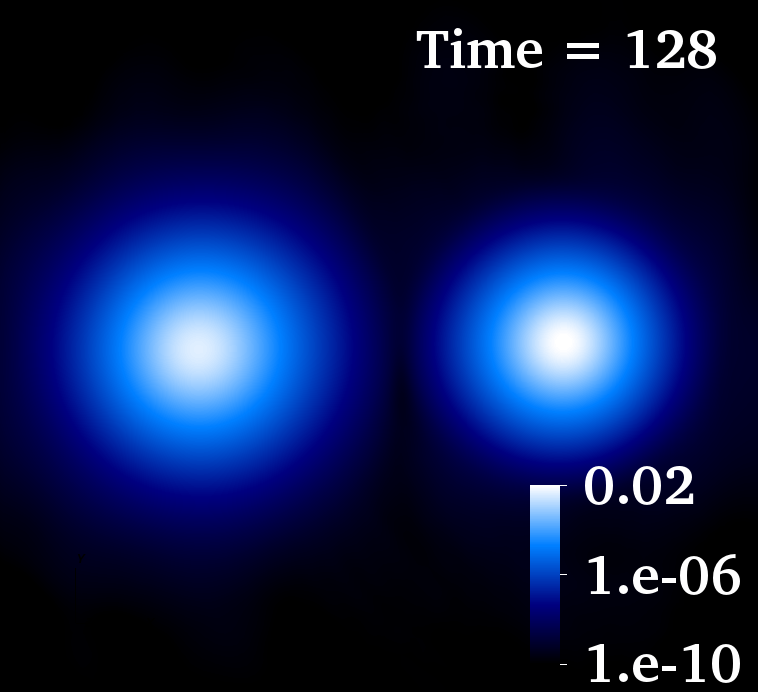}
\includegraphics[width=0.24\linewidth]{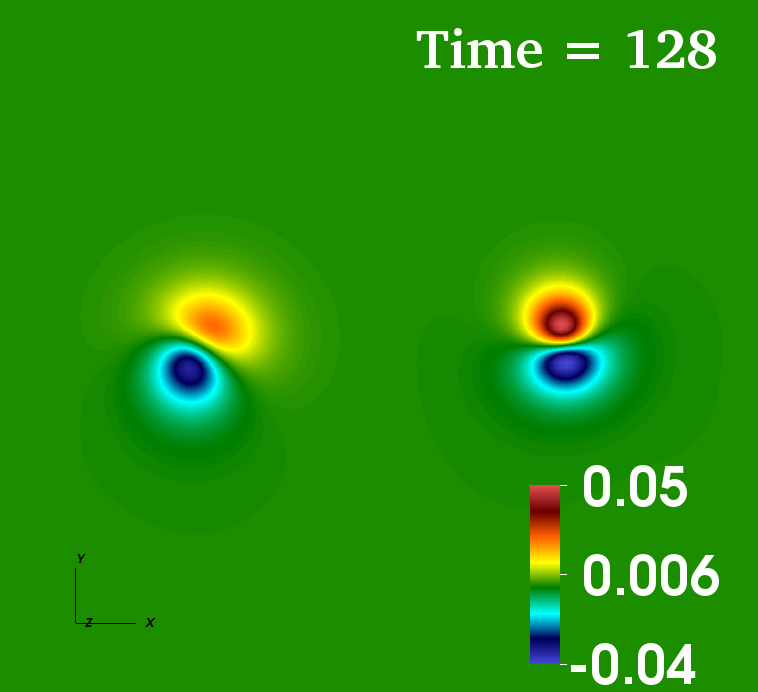}
\includegraphics[width=0.24\linewidth]{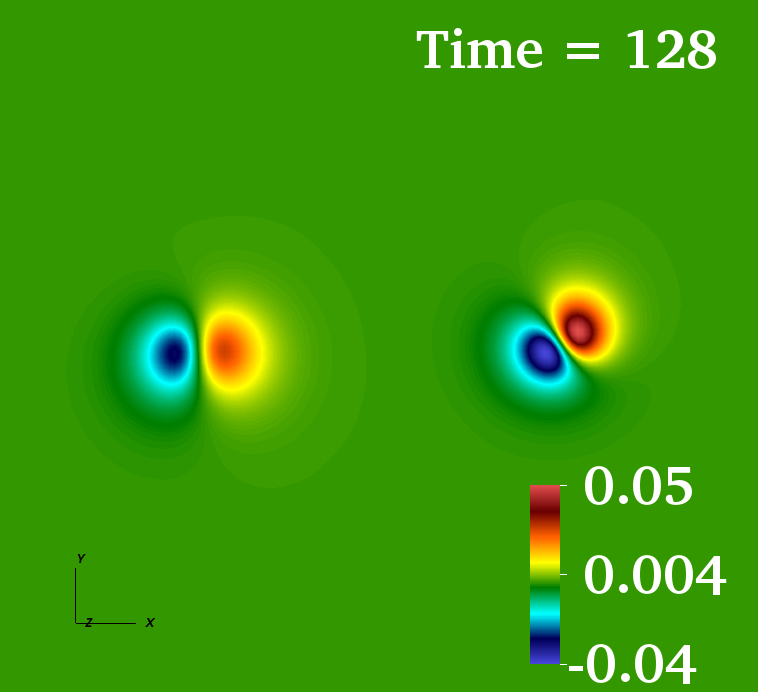}
\includegraphics[width=0.24\linewidth]{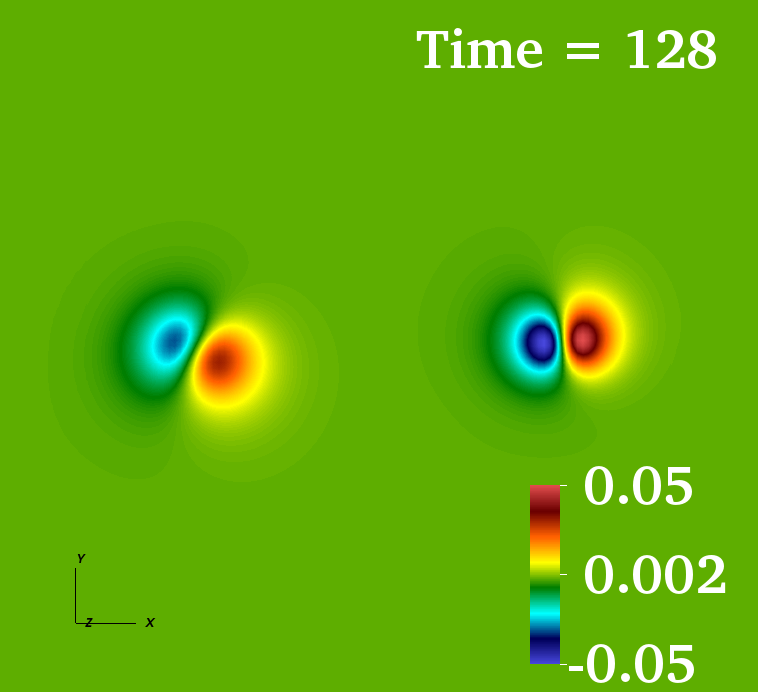}\\%
\includegraphics[width=0.24\linewidth]{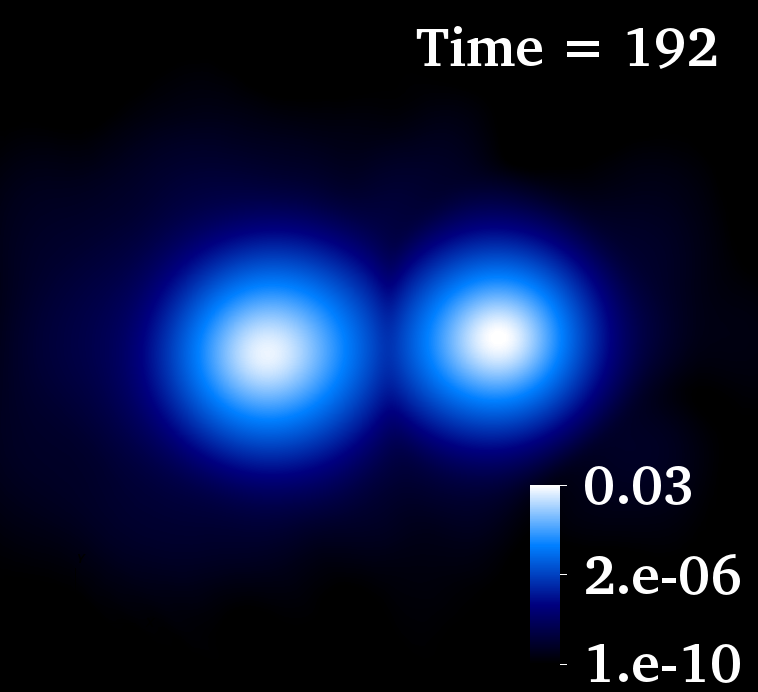}
\includegraphics[width=0.24\linewidth]{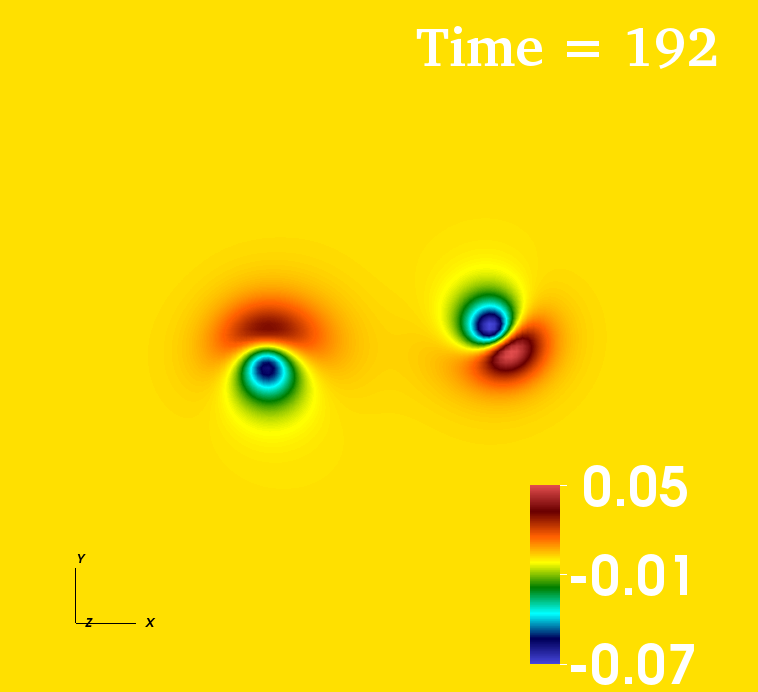}
\includegraphics[width=0.24\linewidth]{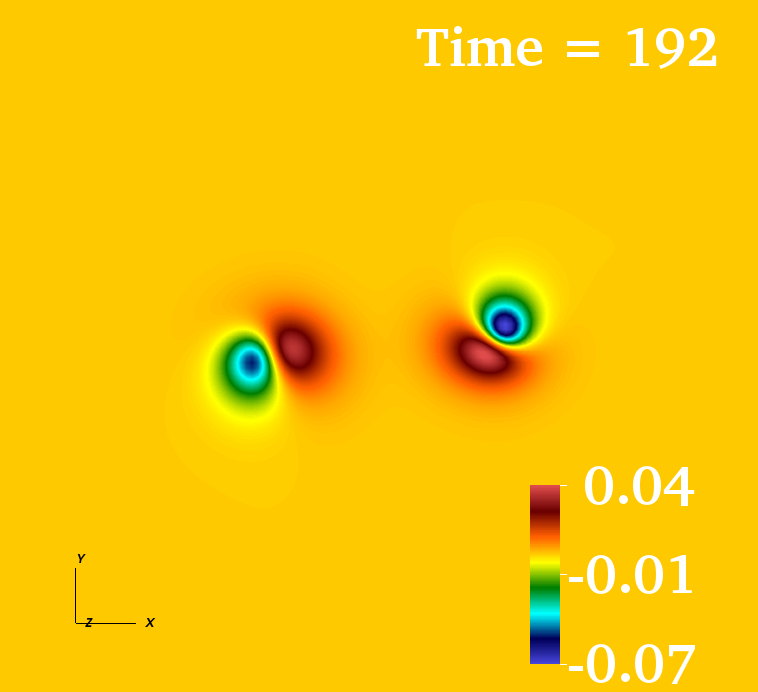}
\includegraphics[width=0.24\linewidth]{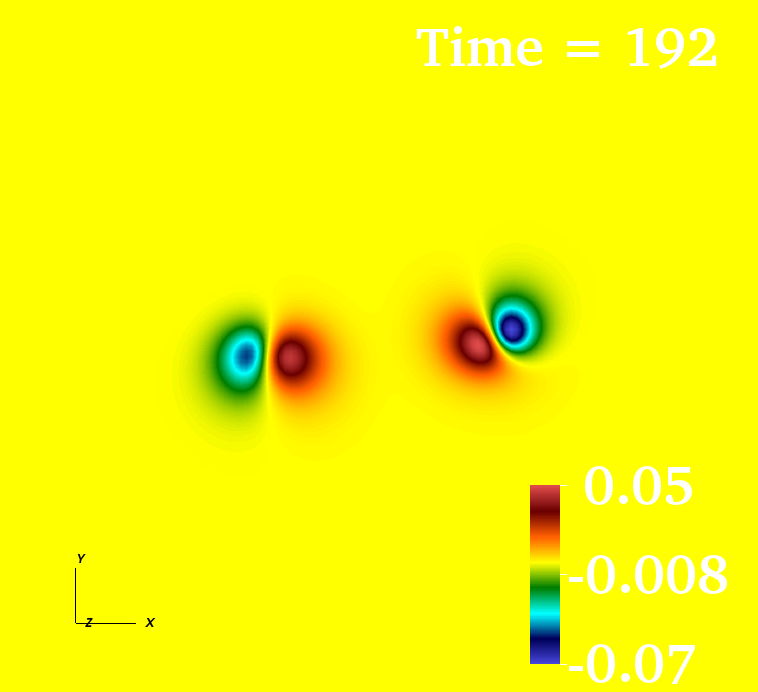}\\%
\includegraphics[width=0.24\linewidth]{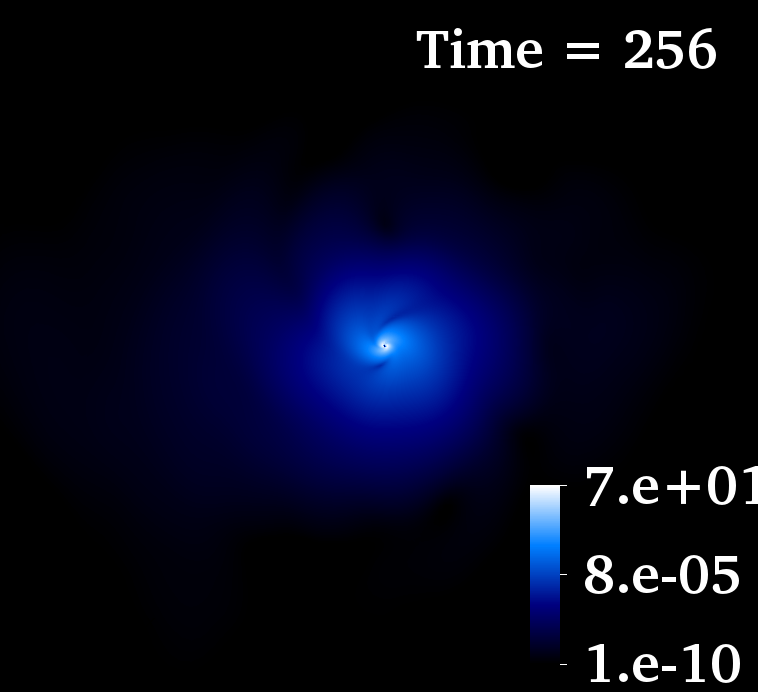}
\includegraphics[width=0.24\linewidth]{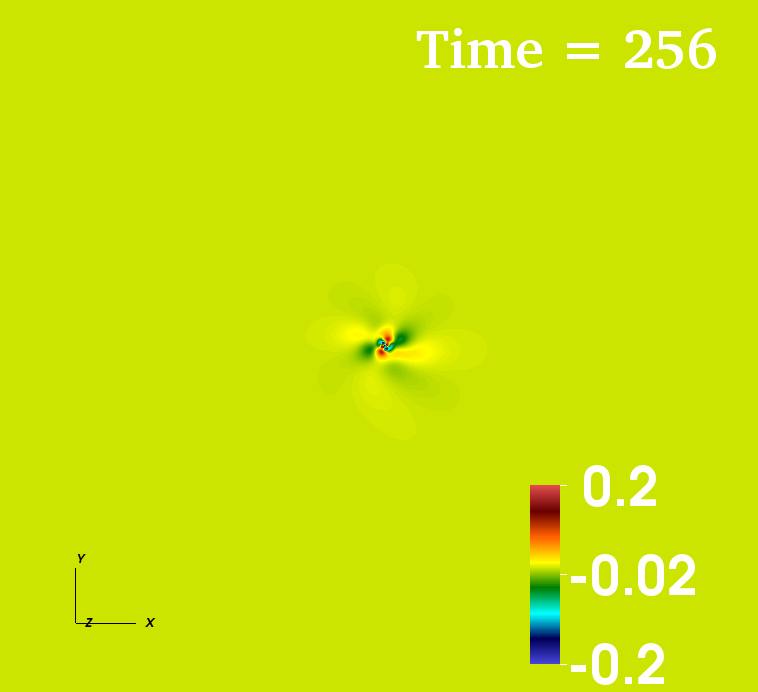}
\includegraphics[width=0.24\linewidth]{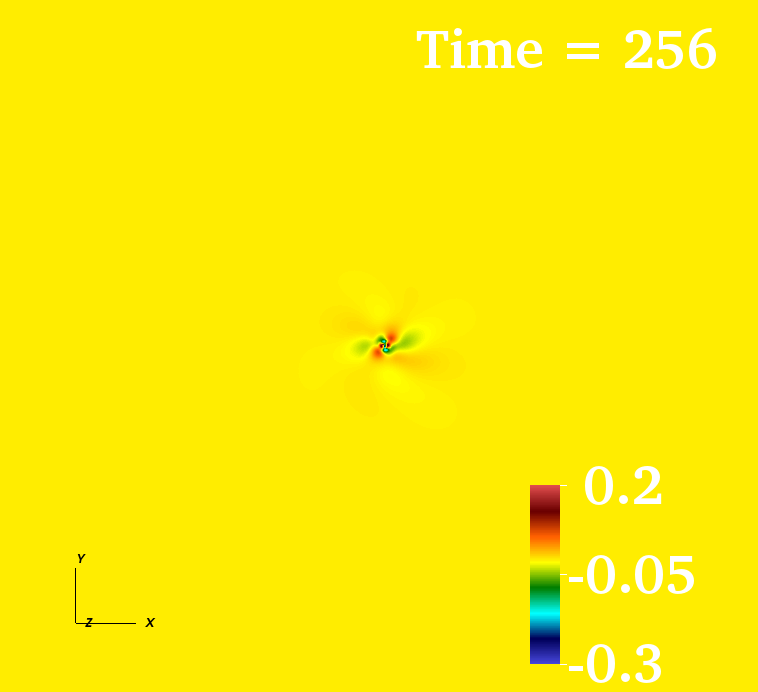}
\includegraphics[width=0.24\linewidth]{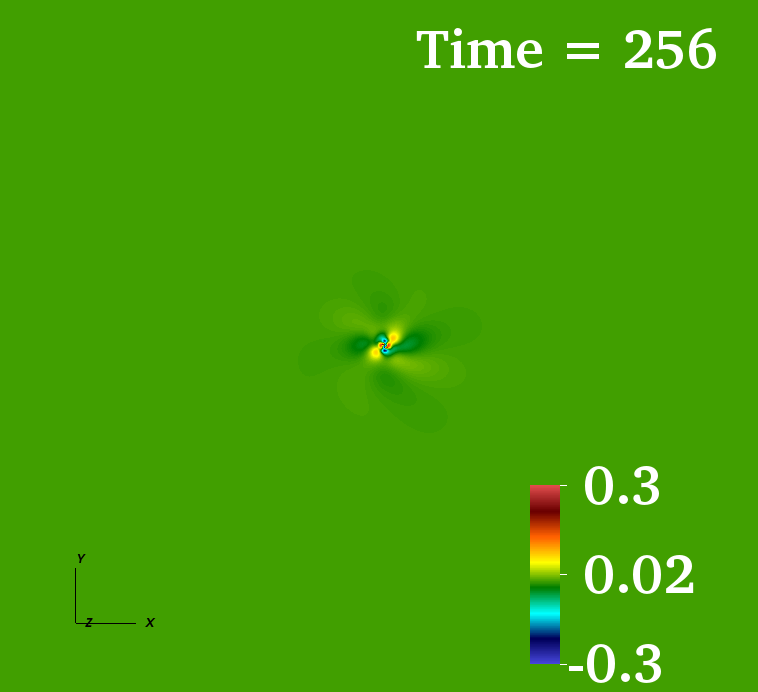}\\%
\includegraphics[width=0.24\linewidth]{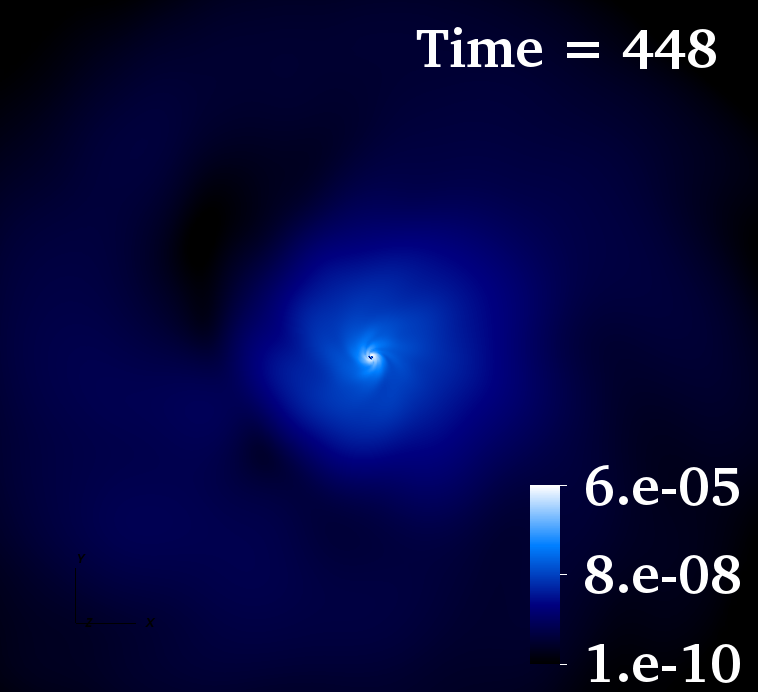}
\includegraphics[width=0.24\linewidth]{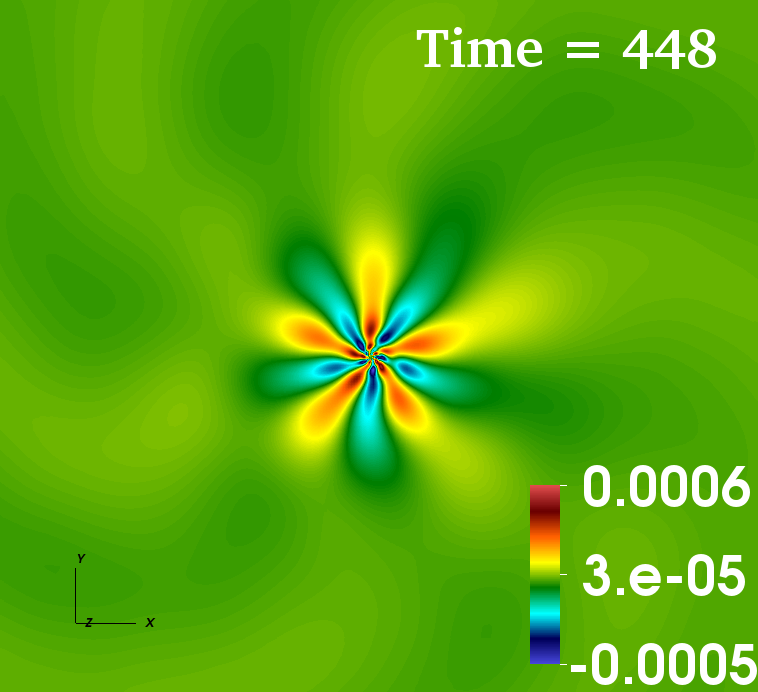}
\includegraphics[width=0.24\linewidth]{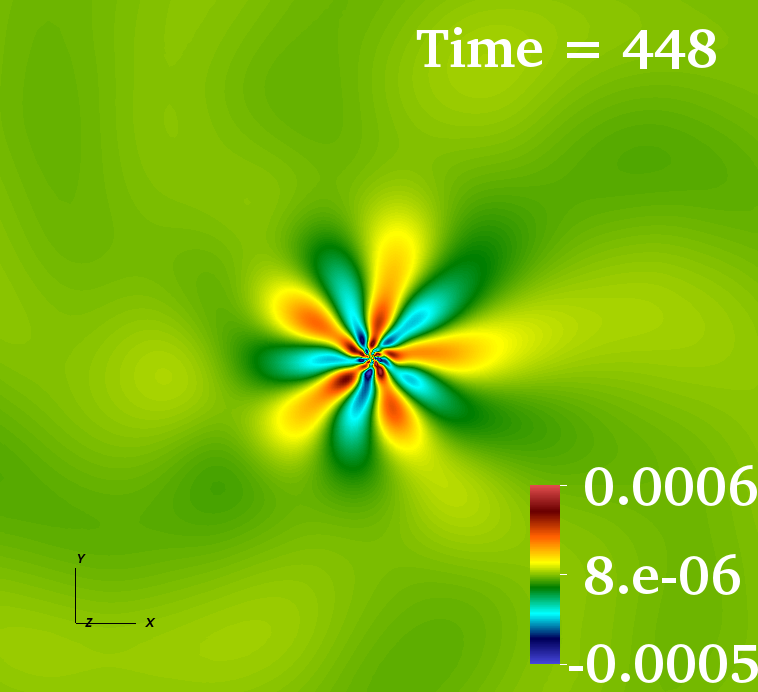}
\includegraphics[width=0.24\linewidth]{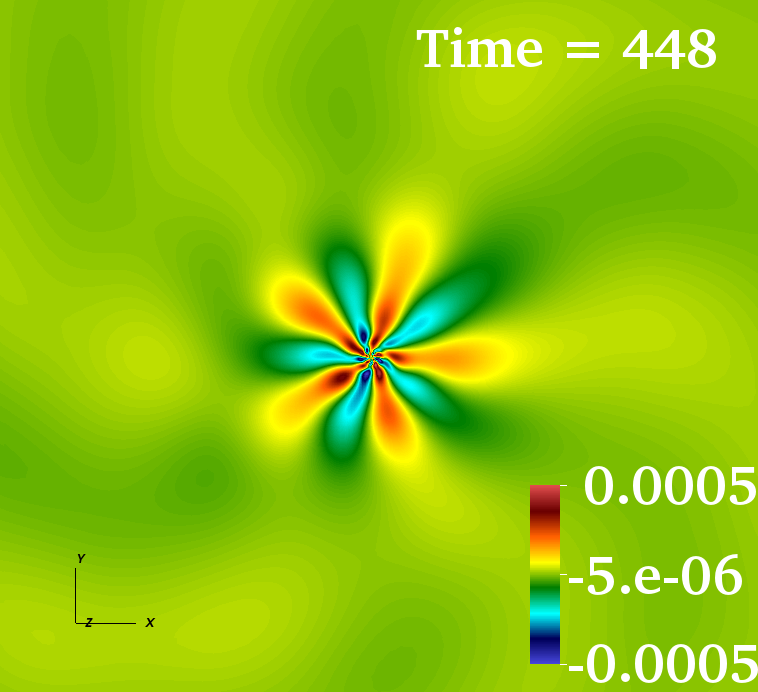}\\%
\includegraphics[width=0.24\linewidth]{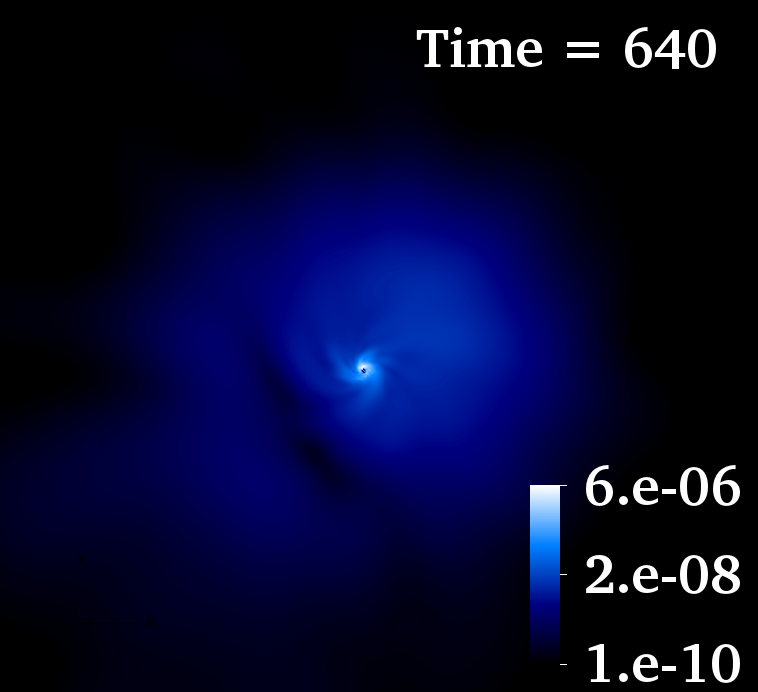}
\includegraphics[width=0.24\linewidth]{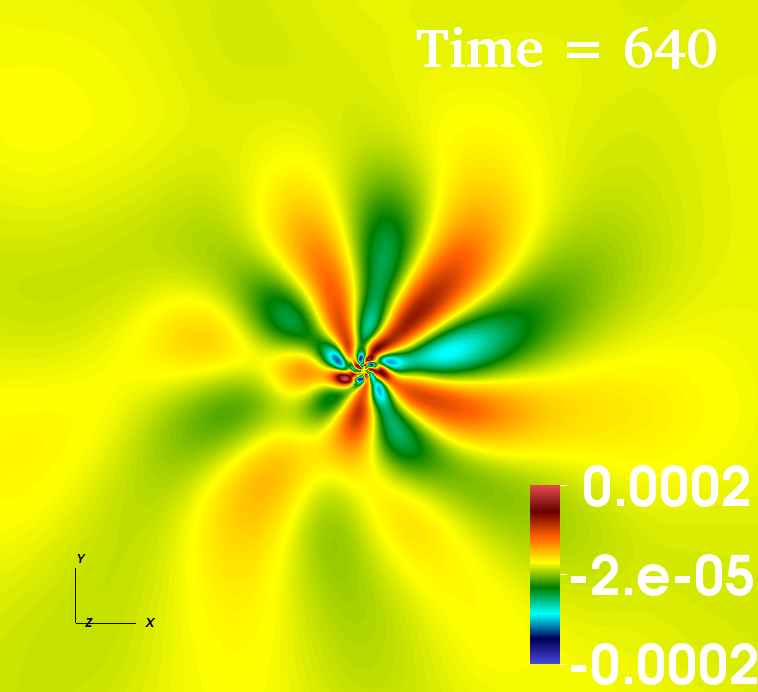}
\includegraphics[width=0.24\linewidth]{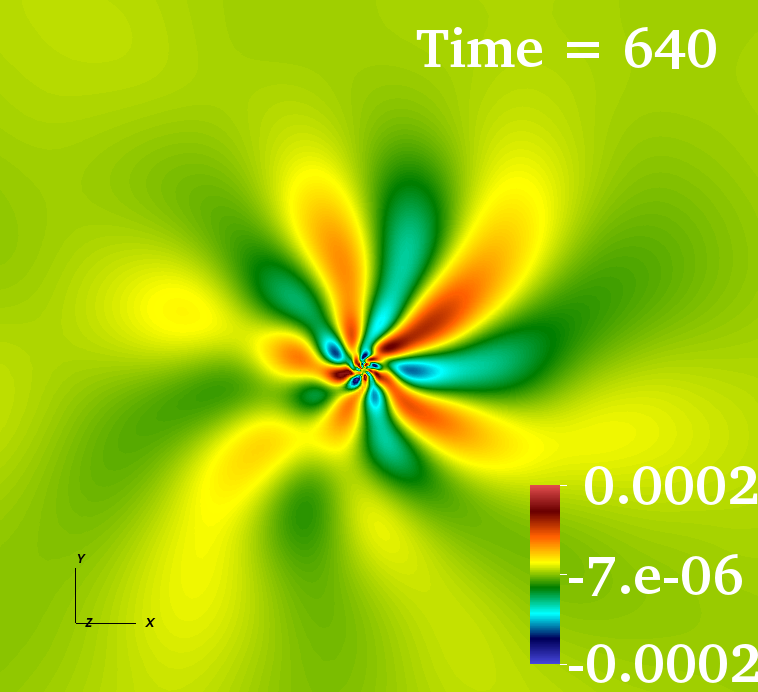}
\includegraphics[width=0.24\linewidth]{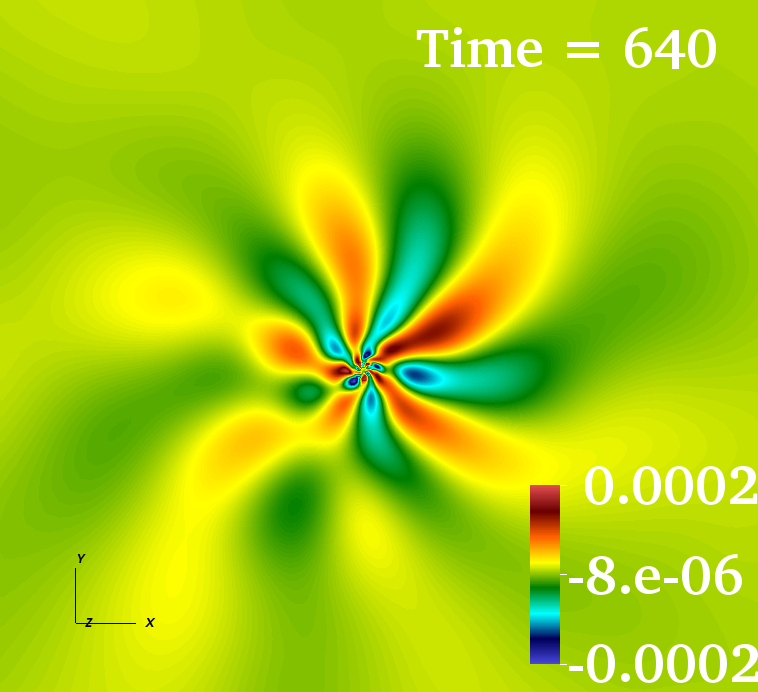}\\%
\caption{Equatorial ($xy$) plane snapshots of the energy density (left column) and the real part of the scalar potential $\mathcal{X}_{\phi}$ (remaining columns) taken during the time evolution of the collisions of spinning PSs with $\omega_1/\mu=0.8000$ and $\omega_2/\mu=0.8450$, changing the initial phase of the stars. Time runs from top to bottom and is given in code units with $G=c=\mu=1$.}
\label{figphases}
\end{figure}
\begin{figure}[t!]
\includegraphics[width=0.98\linewidth]{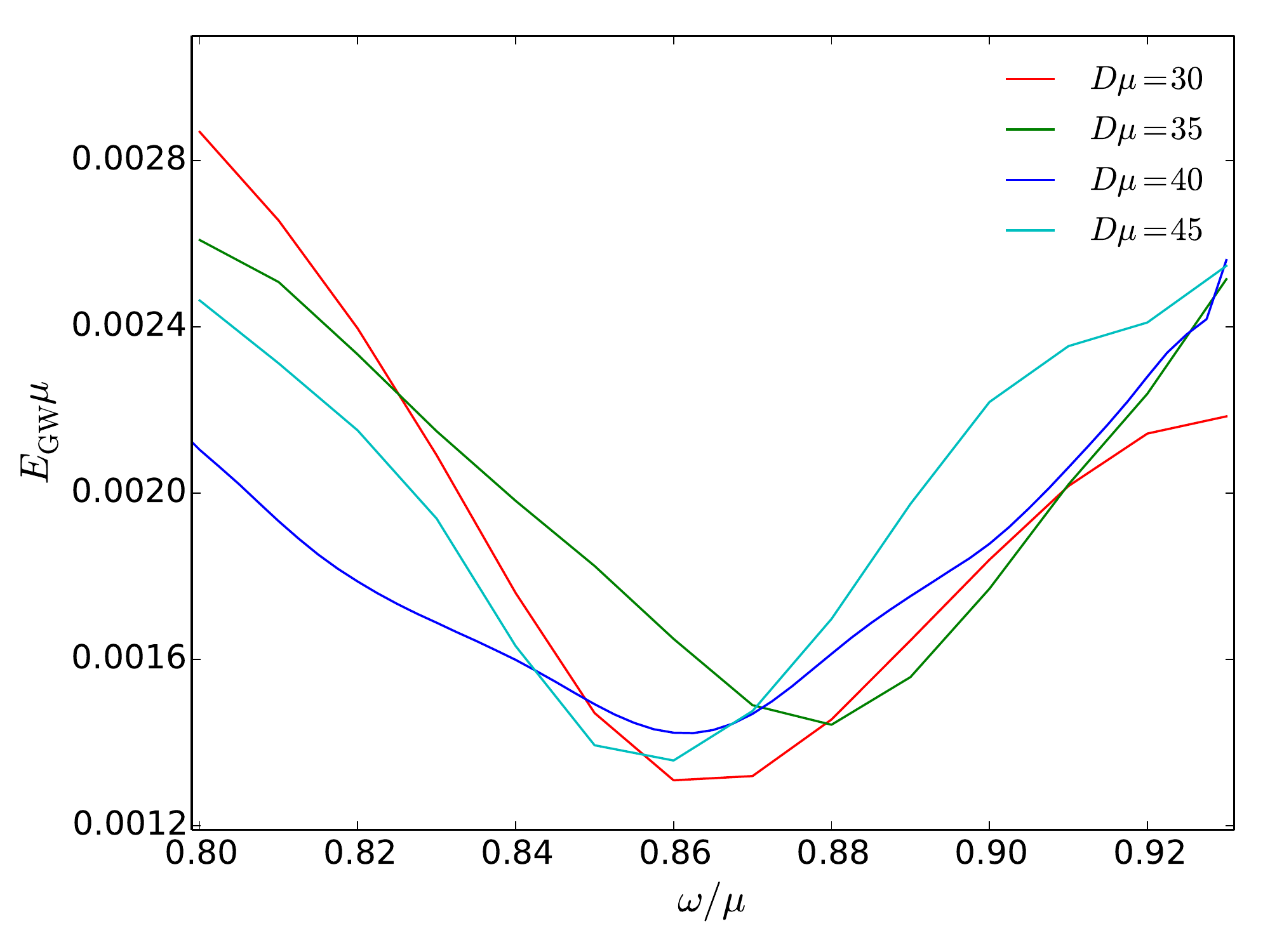}\\
\includegraphics[width=0.98\linewidth]{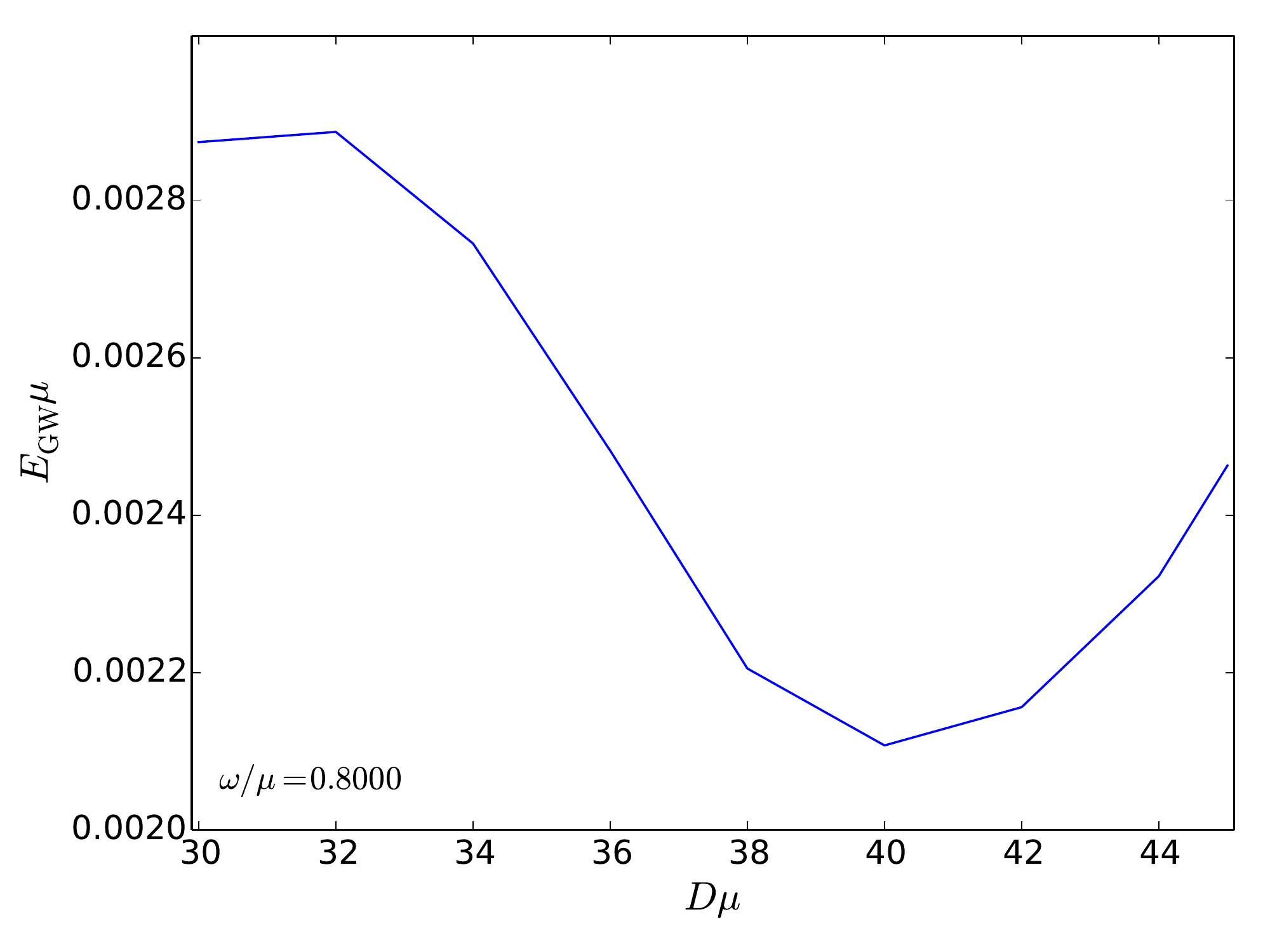}
\caption{Top panel: GW energy as a function of frequency for equal-mass head-on collisions for four different initial distances $D\mu=[30,35,40,45]$. Bottom panel: GW energy as a function of the initial distance $D\mu$ for the equal-mass case with $\omega/\mu=0.8000$.}
\label{figDistances2}
\end{figure}
Gravitational radiation greatly depends on the distribution and amplitude of the energy density. The square of the amplitude of the Proca field is proportional to the energy density (see Eq.~(\ref{tmunu})) and can be related to the amplitude of the GW emission. To illustrate this, Fig.~\ref{figGWenergy2} shows the total energy emitted for the models with fixed $\omega_1/\mu=0.8700$ computed from the simulations together with the estimated value of the Proca field amplitude from Eq.~(\ref{amplitude}) and a variation of Eq.~(\ref{amplitude}) as a function of $\omega_2/\mu$ at $t_{\rm{col}}=210$. Removing the drift of the energy due to dynamics and variations in the total mass, we find an excellent overall agreement, in particular in the location of the maxima and minima. We note that we do not find null GW emission when $\Delta\omega/\mu=(2k+1)0.015$, probably because there is no perfect cancellation of the Proca field during the whole merger process. We stress that while this linear argument is a remarkably good approximation, it is not really valid to explain a complete destructive interference of the stars. 

\subsection{The role of the initial relative distance}

We now explore the impact of (implicitly) varying the relative phase at merger by changing the time of the collision $t_{\rm{col}}\mu$. To this end, we place the stars at two additional initial separations, namely $D\mu=30$ and 45. We repeat the simulations with these setups for the cases of binaries with fixed primary frequency $\omega_1/\mu = 0.8000$ and secondary frequency in the interval $\omega_2/\mu\in [0.8000,0.9300]$  with variations in steps $\Delta\omega_2/\mu=0.0025$. Our results are shown in Fig.~\ref{figDistances1}. The top left panel corresponds to the energy radiated in GWs. This exhibits the same global decreasing trend and periodic oscillations with local maxima and minima as a function of $\omega_2/\mu$ for all values of the initial separation distances. However, $\Delta\omega_{\rm{max}}/\mu $ and $\Delta \omega_{\rm{min}}/\mu$ are found to depend on $D\mu$ (and $t_{\rm{col}}\mu$). The new collision times are $t_{\rm{col}}^{D\mu=30}\mu\sim135$ and $t_{\rm{col}}^{D\mu=45}\mu\sim250$, and from the analysis of the top left panel of Fig.~\ref{figDistances1} we obtain $\Delta \omega_{\rm{max}}^{D\mu=30}/\mu\sim0.046$ and $\Delta \omega_{\rm{max}}^{D\mu=45}/\mu\sim0.025$. These values are exactly what we expect from Eqs.~(\ref{amplitude})-(\ref{minimum}) for such collision times. 

In addition, the top right panel of Fig.~\ref{figDistances1} shows the GW energy emitted by an unequal-mass binary with $\omega_1/\mu=0.8000$ and $\omega_2/\mu=0.8450$ as a function of the initial separation. The GW energy does not depend monotonically with the distance but instead it displays an oscillatory pattern. Moreover, the bottom panels show the $l=m=2$ gravitational waveforms for two unequal-mass cases and three initial separations. These two plots illustrate that the initial distance is an important parameter of the system as it can change the morphology and energy of the emitted GWs for the same binary stars.

\subsection{The role of the initial phases}

The fact that the initial separation plays an important role in the dynamics and interactions of the two PSs raises the question of whether the initial phase of the stars may also cause a similar effect. Note that we keep the same phase {\it for both stars} (zero initial phase difference), as we have focused in the simplest possible scenario. Recall that while the energy density of PSs is axisymmetric, their real and imaginary parts are not. Therefore, different phases lead to different orientations of the real and imaginary parts at the time of the collision, which in turn yields different results that could potentially reveal the inner complex structure of these stars (for instance, the dipolar distribution of the real and imaginary parts of the Proca field for a $m=1$ spinning star). To test this idea, we perform several simulations of a binary with $\omega_1/\mu=0.8000$ and $\omega_2/\mu=0.8450$ varying the initial phase $\epsilon$ in Eq.~(\ref{Paxial}).

To check that the key parameter at play is the \textit{relative phase} of the stars and not their global ones, we first vary the phase of both stars, keeping always the phase difference equal to zero $\Delta\epsilon=0$ with $\epsilon_1=\epsilon_2$. Fig.~\ref{figphases} shows the time evolution of the energy density (leftmost column) and the real part of the scalar potential $\mathcal{X}_\phi$ for different values of the phase $\epsilon=\lbrace0,\pi/4,\pi/2\rbrace$ (remaining columns). 
The first column shows that even when the orientation of the components of the Proca field (in this case the scalar potential) is different, there is no change at the level of the energy density. No differences are found in the dynamics of the binary, the final object, or the gravitational waveform. These are all completely independent of the initial phase. Therefore, the inner structure and dipolar distribution ($\bar m=1$) of the real and imaginary parts of the star do not play a role in the collisions. 
We note that the real part of the scalar potential shows a $\bar m=5$ distribution after the collapse and black hole formation (as discussed in~\cite{sanchis2020synchronized}; see also~\cite{Bezares:2022obu}) that could trigger the development of the superradiant instability depending on the final spin of the black hole. However, this would happen within a timescale beyond current computational capabilities.

Next, we study the effect of varying the initial separation in the equal-mass case. The top panel of Fig.~\ref{figDistances2} shows the total GW energy emitted for an equal-mass PS head-on collision, as a function of the stars frequency and for four initial distances, $D\mu=\lbrace30,35,40,45\rbrace$. Correspondingly, the bottom panel of Fig.~\ref{figDistances2} exhibits the GW energy as a function of distance for an equal-mass binary with $\omega/\mu=0.8000$. In both cases we observe the same trend discussed in the top left panel of Fig.~\ref{figGWenergy}, together with the corresponding oscillatory pattern for fixed $\omega/\mu$. We note that, unlike the unequal-mass case, the top panel of Fig.~\ref{figDistances2} lacks the maxima and minima arising from the constructive and destructive interferences, as in the equal-mass case we always have $\omega_1=\omega_2$. 

\begin{figure}[t!]
\includegraphics[width=1\linewidth]{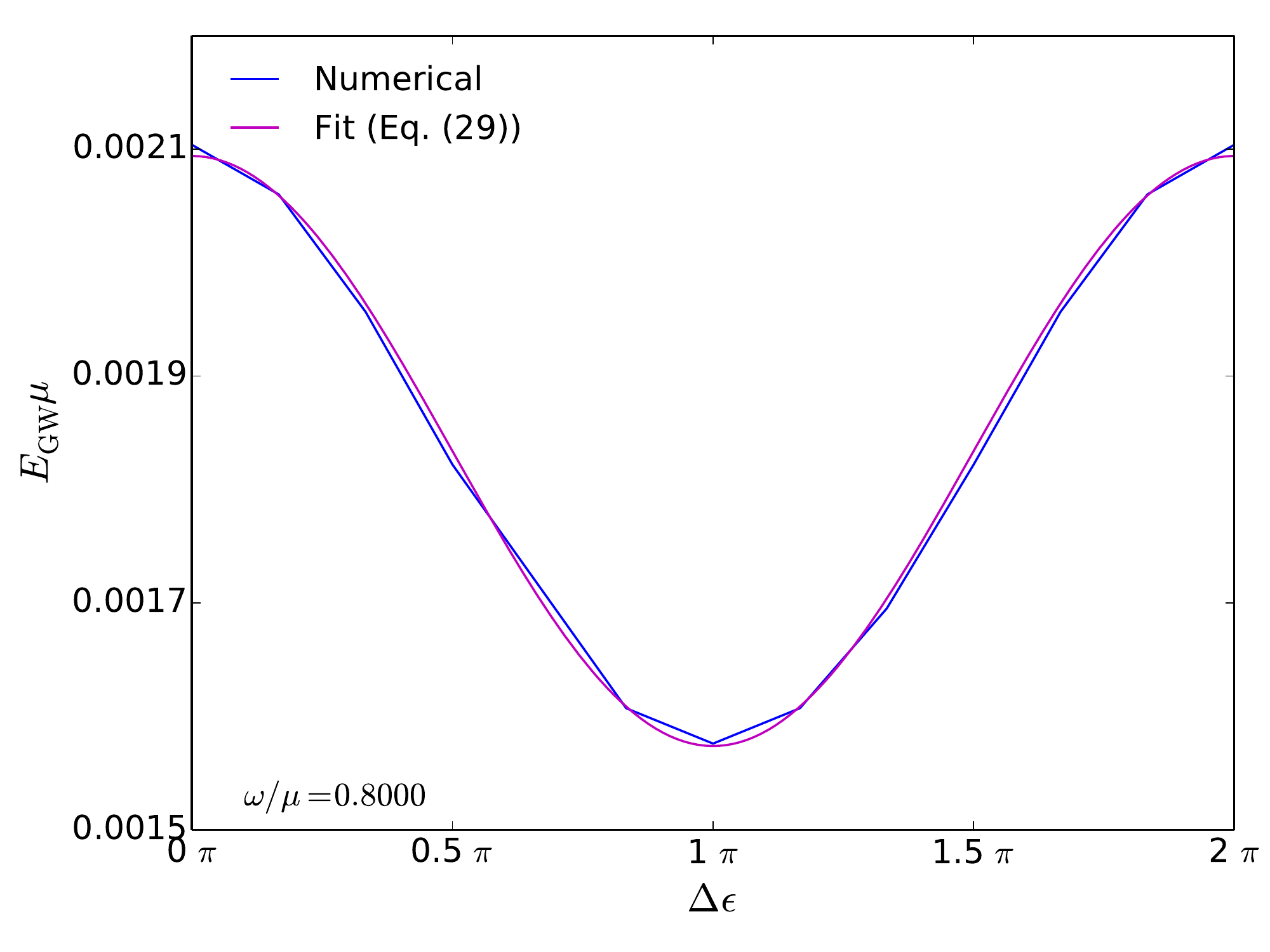}
\includegraphics[width=1\linewidth]{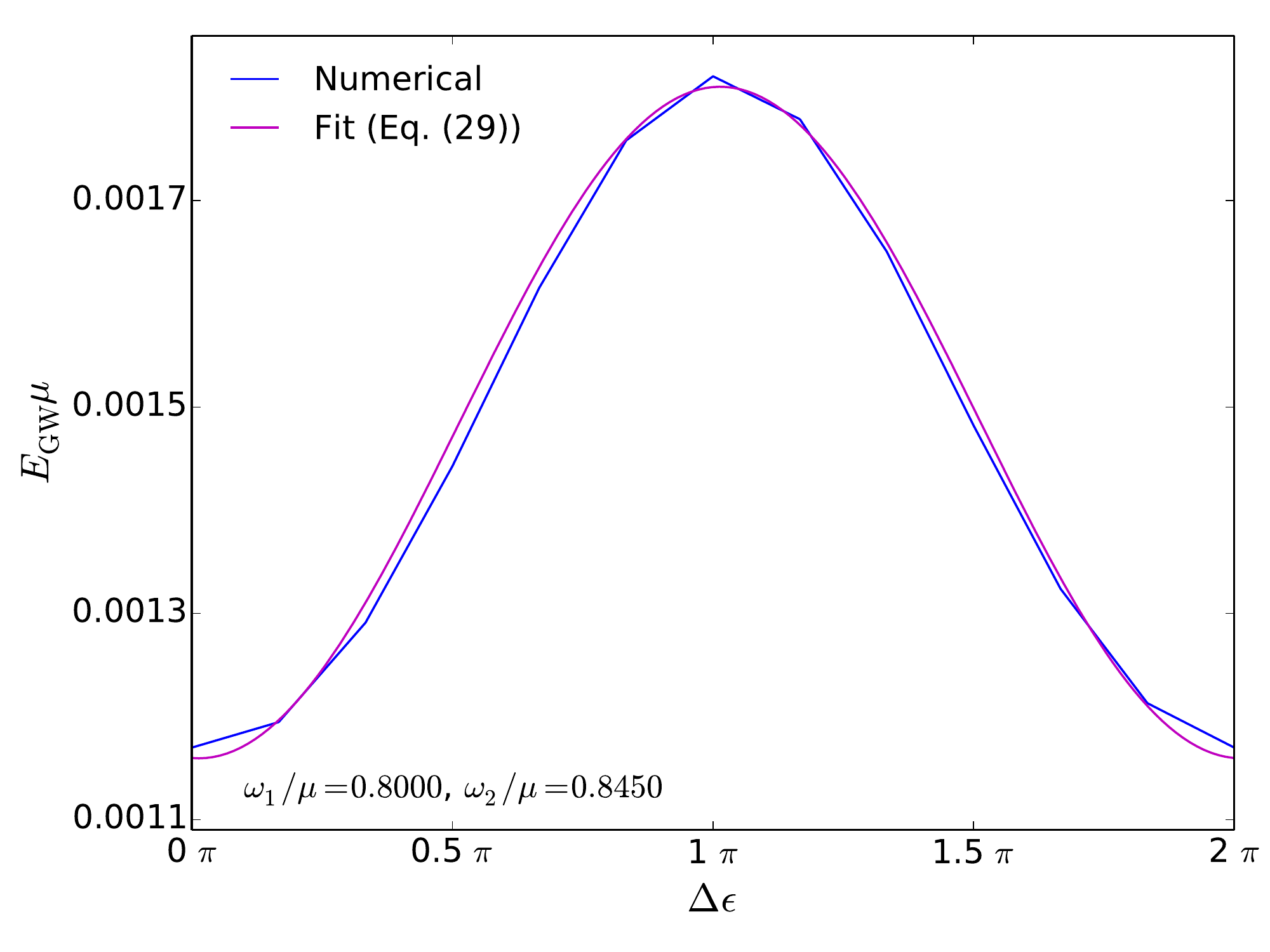}
\caption{Top panel: GW energy for head-on collisions with fixed $\omega/\mu=0.8000$ and non-zero phase difference $\Delta\epsilon$ between the stars. The magenta line depicts the behaviour of the square of the Proca field amplitude with $\Delta\epsilon$ as computed from Eq.~(\ref{phase}). Bottom panel: Same as the top panel but for the unequal-mass case with $\omega_1/\mu=0.8000$ and $\omega_{2}/\mu=0.8450$.}
\label{figGWenergy3}
\end{figure}

Finally, we explore how the relative phase $\Delta\epsilon=|\epsilon_1-\epsilon_2|$ impacts the GW energy. Again, even if we change the initial phase, the initial energy density of the stars is independent of the phase. However, the relative phase will change the interference pattern and the dynamics of the Proca field at the time of the collision. From the amplitude of the Proca field
\begin{eqnarray}\label{phase}
|\mathcal{A}|^2&=&{\rm Re}(\mathcal{A})^2+{\rm Im}(\mathcal{A})^2\sim4\cos^2\biggl(\frac{(\omega_1-\omega_2)}{2}t + \Delta\epsilon\biggl)\nonumber\\
&=&2\bigl[1+\cos\bigl((\omega_1-\omega_2)t+\Delta\epsilon\bigl)\bigl],
\end{eqnarray}
we see that varying $\Delta\epsilon$ produces a similar effect to changing the initial distance separation (and $t_{\rm{col}}$). The stars merge with a different internal configuration producing a different GW emission. This is indeed what we get as shown in Fig.~\ref{figGWenergy3} where we plot the GW energy for one equal-mass case ($\omega/\mu=0.8000$) and one unequal-mass case  ($\omega_1/\mu=0.8000$, $\omega_2/\mu=0.8450$) together with the analytical fit from Eq.~(\ref{phase}) taking into account that there is no perfect destructive interference that would lead to zero emission. Compared to the $\Delta\epsilon=0$ situation, now the most luminous collision emits about 25\% more energy in the form of GWs in the equal-mass case and about 35\% more in the unequal-mass case. 

\begin{figure*}
\includegraphics[width=0.50\linewidth]{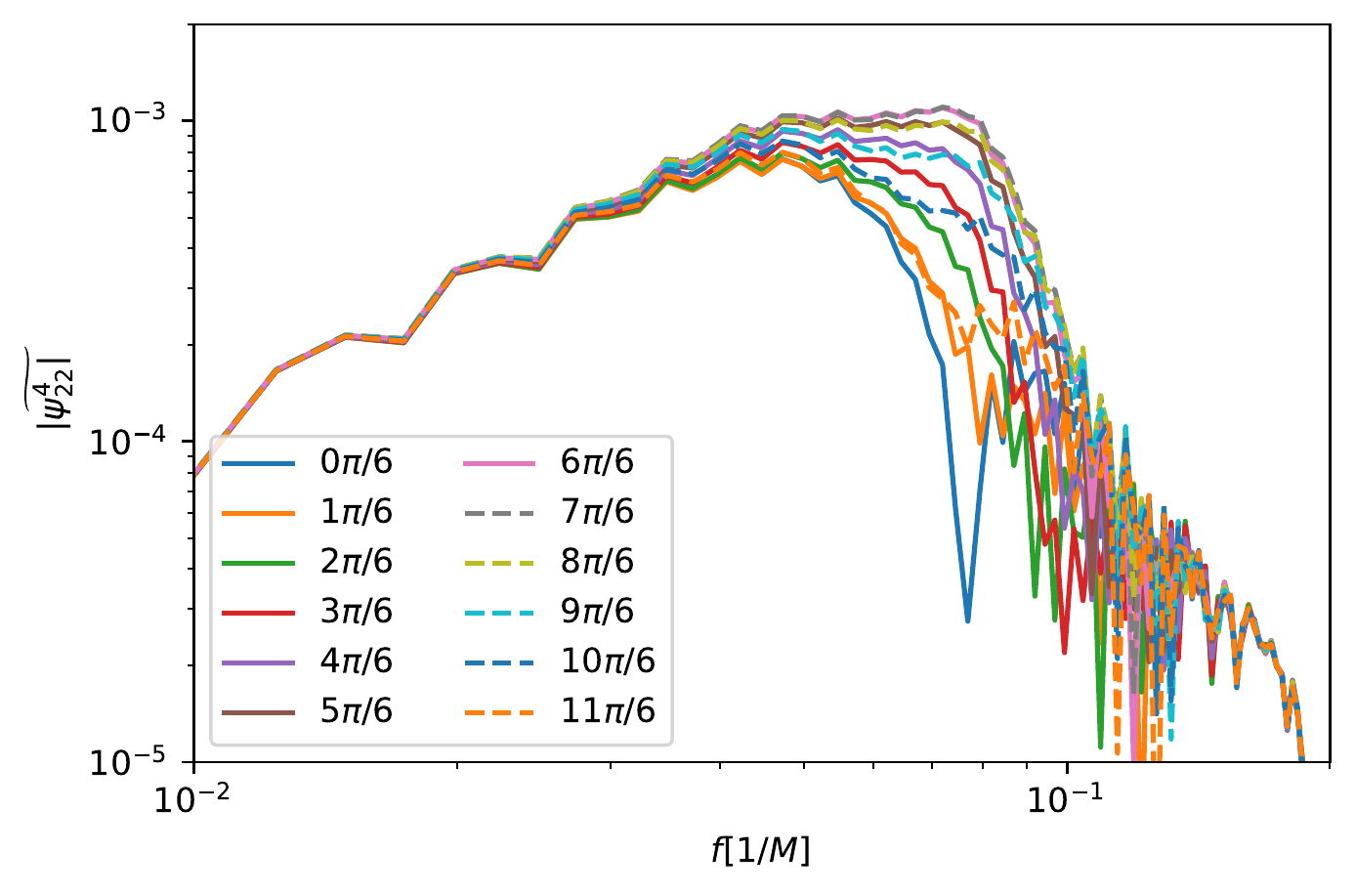}
\includegraphics[width=0.49\linewidth]{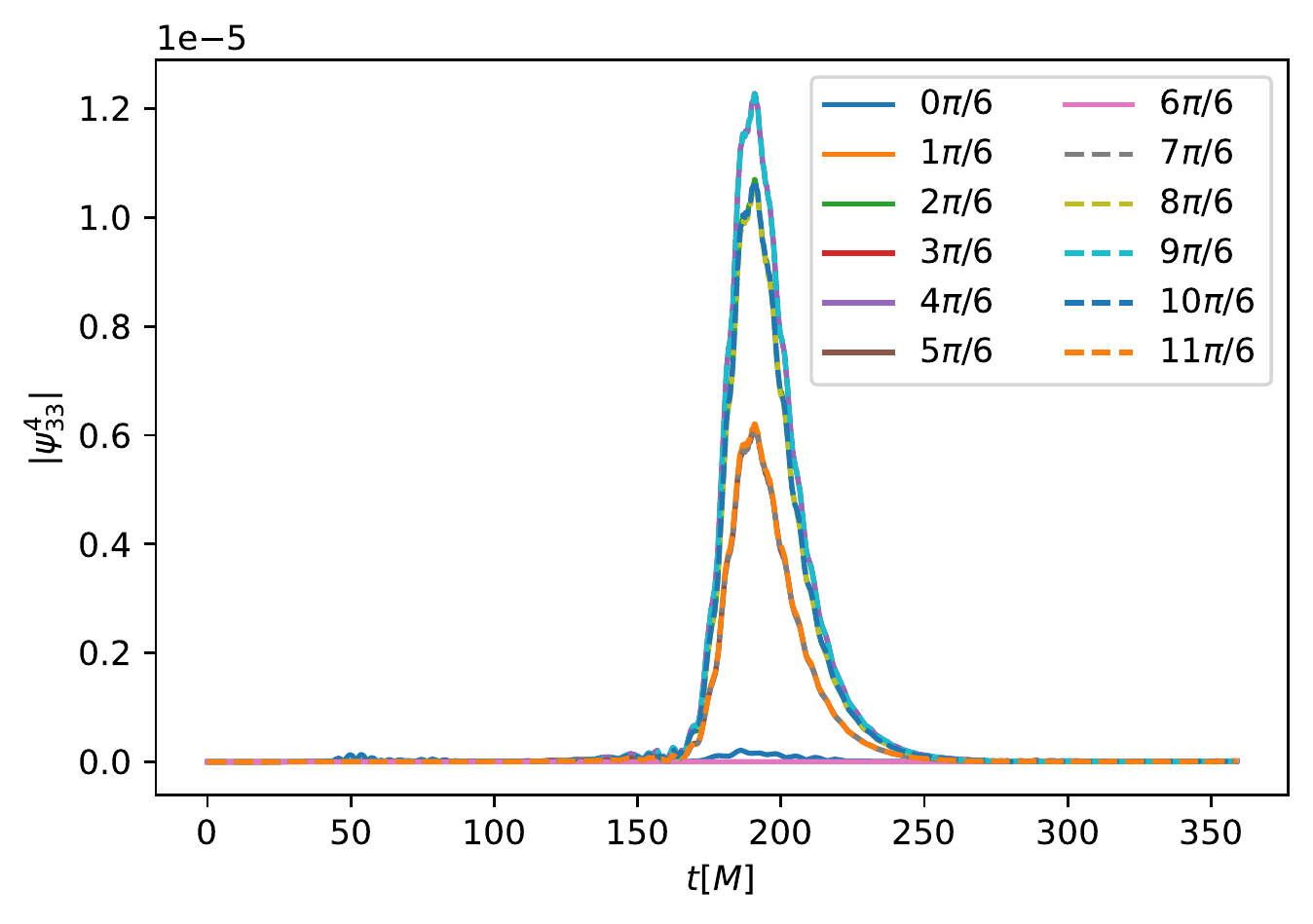}
\caption{Left panel: Absolute value of the Fourier transform of the quadrupole mode of $\Psi_4$ for an unequal-mass PS head-on collision as a function of the relative phase $\Delta \epsilon$ of the two stars. A clear dependence of both the mode amplitude and frequency content on $\Delta \epsilon$ is observed. Right panel: Absolute value of the $\ell=m=3$ mode of $\Psi_4$ for an equal-mass collision as a function of $\Delta \epsilon$. When $\Delta \epsilon = 0$ this mode is completely suppressed (as in the black-hole merger case for equal masses and aligned spins) while variations of this parameter trigger this mode during merger and ringdown. }
\label{figmodesphase}
\end{figure*}

The relative phase $\Delta \epsilon$ also alters the mode-emission structure of the source and the frequency content of the modes (or equivalently, their morphology). In particular, the left panel of Fig.~\ref{figmodesphase} shows the frequency content, by means of the amplitude of the Fourier transform, of the quadrupole $\ell=m=2$ mode of an unequal-mass PS merger as a function of $\Delta \epsilon$. It can be noted how variations of this parameter have an influence not only on the amplitude of the mode, therefore impacting the observability of the source, but also greatly modify its frequency content. This  suggests that this effect (or rather the parameter $\Delta \epsilon$) could actually be measurable in a Bayesian parameter inference framework.

The effect of $\Delta \epsilon$ in equal-mass mergers is particularly useful to understand the potential impact of this parameter in GW data analysis as a possible smoking gun to distinguish PS mergers from vanilla black-hole mergers (equal masses and aligned -- or zero -- spins). In this situation, for the case of black-hole mergers, odd-$m$ emission modes are exactly suppressed due the symmetry of the source. The same is true, as expected, for the case of PSs when we set $\Delta \epsilon = 0$. The right panel of Fig.~\ref{figmodesphase}, however, shows that the introduction of $\Delta \epsilon \neq 0$ activates the $\ell=m=3$ mode during merger and ringdown. This reflects the fact that the phase difference between the stars breaks the symmetry of the source. While at the moment we cannot perform simulations for the case of quasi-circular PS mergers (for lack of constraint-satisfying initial data) we anticipate that this effect would lead 
to an inconsistency between the binary parameters inferred from the inspiral stage and the corresponding ringdown emission modes of the final black hole if the source were assumed at face value to be a black-hole merger. Moreover, such a signature shall represent a smoking gun of the non-black-hole nature of the merging objects. We leave the quantitative exploration of this possibility for future work.

\section{Conclusions}
\label{Sect:conclusions}

Black holes and neutron stars are widely considered the most plausible compact objects populating the Universe. Theoretical proposals for other types of compact objects, dubbed dark or “exotic” compact objects, however, have also been proposed (see e.g.~\cite{Cardoso:2019} and references therein). The brand new field of gravitational-wave astronomy offers, potentially, the intriguing opportunity to probe those theoretical proposals. In particular the study and characterization of the GWs from collisions of ECOs -- the building of waveform template banks -- seems a key requisite towards that goal, as those datasets could allow for direct comparisons with the signals produced in mergers of black holes and neutron stars. The expectation is that the distinct nature of the different families of compact objects is somewhat encoded in the GW signals each member of the class emits, hence offering a way to single them out. In order to identify the specific and subtle signatures of each type of object in their GW emission, it is crucial to produce accurate signal models that can be compared to the data collected by detectors and that can also reveal new specific phenomenology. Presently, numerical relativity offers the most accurate way to do so, particularly in the highly non-linear, strong-gravity situations produced when two compact objects merge. 

In this paper we have presented a catalogue of nearly 800 simulations of head-on mergers of PSs. We recently used this dataset to search for signatures of these objects in existing LIGO-Virgo data~\cite{CalderonBustillo:2020srq,psi4_obs}. Here, we have performed a systematic study of the properties and gravitational-wave emission of these physical systems. Our study has revealed that the relative phase of the two PSs, an intrinsic parameter of bosonic stars that is absent for the case of black-hole mergers, has a strong impact in the GW emission. This parameter, which reflects the wave-like nature of the PSs by controlling the way the Proca field interacts with itself, impacts not only the amplitude of the emission modes (and therefore the total emitted energy) but also the frequency content of the signal and its mode structure. Interestingly, these findings suggest that such an intrinsic parameter of PS binaries could be measurable. As a particular illustration, we have shown here that the asymmetry induced by phase differences in an equal-mass PS head-on collision can trigger odd-parity (odd-$m$) modes during the merger-ringdown stage which are completely suppressed for the case of equal-mass (and equal-spin) binary black hole mergers. We argue that this may evidence the non-black-hole nature of the merging objects.

The LVK event GW190521 has represented the first example of a GW signal that can be explained both in the classic framework of binary black hole mergers {\it and} in the less-common framework of PS mergers~\cite{CalderonBustillo:2020srq}. However, to conclusively probe the existence of the latter class of ECOs will require either the accumulation of small evidences in favour of this scenario through the systematic comparison of signals to waveform catalogs and/or the observation of a signal with distinct signatures that cannot be reproduced by black-hole mergers by current or future LIGO-Virgo-KAGRA detectors or by third-generation detectors, such as the Einstein Telescope~\cite{Hild:2011}. On the one hand, the GW catalogue we have discussed in this paper represents the first step towards such systematic comparisons. On the other hand, our results suggest that the wave-like nature of PSs, via the impact of the relative phase parameter $\Delta \epsilon$ on the GW emission, might serve as a distinct smoking gun for the existence of these objects. 

In this work we have focused on the particular case of head-on collisions due its technical and computational simplicity. In the future we plan to extend the catalogue to eccentric and orbital quasi-circular mergers of bosonic stars. This will help us to firmly establish if the GW interference patterns found are specific to or can be amplified by the geometry of the collisions considered in this paper, and, thus, gauge the potential imprint they may actually have in the GW emission.

\section*{Acknowledgements}

This work was supported by the Center for Research and Development in Mathematics and Applications (CIDMA) through the Portuguese Foundation for Science and Technology (FCT - Funda\c c\~ao para a Ci\^encia e a Tecnologia), references UIDB/04106/2020 and UIDP/04106/2020, by national funds (OE), through FCT, I.P., in the scope of the framework contract foreseen in the numbers 4, 5 and 6 of the article 23, of the Decree-Law 57/2016, of August 29, changed by Law 57/2017, of July 19 and by the projects PTDC/FIS-OUT/28407/2017,  CERN/FIS-PAR/0027/2019, PTDC/FIS-AST/3041/2020 and
CERN/FIS- PAR/0024/2021. This work has further been supported by he Spanish Agencia Estatal de Investigaci\'on (PGC2018-095984-B-I00), by the Generalitat Valenciana (PROMETEO/2019/071), and by  the  European  Union's  Horizon  2020  research  and  innovation  (RISE) programme H2020-MSCA-RISE-2017 Grant No.~FunFiCO-777740. N.S.G. acknowledges financial support by the Spanish Ministerio de Universidades, through a Mar\'ia Zambrano grant (ZA21-031) with reference UP2021-044, within the European Union-Next Generation EU. J.C.B. is supported by a fellowship from “la Caixa” Foundation (ID 100010434) and from the European Union’s Horizon 2020 research and innovation programme under the Marie Sklodowska-Curie grant agreement No 847648. The fellowship code is LCF/BQ/PI20/11760016. J.C.B. is also supported by the research grant PID2020-118635GB-I00 from the Spain-Ministerio de Ciencia e Innovaci\'on. Computations have been performed at the Servei d'Inform\`atica de la Universitat de Val\`encia and the Argus and Blafis cluster at the U. Aveiro. This manuscript has ET preprint number ET-0181A-22. 

\bigskip

\begin{appendix}
\section{Code assessment}\label{appendix}
\begin{figure}
\centering
\includegraphics[height=2.45in]{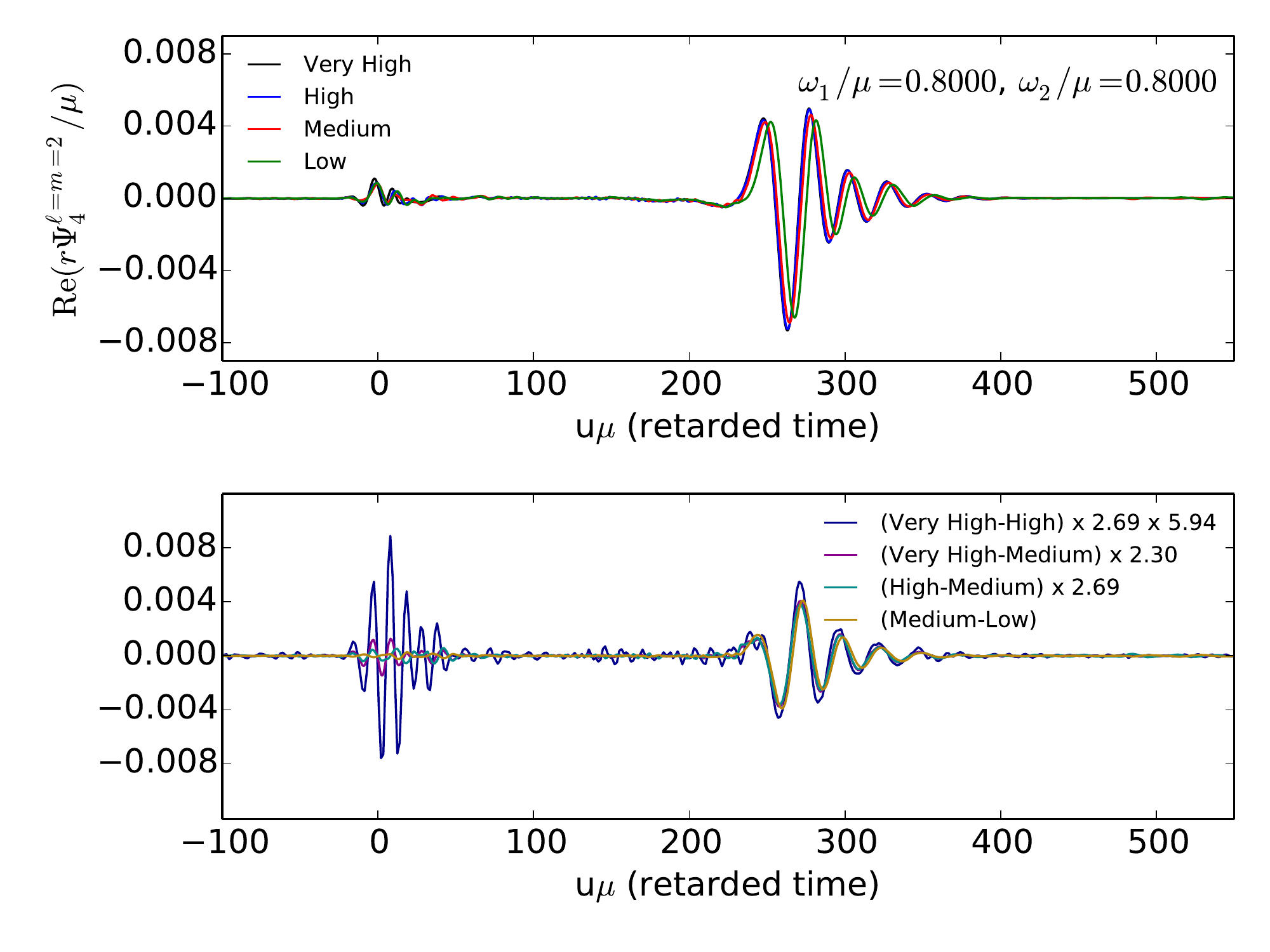}
\includegraphics[height=2.45in]{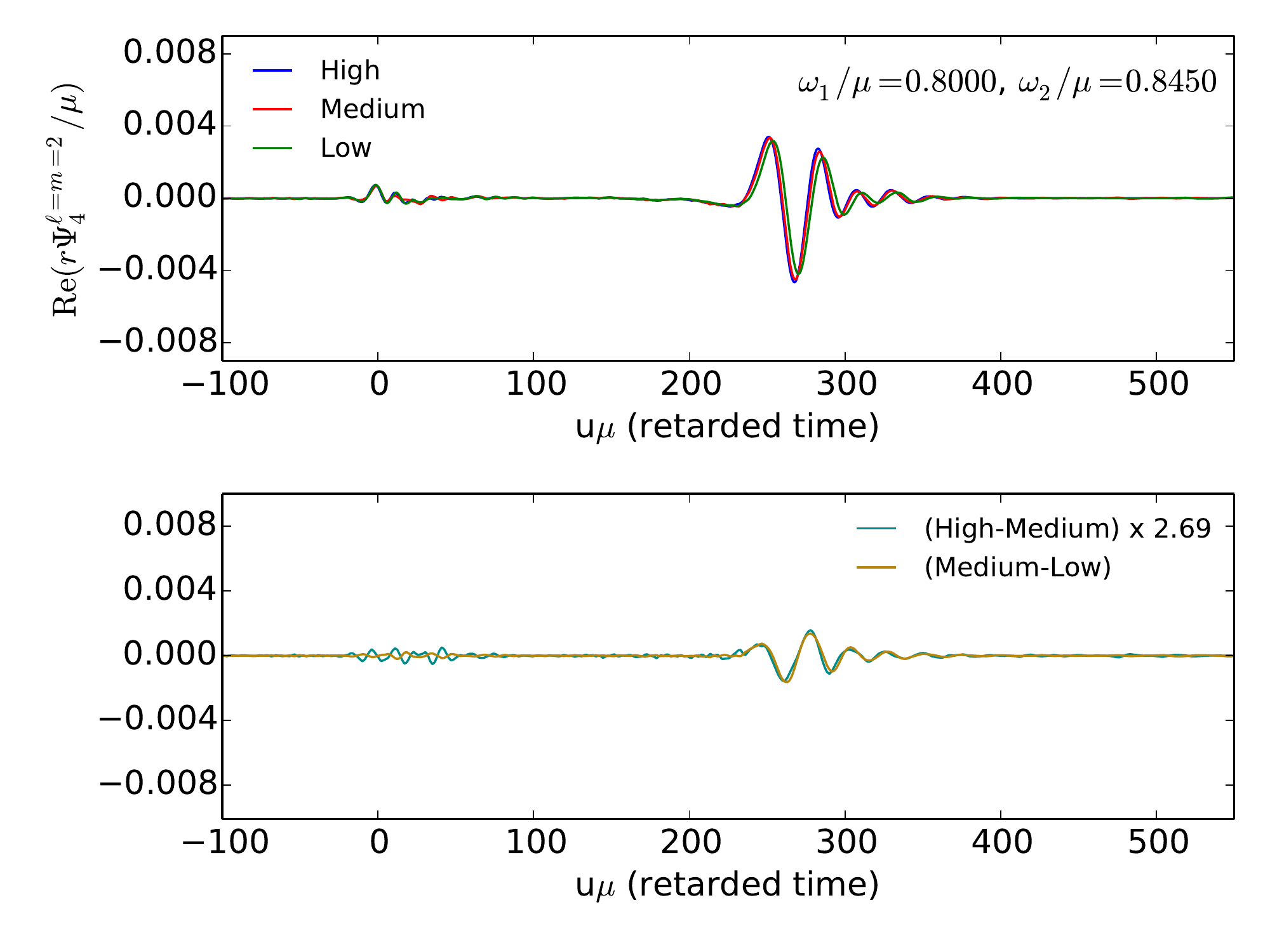}
\caption{Convergence study using for the equal-mass case with $\omega_1/\mu=\omega_2/\mu=0.8000$ and the unequal-mass case with $\omega_1/\mu=0.8000$ and $\omega_2/\mu=0.8450$. The first and third panels show the respective $r\Psi_{4}^{\ell=m=2}$ for different simulation resolution levels. The second and fourth panels show difference between different resolutions scaled for fourth-order convergence.}
\label{figGWconvergence}
\end{figure}

We briefly comment here on the convergence analysis we carried out to assess the quality of our simulations. In Fig.~\ref{figGWconvergence} we plot the gravitational wave from an equal-mass ($\omega_1/\mu=\omega_2/\mu=0.8000$) and an unequal-mass ($\omega_1/\mu=0.8000$ and $\omega_2/\mu=0.8450$) collisions using four different resolutions with ($dx=\lbrace0.046875,0.0625,0.09375,0.125\rbrace/\mu$) in the finest level. We obtain fourth-order convergence. { The initial transient is due to spurious radiation in the initial data, which is not constraint-satisfying, and, therefore, does not converge with resolution.}

\end{appendix}

\bibliography{num-rel2}

\end{document}